\documentclass[smallextended,reqno]{amsart}

\newcommand{\helpers}{helpers}
\usepackage{\helpers/mlmc_style}

\newcommand{\plotBenchmarkingSeries}[4]
{	
	\begin{figure}[H]
		\newcommand{\stdfile}{#1/std_#2.csv}
		\newcommand{\wtdfile}{#1/wtd_#2.csv}
		\centering
		\setlength\fboxrule{3pt}
		\begin{tikzpicture}[scale=0.95]
			\begin{groupplot}[group style={group size=2 by 2, horizontal sep=2.0cm, vertical sep=1.5cm}, height=5.0cm, width=7.0cm, yticklabel style={/pgf/number format/fixed}, scaled y ticks=false]
				\nextgroupplot[tick pos=left, xlabel=strikes, ylabel=prices, legend pos=north east, legend style={font=\scriptsize}]
				\addplot[color=blue, mark=asterisk, dashdotted] table[x=strikes, y=ctmcPrice, col sep=comma]{\stdfile}; \addlegendentry{CTMC}
				\addplot[color=cyan, mark=+, dashed, skip coords between index={0}{1}] table[x=strikes, y=seriesPrice, col sep=comma]{\stdfile}; \addlegendentry{Series}
				
				\nextgroupplot[tick pos=left, xlabel=strikes, ylabel=prices, legend pos=north east, legend style={font=\scriptsize}]	          					   
				\addplot[color=blue, mark=asterisk, densely dashed, opacity=0.8] table[x=strikes, y=ctmcPrice, col sep=comma]{\wtdfile}; \addlegendentry{CTMC}
				\addplot[color=cyan, mark=+, loosely dashed, skip coords between index={0}{1}] table[x=strikes, y=seriesPrice, col sep=comma]{\wtdfile}; \addlegendentry{Series}			
				
				\nextgroupplot[tick pos=left, xlabel=strikes, ylabel=error (bps), legend pos=south east, legend style={font=\scriptsize}, 
				ymin=-1.5, ymax=1.5]	          					   
				\addplot[color=red, mark=asterisk, dashed]  table[x=strikes, y expr=10000*(\thisrow{ctmcPrice} - \thisrow{seriesPrice}), col sep=comma]{\stdfile}; \addlegendentry{Difference}
				\addplot[name path=stddevup,  color=green!50, forget plot] table[x=strikes, y expr =10000*2.576*(\thisrow{ctmcMcStddev} + \thisrow{seriesMcStddev}), col sep=comma]{\stdfile};  
				\addplot[name path=stddevdown,color=green!50, forget plot] table[x=strikes, y expr=-10000*2.576*(\thisrow{ctmcMcStddev} + \thisrow{seriesMcStddev}), col sep=comma]{\stdfile};			
				\addplot[green!30,fill opacity=0.3] fill between[of=stddevup and stddevdown];
				\addlegendentry{$99\%$ CI};	
				
				\nextgroupplot[tick pos=left, xlabel=strikes, ylabel=error (bps), legend pos=south east, legend style={font=\scriptsize}, 
				ymin=-1.5, ymax=1.5]	          					   
				\addplot[color=red, mark=asterisk, densely dashed, opacity=0.8] table[x=strikes, y expr=10000*(\thisrow{ctmcPrice} - \thisrow{seriesPrice}), col sep=comma]{\wtdfile}; \addlegendentry{Difference}
				\addplot[name path=stddevup,  color=green!50, forget plot] table[x=strikes, y expr =10000*2.576*(\thisrow{ctmcMcStddev} + \thisrow{seriesMcStddev}), col sep=comma]{\wtdfile};  
				\addplot[name path=stddevdown,color=green!50, forget plot] table[x=strikes, y expr=-10000*2.576*(\thisrow{ctmcMcStddev} + \thisrow{seriesMcStddev}), col sep=comma]{\wtdfile};			
				\addplot[green!30,fill opacity=0.3] fill between[of=stddevup and stddevdown];
				\addlegendentry{$99\%$ CI};	
			\end{groupplot}		
		\end{tikzpicture}
		\caption{#3 option with strong (left graphs) and weak (right graphs) tail dependences}\label{#4}
	\end{figure}
}

\makeatletter
\define@cmdkey[helper]{plot}{pathconvergence}{}
\define@cmdkey[helper]{plot}{pathapplied}{}
\define@cmdkey[helper]{plot}{modelname}{}
\define@cmdkey[helper]{plot}{xmin}{}
\define@cmdkey[helper]{plot}{xmax}{}
\define@cmdkey[helper]{plot}{ymin}{}
\define@cmdkey[helper]{plot}{ymax}{}
\define@cmdkey[helper]{plot}{bg}{}
\define@cmdkey[helper]{plot}{thecaption}{}
\define@cmdkey[helper]{plot}{label}{}

\pgfset{
	foreach/parallel foreach/.style args={#1in#2via#3}{evaluate=#3 as #1 using {{#2}[#3-1]}},
}

\pgfplotsset{
	discard if not in/.style 2 args={
		x filter/.append code={
			\readlist\mylist{#2}%
			\providecommand{\foundit}{0}
			\renewcommand{\foundit}{0}
			\foreachitem\z\in\mylist[]{%
				\ifdim\thisrow{#1} pt=\z pt
				\renewcommand{\foundit}{1}
				\else				
				\fi
			}
			\pgfmathparse{\foundit == 1? \pgfmathresult : nan}
		}
	},
}

\newcommand{\newplotmlmc}[2]{
	\setkeys[helper]{plot}{#1}
	\def\modelname{\cmdhelper@plot@modelname}
	\def\pathtoconvergence{\cmdhelper@plot@pathconvergence}
	\def\pathtoapplied{\cmdhelper@plot@pathapplied}
	\def\bg{\cmdhelper@plot@bg}{}
	\def\xmin{\cmdhelper@plot@xmin}
	\def\xmax{\cmdhelper@plot@xmax}
	\def\ymin{\cmdhelper@plot@ymin}
	\def\ymax{\cmdhelper@plot@ymax}
	\def\thecaption{\cmdhelper@plot@thecaption}
	\def\figLabel{\cmdhelper@plot@label}
				
	\def\mlmcpathconvergence{\pathtoconvergence/\modelname.csv}
	\def\pathnls{\pathtoapplied/\modelname_nls.csv}
	\def\pathcosts{\pathtoapplied/\modelname_costs.csv}
	\def\costslope{\pgfmathparse{\bg <= 1 ? 0 : -4 * (\bg - 1) / (2 - \bg)}\pgfmathprintnumber[fixed, fixed zerofill, precision=1]{\pgfmathresult}}
	
	\def\mycolours{"blue","red","RubineRed","blue","red","RubineRed","blue","red","RubineRed","blue","red","RubineRed","blue","red","RubineRed","blue","red","RubineRed"}%
	\def\mylinestyles{"loosely dashed","dashed","loosely dotted","dotted","dashdotted","dashdotted","loosely dashed","dashed","loosely dotted","dotted","dashdotted","dashdotted","loosely dashed","dashed","loosely dotted","dotted","dashdotted","dashdotted"}%
	\begin{figure}[H]
		\centering
		\setlength\fboxrule{3pt}
		\begin{tikzpicture}
			\begin{groupplot}[group style={group size=2 by 2, horizontal sep=2.cm, vertical sep=1.3cm}, height=5.3cm, width=6.8cm] 
				\nextgroupplot[tick pos=left, xlabel=level $l$, ylabel=$log_2$ variance, legend pos=south west, legend style={font=\footnotesize, nodes={scale=0.65, transform shape}},
				xtick={0, 2, ..., 20},]	          					   
				\addplot[color=blue, mark=asterisk, only marks] table[x=level, y expr={log2(\thisrow{varlevell})}, col sep=comma]{\mlmcpathconvergence};                            \addlegendentry{$P_l$}
				\addplot[color=cyan, mark=asterisk, skip coords between index={0}{1}, only marks] table[x=level, y expr={log2(\thisrow{vl})}, col sep=comma]{\mlmcpathconvergence}; \addlegendentry{$P_l \!-\! P_{l-\!1}$}
				\addplot[color=red, dotted, skip coords between index={0}{1}] table[x=level, y expr={log2(\thisrow{vl_slope})}, col sep=comma]{\mlmcpathconvergence};\addlegendentry{$\propto$  \pgfmathparse{\bg - 2}\pgfmathprintnumber[fixed, fixed zerofill, precision=2]{\pgfmathresult}}
				
				\nextgroupplot[tick pos=left, xlabel=level $l$, ylabel=$log_2$ $|$mean$|$, legend pos=south west, legend style={font=\footnotesize, nodes={scale=0.65, transform shape}}, 
				xtick={0, 2, ..., 20},]]                
				\addplot[color=blue, mark=asterisk, only marks] table[x=level, y expr={log2(abs(\thisrow{meanlevell}))}, col sep=comma]{\mlmcpathconvergence};                           \addlegendentry{$P_l$}
				\addplot[color=cyan, mark=asterisk, only marks] table[x=level, y expr={log2(abs(\thisrow{ml}))}, col sep=comma, skip coords between index={0}{1}]{\mlmcpathconvergence}; \addlegendentry{$P_l - P_{l-1}$}
				\addplot[color=red, dotted, skip coords between index={0}{1}] table[x=level, y expr={log2(\thisrow{ml_slope})}, col sep=comma]{\mlmcpathconvergence};                    \addlegendentry{$\propto$ \pgfmathparse{\bg/2-1}\pgfmathprintnumber[fixed, fixed zerofill, precision=2]{\pgfmathresult}}

				\nextgroupplot[tick pos=left, xlabel=level $l$, ylabel=$N_l$, ymode=log, yminorticks=false, legend pos=north east, 
				               legend style={font=\footnotesize, nodes={scale=0.6, transform shape}}, 
							   xtick={0, 2, ..., 20},]							   
				\foreach \y [count=\c,
					parallel foreach=\myls in \mylinestyles via \c,
					parallel foreach=\myc in \mycolours via \c] 
					in {#2}{
						\edef\myplot{\noexpand\addplot[color=\myc, mark=asterisk, style=\myls] table[x=levels, y=\y, col sep=comma]{\pathnls}};
						\edef\mylegend{\noexpand\addlegendentry{$\epsilon$=\y}};
						\myplot;
						\mylegend;
				}
				\nextgroupplot[tick pos=left, ymode=log, yminorticks=false, xminorticks=false, xmode=log, ymin=\ymin, ymax=\ymax, xmin=\xmin, xmax=\xmax, xlabel=accuracy $\epsilon$, 
					ylabel=$\epsilon^2 cost$, legend pos=north east, 
					legend style={font=\footnotesize, 
						nodes={scale=0.8, transform shape}, 
					},
				]
				\addplot[color=red, mark=asterisk]	        table[x=rmse, y=costxrmse2stdmc, col sep=comma,	discard if not in={rmse}{#2}]{\pathcosts};  \addlegendentry{Std MC}
				\addplot[color=blue, mark=asterisk, dashed] table[x=rmse, y=costxrmse2, col sep=comma, discard if not in={rmse}{#2}]{\pathcosts};  		\addlegendentry{MLMC}
				\addplot[color=red, dotted]                 table[x=rmse, y=slope, col sep=comma, discard if not in={rmse}{#2}]{\pathcosts};  			\addlegendentry{$\propto\costslope$}
			\end{groupplot}
		\end{tikzpicture}
		\captionsetup{justification=centering}
		\caption{\thecaption}\label{\figLabel}
	\end{figure}	
}
\makeatother

\newcommand{\dataBenchmarkSeries}[1]
{
	\pgfplotstabletypeset[
		font=\footnotesize,
		columns = {strikes, ctmcPrice, seriesPrice, ctmcPriceNoCv, seriesPriceNoCv, ctmcMcStddev, seriesMcStddev, ctmcMcStddevNoCv, seriesMcStddevNoCv},
		col sep	= comma,
		every head row/.style={%
			before row={
				& \multicolumn{4}{c|}{Prices} & \multicolumn{4}{c}{MC Standard Errors} \\
				\hline
				& \multicolumn{2}{c|}{\textit{with control variates}} & \multicolumn{2}{c|}{\textit{without control variates}} & \multicolumn{2}{c|}{\textit{with control variates}} & \multicolumn{2}{c}{\textit{without control variates}}\\
			},
			after row=\hline
	},
	every last row/.style				= {after row=\hline},
	columns/strikes/.style				= {column name=\%Strike, column type=c, fixed, zerofill, precision=2},
	columns/ctmcPrice/.style			= {column name=CTMC, column type=|c, fixed, zerofill, precision=6},
	columns/seriesPrice/.style			= {column name=Series, column type=c, fixed, zerofill, precision=6},
	columns/ctmcPriceNoCv/.style		= {column name=CTMC, column type=|c, fixed, zerofill, precision=6},
	columns/seriesPriceNoCv/.style		= {column name=Series, column type=c|, fixed, zerofill, precision=6},
	columns/ctmcMcStddev/.style			= {column name=CTMC, column type=c, sci, sci zerofill, precision=1},
	columns/seriesMcStddev/.style		= {column name=Series, column type=c|, sci, sci zerofill, precision=1},
	columns/ctmcMcStddevNoCv/.style		= {column name=CTMC, column type=c, sci, sci zerofill, precision=1},
	columns/seriesMcStddevNoCv/.style	= {column name=Series, column type=c, sci, sci zerofill, precision=1},
	]{#1}
}

\usepackage{pifont}
\usepackage[margin=0.75in]{geometry}
\usepackage{listofitems}
\usepackage{tabularx}
\usepackage{flafter}

\numberwithin{equation}{section}

\theoremstyle{plain}
\newtheorem{theorem}{Theorem}[section]
\newtheorem{lemma}[theorem]{Lemma}
\newtheorem{proposition}[theorem]{Proposition}
\theoremstyle{remark}
\newtheorem{remark}{Remark}

\pgfplotsset{compat=newest}
\usepgfplotslibrary{fillbetween}
\usetikzlibrary{decorations.markings}
\usetikzlibrary{arrows.meta}

\usepackage{natbib}
\bibpunct[, ]{(}{)}{;}{a}{}{,}

\usepackage[many]{tcolorbox}
\newtcolorbox{cross}{blank,breakable,parbox=false,
	overlay={\draw[red,dashed,line width=1pt] (interior.south west)--(interior.north east);
		\draw[red,dashed,line width=1pt] (interior.north west)--(interior.south east);}}

\newcommand{\DataForPaper}{\helpers/results}
\newcommand{\pathgilesconvergence}[1]{\DataForPaper/giles_convergence/#1}
\newcommand{\pathgilesapplied}[1]{\DataForPaper/giles_applied/#1}

\def\cgmyO{CGMY\textsuperscript{(i)}}
\def\cgmyI{CGMY\textsuperscript{(ii)}}
\def\cgmyIV{CGMY\textsuperscript{(iii)}}
\def\cgmyVII{CGMY\textsuperscript{(iv)}}
\newcommand{\xmark}{\ding{55}}

\title[A Weak MLMC Scheme for L\'evy-copula-driven SDEs]{A Weak MLMC Scheme for L\'evy-copula-driven SDEs with Applications to the Pricing of Credit, Equity and Interest Rate Derivatives}

\author{Aleksandar Mijatovi\'c}
\address{Department of Statistics, University of Warwick and The Alan Turing Institute, UK}
\email{a.mijatovic@warwick.ac.uk}

\author{Romain Palfray}
\address{Department of Statistics, University of Warwick, UK}
\email{romain.palfray@warwick.ac.uk}
\thanks{We thank Krzysztof \L atuszy\'nski for his useful suggestions with regards to the binary search tree algorithm and Peter Tankov for sharing his code of the series representation algorithm with us.}

\subjclass[2000]{60G51}
\keywords{L\'evy-driven Stochastic Differential Equations (SDEs), L\'evy copula, continuous-time Markov chains, coupling of Markov chains, pricing of derivatives.}

\begin{document}

\begin{abstract}
	This paper develops a novel weak multilevel Monte-Carlo (MLMC) approximation scheme for L\'evy-driven Stochastic Differential Equations (SDEs). The scheme is based on the state space discretization (via a continuous-time Markov chain approximation \cite{MijatovicVidmarJacka12}) of the pure-jump component of the driving L\'evy process and is particularly suited if the multidimensional driver is given by a L\'evy copula. The multilevel version of the algorithm requires a new coupling of the approximate L\'evy drivers in the consecutive levels of the scheme, which is defined via a coupling of the corresponding Poisson point processes. The multilevel scheme is weak in the sense that the bound on the level variances is based on the coupling alone without requiring strong convergence. Moreover, the coupling is natural for the proposed discretization of jumps and is easy to simulate. The approximation scheme and its multilevel analogous are applied to examples taken from mathematical finance, including the pricing of credit, equity and interest rate derivatives.  
\end{abstract}
	
\maketitle

\section{Introduction}
\subsection{The problem}\label{Introduction:Problem}
L\'evy processes have applications in different areas: quantum physics, financial mathematics, biology, actuarial science, etc (see \cite{Applebaum2004} for some examples). In mathematical finance, they have been widely used to capture desirable characteristics of stock returns (namely fat-tailed and skewed distributions) and to model credit events and insurance risks. Merton was the first to propose a jump-diffusion model to price financial derivatives in the seminal paper \cite{Merton76} for stock returns. Numerous other models were subsequently proposed (for example \cite{Bates96}, \cite{VG}, \cite{CGMY02}, \cite{Kou12}, \cite{BarndorffNielsen07}, \cite{EberleinPrause13} and \cite{Schoutens19}), extending the scope to other types of L\'evy processes and financial underlyings.
The use of closed-form formulas for vanilla financial instruments is key to a fast and efficient calibration of the pricing model to market prices. Numerical methods such as the Laplace \cite{Raible00} and Fourier \cite{CarrMadamFFT} transforms or the COS method \cite{DBLP:journals/siamsc/FangO08} can be successfully applied for some vanilla payoffs, but the pricing of exotic products generally requires a more generic methodology such as Monte-Carlo methods. These problems are even more pronounced when dealing with general L\'evy-driven SDEs; therefore, it is crucial to be able to compute an expectation of the type \(\E{f(Y)}\) with Monte-Carlo methods, where \(Y\) is defined by a L\'evy-driven SDE and \(f\) is a payoff functional (with the notation \(f(Y) = f\Par{Y_t, t \in [0,T]}\) where \(T>0\) is a time horizon).

This paper starts with the study of the Monte-Carlo approximation problem for the expectation \(\E{f(Y)}\) where $f$ is a Lipschitz path functional of a L\'evy process \(Y\) in \(\Rd\). The proposed scheme turns out to be directly applicable to L\'evy-driven SDEs: when \(Y\) solves a L\'evy-driven SDE, this problem has received much attention over the past two decades, see e.g. the review article \cite{KohatsuNgoWeakApproximationSDE} and the references therein. In broad terms, the studies analyzes either the Euler scheme approximation to the solution of the SDE \cite{ProtterTalayEulerSchemeSDE,JacodEulerSchemeSDE} where it is assumed that the increments of the driving L\'evy process can be simulated directly, or the so-called approximate or weak Euler scheme, where the increments are approximated by IID random vectors that can in principle be simulated \cite{Rubenthaler2022,JacodKurtzMeleardProtter2005,DereichHeidenreich2011,Dereich2011}.

The main difficulty one faces when approximating increments of a L\'evy process lies in the simulation of the increments of the pure-jump component of the process. In the case of interest, when the jumps of the L\'evy process are of infinite activity (cf. \cite{Jacod2013Limit}), two issues arise in this context: for a small \(h > 0\), the simulation of (i) the jumps smaller than \(h\) and (ii) jumps greater than \(h\). Since the number of jumps is infinite in any compact time interval, it has been suggested to simulate all jump times for jump sizes greater than \(h\) and use a Gaussian approximation, first proposed in \cite{AsmussenRosinski01}, to mimic the jumps of size less than \(h\) by a Brownian motion (see \cite{Fournier12,Dereich2011,JacodKurtzMeleardProtter2005}). This approach to issue (i) has been generally accepted in the literature: for example, all the weak Euler schemes described in \cite{KohatsuNgoWeakApproximationSDE} use it.

As far as (ii) is concerned, it is usually assumed that the jumps larger than \(h\) can be simulated in constant time (see the first assumption for the cost of the algorithm in \cite{DereichHeidenreich2011,Dereich2011}). Put differently, this means that random variates with distribution \(F_h\), 
\begin{equation}\label{Introduction:Problem:distribution}
	F_h \propto \One{\Rd \setminus [-h, h]^d} \lambda
\end{equation}
proportional to the restriction of the L\'evy measure \(\lambda\) of the driving L\'evy process to the complement of a small neighborhood \([-h, h]^d\) of the origin in \(\Rd\), can be simulated in constant time. However, the simulation from the distribution \(F_h\) can be quite challenging, particularly in dimensions greater than one. To see this, let us look at some examples.

Consider first the case of a one-dimensional tempered L\'evy process, one of the simplest and most widely used infinite activity pure-jump L\'evy processes in mathematical finance \cite{CGMY02,PoirotTankov06} and other fields such as statistical physics, financial econometrics and mathematical biology, see \cite{KawaiMasuda11} and the references therein. The problem of simulation of tempered stable processes is analyzed in \cite{KawaiMasuda11}. When discussing the simulation of the decomposition of the process into jumps smaller and greater than \(h\), the authors remark that this is `\textit{never as handy as often claimed in the literature}' (see \cite[Section 4.1.1]{KawaiMasuda11}): if the truncation level \(h\) is chosen very small, which is necessary for the Gaussian approximation to work well, the intensity of jumps of size greater than \(h\) explodes. In particular, the proportionality constant in \eqref{Introduction:Problem:distribution} is given by \(1/\lambda\left(\Rd \setminus \SquaredBracket{-h, h}^d \right)\) which makes the simulation of jumps from the distribution $F_h$ difficult even if the constant is computable. The remedy in the case of the tempered stable L\'evy process suggested in \cite{KawaiMasuda11} is to apply an acceptance-rejection sampling algorithm using a Pareto proposal with a tractable stochastic representation based on a uniform random variable, see \cite[eq. (4.2)]{KawaiMasuda11}. However this approach (a) only applies in the special case when the one-dimensional L\'evy measure is tempered stable and (b) even in this case the acceptance probability is proportional to \(1/\lambda\left(\Rd \setminus \SquaredBracket{-h, h} \right)\), making the algorithm numerically inefficient: on average one has to simulate \(\lambda\left(\Rd \setminus \SquaredBracket{-h, h} \right)\) (\(\rightarrow \infty\) as \(h \downarrow 0\)) Pareto random variables to accept a single jump of size at least \(h\) of a tempered stable L\'evy process.

The simulation issues to do with truncated jumps become much more severe in the multidimensional setting (\(d > 1\)). To start with, there is no easy parametric way to build multidimensional L\'evy processes. A generic method adopted in the literature is via the \textit{L\'evy copula} \cite{TankovKallsen}. In this context, given a L\'evy 
copula \(\mathcal{F}\) and the tail integrals \(U_i(x) := sign(x) \lambda_i(\mathcal{I}(x))\) of the marginal one-dimensional L\'evy measure \(\lambda_i\) for \(i=1,\dots, d\), where \(\mathcal{I}(x) := (x, +\infty)\) if \(x \geq 0 \) and \(\mathcal{I}(x) := (-\infty, x]\) if \(x < 0 \), one constructs the tail integral of \(\lambda\) on \(\left( \R \setminus \{0\} \right)^d\) by:
\[ \lambda \Par{ \mathcal{I}(x_1)  \times \dots \times \mathcal{I}(x_d) } := \mathcal{F} \Par{U_1(x_1), \dots, U_d(x_d)}, \quad x_i \in \R \setminus \{0\} \;\;\text{for } i=1,\dots,d\]
this formula, together with analogous marginal tail integrals, uniquely determines the corresponding L\'evy measure \(\lambda\) (see \cite[Theorem 3.6]{TankovKallsen}, which (if \(\mathcal{F}\) is sufficiently smooth) can be viewed as follow:
\begin{equation}\label{Introduction:Problem:Copula}
	\lambda \Par{dx_1 \otimes \dots \otimes dx_d } = \frac{\partial^d \mathcal{F}}{\partial_1 \dots \partial_d} \Par{U_1(x_1), \dots, U_d(x_d))} \lambda_1(dx_1) \otimes \dots \otimes \lambda_d(dx_d)
\end{equation}
The only general method for the simulation of an IID sample from the distribution \(F_h\) in \eqref{Introduction:Problem:distribution} (with \(\lambda\) from \eqref{Introduction:Problem:Copula}) appears to be a version of the acceptance-rejection algorithm. However, such an approach is not feasible in general as the L\'evy measure \(\lambda\) in \eqref{Introduction:Problem:Copula} depends on the \(d\)-th  derivative of a L\'evy copula and the rejection sampling requires analytical control of the target distribution, typically not afforded by higher derivatives, in order to construct a suitable proposal distribution. Furthermore, even if a suitable proposal distribution were feasible, the issues concerning the normalizing constant for \(F_h\), mentioned in the one-dimensional case above, persist making the acceptance-rejection algorithm numerically inefficient.

Our approach to this problem is based on the state space discretization of the pure-jump L\'evy process developed in \cite{MijatovicVidmarJacka12}. Before describing the Monte-Carlo algorithm in the next section, we remark that the simulation of the truncated jumps with law \(F_h\), where the L\'evy measure is given by a L\'evy copula, is made possible by the fact that the mass under \(\lambda\) of any rectangular set \([a_1, b_1]\times\dots \times[a_d,b_d]\) in \(\Rd\) can be expressed as a linear combination of terms \(\mathcal{F}\Par{U_1(x_1),\dots,U_d(x_d)}\) (one for each ``corner" of the set). This makes the L\'evy copula representation\footnote{Note that by \cite[Theorem 3.6]{TankovKallsen}, there is a one-to-one correspondence between the L\'evy measures and L\'evy copulas equipped with a family of one-dimensional (marginal) L\'evy measures. This allows the modeler to construct a \(d\)-dimensional L\'evy measure with arbitrary jump-structures of the components and arbitrary dependence between them.} of \(\lambda\) particularly well suited for our simulation algorithm.

\subsection{Aims and the main results.}
This paper aims to do the following:
\begin{enumerate}[label=(\Roman*)]
	\item develop a Monte-Carlo scheme for the path-functionals of a L\'evy process that is able to simulate jumps of size greater than \(h\) of the L\'evy driver (cf. Section \ref{Introduction:Problem}) for an arbitrary L\'evy measure, assuming only that the L\'evy copula of the process can be evaluated;
	\item analyze the computational complexity of the algorithm in (I), taking into account the complexity of sampling jumps of magnitude greater than \(h\);
	\item develop a multilevel version of the Monte-Carlo algorithm in (I) and analyze its complexity;
	\item extend the multilevel Monte-Carlo algorithm to path-functionals of L\'evy-driven SDEs (cf. Section \ref{subsec:mlmcAlgorithm:LevyDrivenSDE});
	\item apply the Monte-Carlo and its multilevel analogous to numerical examples and illustrate the consistency of the convergence rates of the cost of the multilevel Monte-Carlo with the theoretical bounds of equation \eqref{eq:mlmcCostBounds} in Section \ref{subsec:mlmcAlgorithm:Algo} below. 
\end{enumerate}
We build on the scheme for the L\'evy-driven SDEs described and analyzed in \cite{Dereich2011}. In particular, as in \cite{Dereich2011}, we use the Gaussian correction for the jumps of size less than \(h\). Our aim in a word is to introduce a simulation algorithm for jumps greater than \(h\) into this scheme and analyze how it affects the overall computational complexity. In contrast, the analysis in \cite{Dereich2011} is carried out under the assumption that jumps of size greater than \(h\) can be simulated in constant time for any L\'evy measure of the driving L\'evy process. We are interested in what the computational cost of this assumption is, i.e. how much worse the convergence rate of our algorithm is compared to \cite{Dereich2011} and how it behaves as a function of the Blumenthal-Getoor index (see \eqref{def:bg} below for definition), which measures the ``frequency" of the jumps. In order to emphasis the simulation scheme itself, in this paper we focus primarily on the multilevel Monte-Carlo algorithm for functionals of L\'evy processes. However, as we shall see below, a standard jump-adapted time discretization makes our scheme directly applicable to L\'evy-driven SDEs (see Section \ref{NumApp:subsec:mlmcLLM} for a numerical application to the Forward Market Model and Section \ref{subsec:mlmcAlgorithm:LevyDrivenSDE} for the relevant theory).

\subsubsection{The simulation scheme and its computational complexity.}\label{Introduction:SimluationSchemeAndComplexity}
Let \(X\) be a L\'evy process equal to \(Y\) with jumps truncated at some very large level \(R\). Hence, the L\'evy measure \(\lambda\) of \(X\) has support contained in the box \(\SquaredBracket{-R, R}^d \subset \Rd\). The bias introduced by the truncation is under \eqref{def:AS1} easily incorporated in the complexity analysis of the Monte-Carlo algorithm \ref{algo:MCAlgorithm} for \(Y\). The proposed simulation \eqref{Scheme} below is based on the weak approximation, introduced in \cite{MijatovicVidmarJacka12}, of the pure-jump part \(X^{\lambda}\) of \(X\) (see \eqref{eq:decomposition} for the definition of \(X^{\lambda}\)) by a continuous Markov chain \(X^{h(\lambda)}\). This weak approximation, briefly described in Section \ref{subsec:CTMC} below, uses the generator of \(X^{\lambda}\) to define a \(Q\)-matrix for the approximating chain \(X^{h(\lambda)}\) with state space \(h\Zd\), which best mimics the behavior of \(X^{\lambda}\). The process \(X^h\), defined in \eqref{Scheme}, is a jump-diffusion process with a Gaussian component compensating for the jumps of \(X\) of size less than \(h\) and a jump component consisting of discretized jumps. Each jump of \(X^h\) takes only finitely many values since the L\'evy measure \(\lambda\) has compact support. This circumvents the problem of the simulation of jumps of size greater than \(h\), as it allows the application of the complete binary tree methods in \cite[Section III.2.3]{Devroye} for finite random variables (note that some methods do not require the direct computation of the normalization constant in \eqref{Introduction:Problem:distribution}). Furthermore, since the support of \(\lambda\) lies in a box of size proportional to some large value \(R > 0\), each jump of \(X^h\) can take at most \(\bigO{\Par{R/h}^d}\) values. Hence the simulation algorithm in \cite[Section III.2.3]{Devroye} requires at most \(\bigO{d \log R + d \log(1/h)}\) steps for a single draw from the jump distribution of \(X^h\). We stress that this algorithm is linear in the dimension \(d\) and logarithmic in \(R\) (i.e. the diameter of the support of \(\lambda\)) and \(1/h\). We show that the computational complexity \(\mathcal{C}(\epsilon)\) (i.e. the number of steps) for achieving root mean-square error of size at most \(\epsilon\) in the Monte-Carlo algorithm \eqref{def:mcestimator} is bounded above by:
\[\mathcal{C}(\epsilon) \leq D_2 \epsilon^{-2 -2\beta/(2 - \beta)} \SquaredBracket{\log(1/\epsilon) + D_3},\]
where the constants are given explicitly in terms of the underlying process in Section \ref{subsec:MCComplexity}.

We remark that the intensity of jumps in \eqref{Scheme} equals \(\bigO{\lambda\Par{\Rd \setminus \SquaredBracket{-h, h}^d}}\) and requires the simulation of the Poisson random variable with this intensity for any \(h > 0\). A naive simulation would require \(\bigO{\lambda\Par{\Rd \setminus \SquaredBracket{-h, h}^d}}\) steps. However, there exist simple algorithms (see e.g. \cite{AhrensDieter1980Sampling}), requiring \(\bigO{1}\) steps uniformly for all intensities. Furthermore, in \cite{Devroye}, such an algorithm is presented with decreasing complexity for increasing intensity. This is in particular very important for our multilevel Monte-Carlo algorithm, described in the next section, where simulation for increasingly small \(h\) is required.

\subsubsection{A weak multilevel Monte-Carlo algorithm and its complexity.}\label{Introduction:ProblemWeakMLMC}
The idea of the multilevel Monte-Carlo (MLMC) was introduced in \cite{Giles} (see also the overview paper \cite{Giles15}). As observed in \cite{Giles15}, this framework can be applied to an arbitrary Monte-Carlo scheme in order to reduce the computational cost for a given level of variance of the estimator, as long as one can simulate jointly the random variables arising in the neighboring levels of the MLMC estimator and provide a suitable bound on the level variance. This approach works well for diffusion \cite[Theorem 3.1]{Giles}.

\begin{figure}
	\centering
	\begin{tikzpicture}
		\begin{axis}[
			unit vector ratio*=1 1.4 1,
			scale       = 1.3,
			axis x line	= left,
			axis y line	= left,
			xmin = 0, xmax = 2.1,
			ymin = 0, ymax = 0.6,
			xtick 				= {0,0.2,...,2},
			ytick 				= {0,0.1,...,0.5},
			xlabel 				= {Blumenthal-Getoor index},
			ylabel 				= {Order of convergence},
			label style			= {font=\tiny},
			tick label style 	= {font=\tiny},
			legend style	 	= {font=\tiny},
			legend cell align	= left,
			legend pos		 	= north east
			]
			\draw[dashed]  (0.0, 0.5) -- (2.0, 0.0); \addlegendimage{line width=0.1mm, dashed}	\addlegendentry{MC};
			\draw[solid]  (0.0, 0.5) -- (1.0, 0.5);  \addlegendimage{line width=0.1mm, solid}	\addlegendentry{MLMC};
			\draw[domain=1:2, variable=\x] plot ({\x}, {1/\x-1/2});
			\draw[dotted, domain=1:2, variable=\x] plot ({\x}, {(4 - \x)/(6*\x)}); 	\addlegendimage{line width=0.1mm, dotted}	\addlegendentry{MLMC (njd)};
		\end{axis}
	\end{tikzpicture}	
	\caption{The convergence rate \(\beta \rightarrow \rho(\beta)\), as a function of the Blumenthal-Getoor index \(\beta\) of the L\'evy process \(Y\), of the MC \ref{def:mcestimator} and the MLMC \ref{def:mlmcEstimator} algorithms as well as that of the MLMC (njd) algorithm in \cite{Dereich2011} with ``no jump-discretization", is presented. More precisely, for a given \(\beta\), any of the three simulation algorithms has the root mean-square error of size \(\bigO{\tau^{-\rho(\beta)}}\) for a prescribed number of computational steps \(\tau\). \(\rho(\beta)\) for MC and MLMC are implied by the computational complexities in Section \ref{Introduction:SimluationSchemeAndComplexity} and \ref{Introduction:ProblemWeakMLMC} respectively. This figure sheds light on the cost of the assumption in \cite{Dereich2011} that jumps of size at least \(h\) can be simulated in \(\bigO{1}\) steps for all \(h\) and dimension \(d\).}
	\label{Introduction:figure:orderOfConvergence}
\end{figure}
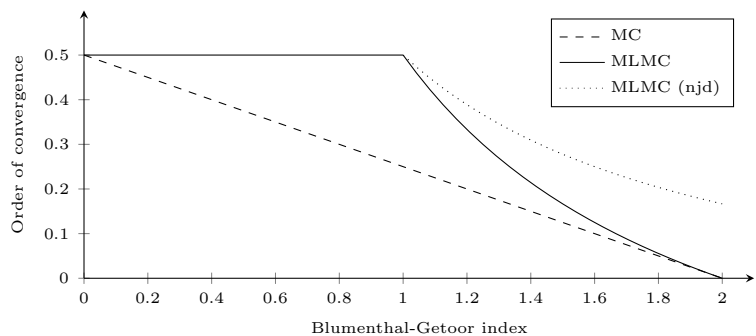

In the context of MLMC, the case of L\'evy-driven SDEs was analyzed in \cite{DereichHeidenreich2011,Dereich2011}. The key idea in these papers is to construct the multilevel estimator by varying the truncation level \(h\). Since it is assumed that one can simulate from the distribution \(F_h\) in \eqref{Introduction:Problem:distribution}, proportional to the truncated L\'evy measure, in \(\bigO{1}\) steps independently of the dimension of the state space and the truncation level \(h\), it is natural to couple the neighboring levels \(h\) and \(2h\) in the MLMC estimator by simulating the jump of size at least \(h\) and accepting it if and only if its size is at least \(2h\). Since the distribution \(F_{2h}\) is by \eqref{Introduction:Problem:distribution} proportional to the truncation of the L\'evy measure at the level \(2h\), this construction yields a coupling of the neighboring levels and facilitates the analysis of the computational complexity of the MLMC estimator. Under the assumption that random vectors with distribution \(F_h\) can be simulate in \(\bigO{1}\) steps for all \(h\), the order of convergence of the MLMC algorithm in \cite{Dereich2011}, as a function of the Blumenthal-Getoor index \(\beta\), is plotted in Figure \ref{Introduction:figure:orderOfConvergence} (see dotted line for MLMC (njd)). In particular, it is equal to \(1/2\) for \(\beta \in [0,1]\) and decreases to \(1/6\) over the interval \([1,2]\).

In order to develop a MLMC algorithm based on \eqref{Scheme}, we are faced with a key problem of defining the process \(X^h\) and \(X^{2h}\) on the same probability space so that they stay close to each other with high probability and thus making the level variances sufficiently small. As in the MLMC algorithm in \cite{Dereich2011}, this problem reduces to the coupling of the pure-jump process \(X^{h(\lambda)}\) and \(X^{2h(\lambda)}\), approximating the pure jump component \(X^{\lambda}\) of the driver (see Section \ref{Introduction:SimluationSchemeAndComplexity} above). The assumption in \cite{Dereich2011} that one can simulate from \(F_h\) in \eqref{Introduction:Problem:distribution} for any \(h > 0\) in \(\bigO{1}\) steps and the fact that \(F_{2h}\) and \(F_h\) are proportional on \(\Rd \setminus \SquaredBracket{-2h, 2h}^d\) provides a simple and natural coupling of jumps, applied in \cite{Dereich2011}, across two levels \(h\) and \(2h\). In contrast, in our setting the jumps in \eqref{Scheme} at the level \(2h\) are supported in \(2h\Zd\) and their distribution is \textbf{not} proportional to the restriction (to \(2h\Zd\)) of the distribution of the jumps of \(X^{h(\lambda)}\).

The main theoretical result of the paper (see Theorem \ref{thm:coupling}) defines a novel coupling between the continuous-time Markov chains \(X^{h(\lambda)}\) and \(X^{2h(\lambda)}\) and establishes a bound on the level variances (see Section \ref{sec:coupling}), facilitating the analysis of the computational complexity of the corresponding MLMC estimator. The coupling is based on the classical Marking and Mapping Theorems for Poisson point processes (see \cite{Kingman}) and the structure of the stochastic approximation of the pure-jump L\'evy process \(X^{\lambda}\). Simply put, in order to get a joint path of \(X^{h(\lambda)}\) and \(X^{2h(\lambda)}\), one simulates each jump of \(X^{h(\lambda)}\) and, conditional on the value of the jump, draws a finite valued random variable with an explicit mass function to determine the jump of \(X^{2h(\lambda)}\) (see Section \ref{subsec:mlmcAlgorithm:Intuitive} for the intuitive description of the coupling and Section \ref{sec:coupling} for its full construction and the bound on the level variances). The simplicity of the coupling is crucial for the feasibility of our MLMC scheme.

Note that a straightforward coupling of \(X^{h(\lambda)}\) and \(X^{2h(\lambda)}\), where one simulates the jumps from the truncated L\'evy measure at level \(h\), and then deterministically maps the jump sizes to the closest lattice points in \(h\Zd\) and \(2h\Zd\) to obtain a jump of \(X^{h(\lambda)}\) and \(X^{2h(\lambda)}\), respectively, cannot be used as it does not avoid simulation from the distribution of \(F_h\) in \eqref{Introduction:Problem:distribution}.

We stress that our MLMC algorithm is genuinely weak in the sense that the approximating process \(X^h\) and the limiting L\'evy driver \(X\) need not to be defined on the same probability space (see \cite{MijatovicVidmarJacka12} and \cite[Corollary 2.4]{JackMijatovic12}). The absence of strong convergence of the approximation in \eqref{Scheme} makes the analysis of the level variances depend solely on the coupling mentioned above. It is worth noting here that typically the level variances in MLMC are bounded by the second moment of the difference between each level \(X^h\) and the limiting process \(X\) (see e.g. \cite[Theorem 2]{DereichHeidenreich2011} and \cite[Section 2]{Giles}), in effect requiring the strong (i.e. almost sure) convergence of the underlying scheme, even though by definition the MLMC estimator only requires weak convergence. In the case of the MLMC algorithm defined in Section \ref{sec:mlmcAlgorithm}, the bound on the level variance established in Proposition \ref{prop:MLVariance} below, and hence the entire complexity analysis, depends solely on the weak convergence and the coupling between the levels. The computational complexity of our MLMC algorithm (for a root mean-square error less than \(\epsilon\)) as a function of the Blumenthal-Getoor index \(\beta\) takes the form:
\[
\mathcal{C}^{ML}(\epsilon) \leq \bar{D} \epsilon^{-2} (\log(1/\epsilon) + D_3) \times	
\begin{cases}
	D_5, 									& 0 \leq \beta < 1\\
	\log^2(1/\epsilon), 					& \beta = 1\\
	D_6 \epsilon^{-4(\beta-1)/(2-\beta)},	& 1 < \beta < 2.
\end{cases}
\]
The constants are given explicitly in terms of the process following display \eqref{eq:mlmcCostBounds} in Section \ref{subsec:mlmcAlgorithm:Algo} below.

In order to compare our MLMC scheme with the MC algorithm from the previous section and the MLMC scheme in \cite{Dereich2011}, we plot in Figure \ref{Introduction:figure:orderOfConvergence} the rate of convergence, as a function of the Blumenthal-Getoor index \(\beta\), of all three algorithms. The curve corresponding to our MLMC algorithm is implied by the bound on its complexity from the previous paragraph. It is clear that our MLMC estimator outperforms the MC estimator for all values of \(\beta\). Perhaps more interesting is the qualitative difference in the behavior of our MLMC scheme compared to that of \cite{Dereich2011}: taking into account the actual cost of the simulation of the jumps of size greater than \(h\) makes our MLMC algorithm arbitrarily slow as the Blumenthal-Getoor index approaches 2, while the algorithm in \cite{Dereich2011}, which assumes that this cost is of order \(\bigO{1}\), retains a convergence rate of at least \(1/6\) for all \(\beta\) in \((1,2]\). Furthermore, if \(\beta \in [0,1]\), the cost of the simulation of the jumps of size at least \(h\) does not feature, at least as far as the rate of convergence of the two MLMC schemes is concerned.

\subsection{Organization of the paper.}
In Section \ref{NumericalApplications}, we present numerical applications of the (\ref{Scheme}) and its multilevel analogous for one-dimensional L\'evy processes, L\'evy copulas and L\'evy-driven SDEs. 
Section \ref{sec:mc} defines the Continuous-Time Markov Chain approximation (\ref{Scheme}) and analyzes the complexity of the corresponding Monte-Carlo algorithm.
Section \ref{sec:mlmcAlgorithm} describes our results on multilevel Monte-Carlo. More precisely, in Section \ref{subsec:mlmcAlgorithm:Background} we  expose the role of the coupling between the levels in controlling the variances. Section \ref{subsec:mlmcAlgorithm:Intuitive} gives an intuitive description of the coupling developed in this paper. Section \ref{subsec:mlmcAlgorithm:Algo} describes our MLMC algorithm and its complexity. Section \ref{subsec:mlmcAlgorithm:LevyDrivenSDE} describes the version of our MLMC for L\'evy-driven SDEs.
Proofs of all results are given in Appendix \ref{appendix:proofs}. Appendix \ref{appendix:numData} contains data used in numerical examples of Section \ref{NumericalApplications}.

\section{Numerical applications}\label{NumericalApplications}
We illustrate the application of the CTMC (\ref{Scheme}), described in Section \ref{sec:mc}, and its multilevel analogous (see Section \ref{sec:mlmcAlgorithm}) with examples taken from mathematical finance, spanning different asset classes (credit, equity, interest rates), payoffs and pricing models based on both one-dimensional and multidimensional L\'evy drivers.
In Section \ref{NumApp:subsec:GarreauKercheval}, we recall the modeling framework introduced in \cite{GarreauKercheval} which relies on the modeling of the times to default via a L\'evy copula yielding a closed-form formula for a First-to-Default Credit Default Swap (FtD CDS). The CTMC (\ref{Scheme}) naturally fits this modeling framework and also allows the pricing of more complex derivatives.
In Sections \ref{NumApp:subsec:mlmcLevyProcesses} and \ref{NumApp:subsec:mlmcLLM}, we give numerical applications of the MLMC for one-dimensional L\'evy processes, L\'evy copulas and L\'evy-driven SDEs and different types of payoffs: equity put and best-of put options, as well as interest rates swaptions; the first two examples are based on L\'evy processes with infinite activity and infinite variation (that is \(\beta\) between 1 and 2), which corresponds to the most challenging case numerically and also the most ``interesting" case with regard to the convergence rates of the cost of the MLMC algorithm (see Equation \eqref{eq:mlmcCostBounds}).
In Section \ref{NumApp:subsec:SeriesRepresentation}, we benchmark the CTMC (\ref{Scheme}) against the simulation algorithm based on a series representation proposed in \cite{TankovSimulation2}.

\remark{\
	\begin{enumerate}[label=(\roman*)]
		\item In the following numerical applications, the cut-off value \(R\), determined by Equation \eqref{def:hRn}, is computed in an alternative way, independent from the root mean square error \(\epsilon\) and the constant \(D_T\). In the one-dimensional case, it is chosen such that the ratio of the masses under \(\lambda\) of the intervals \(\SquaredBracket{-R, -\frac{h}{2}} \cup \SquaredBracket{+\frac{h}{2}, +R} \) and \(\SquaredBracket{-\infty, -\frac{h}{2}} \cup \SquaredBracket{+\frac{h}{2}, +\infty} \) is at least equal to a threshold, set to \(99.999\%\). In the multidimensional case, the cut-off \(R\) is chosen as the largest computed value over each axis.
		\item The code used to generate the numerical examples of this section is available in the Github repository \cite{PalfrayMLMCsimulation2022}.
	\end{enumerate}
}

\subsection{The CTMC scheme applied to the Garreau-Kercheval Credit Model}\label{NumApp:subsec:GarreauKercheval}
In \cite{GarreauKercheval}, the authors model the bankruptcy of a company via the jumps in its stock price \((S_t)_t\): the time to default \(\tau\) is defined as the first time the instantaneous log-return goes below a possibly stochastic threshold \(a_t < 0\):
\[\tau := \inf \left\{ t > 0, \log\left(\frac{S_t}{S_{t^-}}\right) \leq a_t \right\}\]
The stock price \(S_t\) is modeled as the exponential of a L\'evy process \(Y_t\): \(S_t := S_0 \exp \{(r + \psi(-\textbf{i})t + Y_t) \}\) (where \(\psi\) is the characteristic exponent of \(Y\)). Therefore the time to default \(\tau\) corresponds to the first occurrence of a negative jump of \(Y_t\) below the threshold \(a_t\). The authors show that this framework yields a simple pricing formula for the survival probability \(\Prob{\tau > t}\), which only depends on the tail integral of the L\'evy process.\\
This modeling can be extended to multidimensional L\'evy processes via the L\'evy copula and, in this case, the survival probability, that is the probability that none of the companies has bankrupted, can be calculated from the tail integrals of the L\'evy copula and its marginal tail integrals. In particular, for a constant (or piecewise constant) process \(a_t=a\) and if the tail integrals can be computed in closed-form, it yields a closed-form pricing formula for a FtD CDS.

\subsubsection{Pricing of a single-name CDS}\label{subsubsec:CSDPricing}

In the one-dimensional case, the survival probability is given by \cite[Proposition 2.2]{GarreauKercheval}:
\begin{equation}\label{eq:survivalProbability}
	\Prob{\tau > t} = \E{\exp\left\{-\int_0^t \Lambda(a_u) du\right\}}
\end{equation}
where \(\Lambda(a) = \int_{(-\infty, a]} \lambda(du)\) is the tail integral of the process \(Y\).\\
If \((a_t)_t\) is a constant process equal to \(a\) then the formula simplifies to \(\Prob{\tau > t} = \exp\left\{-\Lambda(a) t \right\}\).
The default and fixed legs of a CDS are respectively given by:
\[DL_t = (1 - R) \int_0^t e^{-rs} d\Prob{\tau \in ds} \quad \text{and} \quad FL_t = m \int_0^t e^{-rs} \Prob{\tau > s} ds \]
Integration by part also gives:
\[DL_t = (1-R) \left[ e^{-rt}(1 - \Prob{\tau >t}) - (e^{-rt} - 1) -r \int_0^t e^{-rs} \Prob{\tau >s} ds \right]\]
Given the values of \(R\), \(m\), and the above expression \eqref{eq:survivalProbability} of the survival probability, the level \(a\) (or equivalently the survival probability at \(t\)) is implied from the price \(DL_t - FL_t\) of the CDS.\\
In the following application displayed in Table \ref{tbl:singlenamecds}, we simulate the time of default via the CTMC and then compute the Present Value (PV) of the CDS, whose spread is implied from the pricing equation \(PV = DL_t - FL_t\), and, in the same way, one can imply a \(99\%\) confidence interval for this CDS spread.\\
In this numerical example, the default time is simulated via the exponential of a HEM model (see \cite{Kou12}) whose parameters are defined in Appendix \ref{appendix:ModelParameters}. The maturity of the CDS is \(T=6M\) and its recovery rate is set to \(R=0.4\). The Monte-Carlo engine uses 1 million paths for different CDS spread values and the spatial step in the CTMC grid \(h\) is set to a small value: \(h = \num{1e-6}\).

\begin{table}[H]
	\centering	
	\small
	\csvreader[
	tabular			= |c|c|c|c|c|,
	table head		= \hline CDS spread & Level \(a\) & Implied Survival  & MC Implied       & 99\% Confidence \\
	                            (bps)   &             &	Probability       & CDS Spread (bps) & Interval (bps)  \\\hline\hline,
	late after line	= \\\hline,
	separator		= comma,
	]%
	{\DataForPaper/credit/cds_hem.csv}
	{theoreticalspread=\theoreticalspread, levela=\level, survivalprobability=\prob,
		impliedspread=\impliedSpread, spreadstddevm=\spreadm, spreadstddevp=\spreadp}%
	{\theoreticalspread & \num[round-precision=3]{\level} & \pgfmathparse{\prob*100}\pgfmathprintnumber{\pgfmathresult}\% & \num[round-mode=places,round-precision=2]{\impliedSpread}  & [\spreadm, \spreadp]} 	
	\caption{Single-name CDS Pricing with the HEM model}
	\label{tbl:singlenamecds}
\end{table}

As expected, the CDS spread given by the Monte-Carlo simulation is close to the theoretical spread, the latter being inside the 99\% confidence interval.

\subsubsection{Pricing of a First-to-Default CDS}

In the \(d\)-dimensional case, the times to default are defined by \cite[Equation (58)]{GarreauKercheval} (for constant values \(a_i\)):
\[\tau_i := \inf \left\{ t >0, \log\left( \frac{S_t^i}{S_{t_-}^i} \right) \leq a_i \right\},\quad i=1, \dots,d,\]
where \(S_t^i\) denotes the \(i\)-th underlying stock price.

The survival probability can be computed in closed-form and is given by (in the two-dimensional case):
\[\Prob{\tau^{(1)} > t} = \exp \left( - t \left \{ \Lambda_1(a_1) + \Lambda_2(a_2) - \rho(\overline{\Lambda_1}(a_1), \overline{\Lambda_2}(a_2)\right\} \right)\]
where:
\begin{itemize*}
	\item \(\tau^{(1)}\) is the first time to default: \(\tau^{(1)} = \min \left(\tau_1, \tau_2 \right)\) 
	\item \(\Lambda_1\) and \(\Lambda_2\) are the margin tail integrals
	\item \(\overline{\Lambda_i}(x) = sgn(x) \Lambda_i(x)\)
	\item \(\rho\) is the L\'evy copula given by the Sklar's theorem for the L\'evy copula (see \cite[Theorem 3.6]{TankovKallsen}.\\
\end{itemize*}

We can benchmark the CTMC \eqref{Scheme} against this closed-form formula (which can be extended to the general \(d\)-dimensional case) as we did in the previous Section \ref{subsubsec:CSDPricing} for the single-name CDS pricing: for each margin, we compute the level \(a_i\) corresponding to a given CDS spread \(m_i\) (or equivalently to a given survival probability); the values of the single-name CDS spreads \(m_i\) are chosen the same for all marginals for simplicity. Then we simulate the \(d\)-dimensional process via the CTMC using Monte-Carlo and compute the PV of the FtD CDS; the single-name CDS are used as control variates. The final FtD CDS spread is implied by the CDS pricing equation of Section \ref{subsubsec:CSDPricing}. 
In the next numerical example displayed in Table \ref{tbl:ftdcds}, the maturity of the FtD CDS is \(T=6M\) and the three margins are defined by exponentials of HEM models with equal parameters:
\(\eta_1=20, \eta_2=25, p=0.6, \sigma = 5\%\) but different intensities: \(\lambda_1 = 5\), \(\lambda_2 = 10\) and \(\lambda_3 = 20\). The Monte-Carlo engine uses 1M paths, the recovery rate is chosen to be \(R= 0.4\) and \(h_0\) is set to \(\num{1e-6}\).
The dependence in the L\'evy copula is characterized by the Clayton copula \eqref{eq:ClaytonCopula} with \(\theta = 0.7\) and \(\eta = 0.3\).

\begin{table}[H]	
	\centering
	\small
	\csvreader[
	tabular			= |*{9}{c|},
	table head		= \hline Individual CDS  & Survival    &  Theoretical FtD  &  MC Implied FtD   & 99\% Confidence  \\
							 spread (bps)    & Probability & CDS spread (bps)  &  CDS Spread (bps) & Interval (bps)   \\ 	\hline\hline,
	late after line	= \\\hline,
	separator		= comma,
	]%
	{\DataForPaper/credit/first_to_default_hem_hem_hem.csv}
	{individualspread=\individualspread, theoreticalspread=\theoreticalspread, levela=\level, survivalprobability=\prob,
		impliedspread=\impliedSpread, spreadstddevm=\spreadm, spreadstddevp=\spreadp}%
	{\individualspread & \pgfmathparse{\prob*100}\pgfmathprintnumber{\pgfmathresult}\% & \theoreticalspread & \impliedSpread  & [\spreadm, \spreadp]} 
	\caption{FtD CDS pricing with a 3d L\'evy copula model}
	\label{tbl:ftdcds}
\end{table}

\remark{Within the framework of the Garreau-Kercheval model, the CTMC (\ref{Scheme}) naturally allows the pricing of a general \(n\)-to-default CDS (for which there is no known closed formula when \(n>1\)). Furthermore, the CTMC also provides the knowledge of the spot underlyings path and therefore allows the pricing of Equity/Credit multidimensional payoffs.}

\subsection{MLMC applied to L\'evy processes and L\'evy copulas: pricing of equity put and best-of put options}\label{NumApp:subsec:mlmcLevyProcesses}
We consider the pricing of an at-the-money (ATM) put option where the spot underlying is modeled as the exponential of a CGMY L\'evy process. In our framework, the convergence rates depend on the value of the Blumenthal-Getoor index \(\beta\) which is, in this case, directly given by the CGMY model parameter \(Y.\) In the following example \(\beta\) is equal to 1.5 (see Appendix \ref{appendix:ModelParameters} for all other model parameters), which corresponds to the case of a L\'evy process with infinite activity and infinite variation.\\
The results of the numerical applications are presented in the same fashion as in \cite{Giles}:
in Figure \ref{fig:MLMC1d} (and subsequent Figures \ref{fig:MLMC3d} and \ref{fig:MLMCSde}), the two top graphs display respectively the log-variances and the log-means of the estimator \(P_l\) and of the correction term \(\Abs{P_l - P_{l-1}}\) at each level \(l\). The convergence rates of the log-means and log-variances of \(\Abs{P_l - P_{l-1}}\) are bounded respectively by \(1 - \frac{\beta}{2}\) and \(2 - \beta\) (see Propositions \ref{prop:MeanBound} and \ref{prop:VarianceBound}, note that we consider only the leading terms), represented by the dotted red line. The Monte-Carlo engine uses \(100,000\) paths for each level and \(h_0\) is set to \(h_0 = 0.01\).\\
The two bottom graphs correspond to the application of the MLMC algorithm for different root-mean square errors \(\epsilon\):
\begin{itemize*}
	\item the bottom left graph displays the optimal numbers of Monte-Carlo paths \(N_0, N_1, \dots, N_L\) for a given root-mean square error \(\epsilon\) (see \cite{Giles} for the expression of \(N_l\)); \(L\) is the final level, that is the minimum between the bound given by equation \eqref{eq:mlmcMaxLevel} and the level for which the convergence criteria was numerically met;
	\item the bottom right graph displays the corresponding computing costs, in function of the root mean square error \(\epsilon\), for the multilevel and standard Monte-Carlo engines. The theoretical upper-bound convergence rate for the quantity \(\epsilon^2 \times\) \textit{cost of the MLMC} is given by the dotted red line (neglecting the log-term).
\end{itemize*}
As in \cite{Giles15}, the cost of the standard Monte-Carlo is approximated by the quantity $\epsilon^{-2} V_0 C_L$.\\

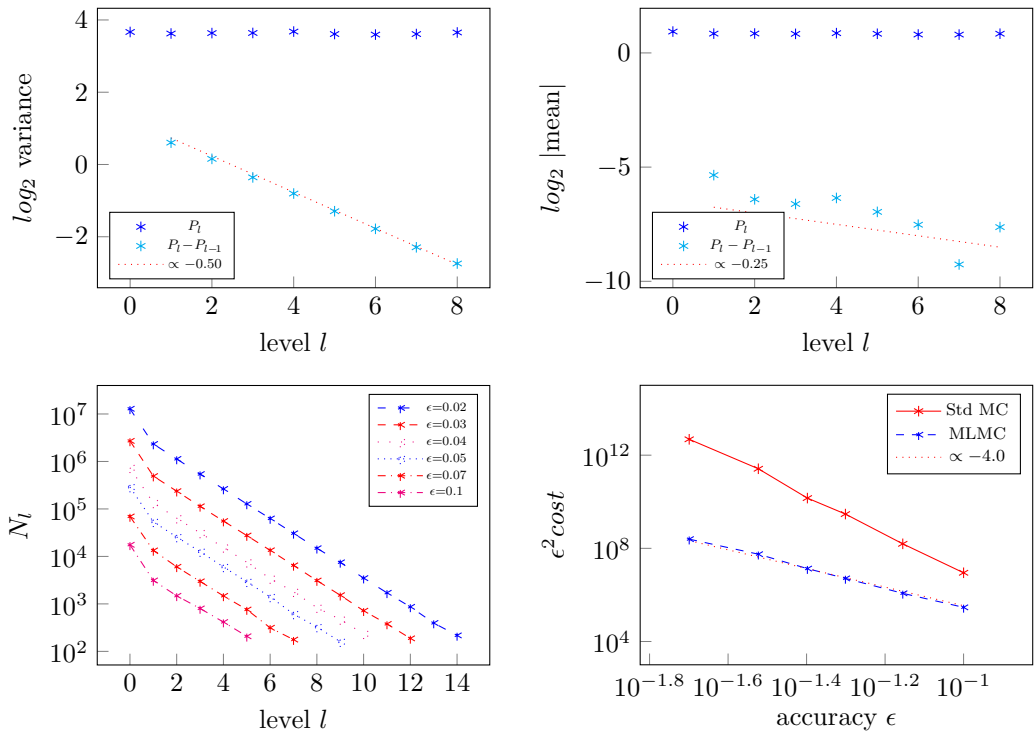
\def\mycaptionOneD{MLMC applied to the pricing of an ATM Put option with the \hyperref[appendix:ModelParameters]{\cgmyVII} model, \(\beta=1.5\)}
\newplotmlmc{pathconvergence=\pathgilesconvergence{1d}, pathapplied=\pathgilesapplied{1d}, modelname=cgmy15, bg=1.5, xmin=0.015, xmax=0.15, ymin=10e2, ymax=10e14, thecaption=\mycaptionOneD, label=fig:MLMC1d}{0.02,0.03,0.04,0.05,0.07,0.1}  

The next set of graphs illustrate the \(d\)-dimensional L\'evy copula case. The numerical application is based on the pricing of a best-of put option on three equity spot processes, with strike \(K=1\), that is a put option over the best performance of the three underlying spots. The margins of the L\'evy copula are modeled by exponentials of a L\'evy process and the dependence is characterized by the Clayton copula \eqref{eq:ClaytonCopula} with \(\theta = 0.7\) and \(\eta = 0.3\); model parameters are given in Appendix \ref{appendix:ModelParameters}. The margins of the multidimensional L\'evy process are given by different L\'evy models to show the flexibility of the L\'evy copula framework; the Blumenthal-Getoor index of the Levy copula is 1.1 which corresponds to the infinite activity/infinite variation case.

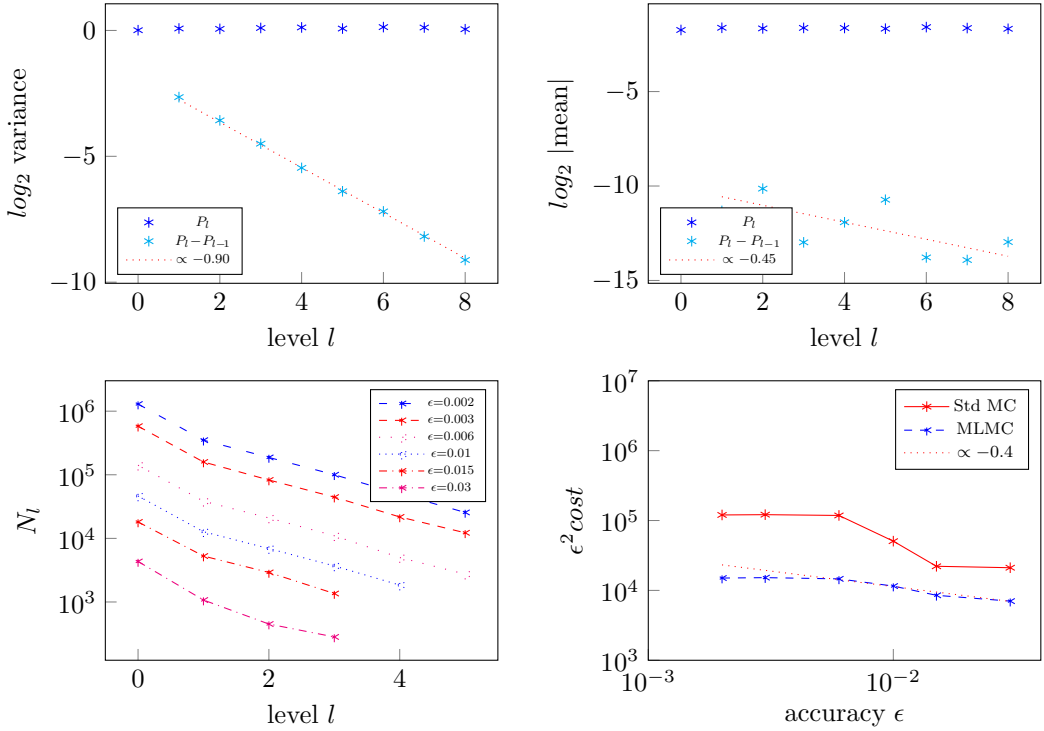
\def\mycaptionThreeDTwoA{MLMC applied to the pricing of a Best-of Put option with a L\'evy copula model characterized by the margins \hyperref[appendix:ModelParameters]{HEM/VG/\cgmyIV}, \(\beta=1.1\)}
\newplotmlmc{pathconvergence=\pathgilesconvergence{copulas}, pathapplied=\pathgilesapplied{copulas}, modelname=hem_vg_cgmy11, bg=1.1, xmin=0.001, xmax=0.040, ymin=10e2, ymax=10e6, thecaption=\mycaptionThreeDTwoA, label=fig:MLMC3d}{0.002,0.003,0.006,0.01,0.015,0.03}

\remark{The Blumenthal-Getoor index of a multidimensional L\'evy process is equal to the maximum of the Blumenthal-Getoor indices of its marginals, as shown in Appendix \ref{proof:bg}.}

\subsection{MLMC applied to L\'evy-driven SDEs: the L\'evy Forward Market Model}\label{NumApp:subsec:mlmcLLM}

\subsubsection{The Forward Market Model dynamics and its extension}
The Libor Market Model (LMM), introduced in \cite{BraceGatarekLMM97}, describes the joint distribution of a set of forward Libor rates. The L\'evy Libor Model is an extension of the LMM to account for jumps in the dynamics; it has been notably studied by Jamshidian, Glasserman and Kou, and Eberlein and \"Ozkan (\cite{Jamshidian1999libor, GlassermanKou2003TermStructure, EberleinOzkan2005LevyLibor}); the book \cite{BrigoMercurioBook} covers the modeling of the LMM and its extensions with a lot of practical details.

Following the IBOR transition (see the letter from the PRA \cite{IBORTransionLetterCEOS}) and the introduction of new benchmark risk free rates (RFR) replacing Libor rates, the relevance of the LMM is now questionable, but Lyashenko and Mercurio have shown in \cite{LyashenkoMercurio2019ModelingFramework} that this framework can be extended to the Forward Market Model (FMM) which is a richer model allowing for the modeling of the dynamics of both the term and individual RFRs. We propose to extend the FMM the same way the L\'evy Libor Model extends the LMM.

In the following, we use the same notations as in \cite{LyashenkoMercurio2019ModelingFramework}. Given a time structure \(T_0, \dots, T_n\), we denote by \(R(T_{j-1}, T_j)\) the daily-compounded setting-in-arrears rate over \([T_{j-1}, T_j)\):
\[R(T_{j-1}, T_j) = \frac{1}{\tau_j} \Par{ e^{ \int_{T_{j-1}}^{T_j} r(u) du } - 1}\]
where \(r\) represents the risk-free rate and \(\tau_j\) the year fraction for the interval \([T_{j-1}, T_j)\).
We denote by \(F(T_{j-1}, T_j)\) the forward-looking rate at time \(T_{j-1}\) maturing at time \(T_j\) (that is the analogous to the Libor rate).

Let now consider the backward-looking forward rate denoted \(R_j(t)\) corresponding to the value at time \(t\) of the strike \(K_R\) of a swaplet paying \(\tau_j \Par{R(T_{j-1}, T_j) - K_R}\) such that the value of the swaplet is 0; and the forward-looking forward rate denoted \(F_j(t)\) corresponding to the value at time \(t\) of the strike \(K_F\) of a swaplet paying \(\tau_j \Par{F(T_{j-1}, T_j) - K_F}\) such that the value of the swaplet is 0.
Mercurio and Lyashenko show in \cite{LyashenkoMercurio2019ModelingFramework} that both the backward-looking forward rate \(R_j(t)\) and the forward-looking forward rate \(F_j(t)\) coincide for \(t \leq T_{j-1}\).

The dynamics of the FMM is then defined by the set of equations:
\[dR_j(t) = \sigma_j(t) g_j(t) dW_j(t), \quad j=1,\dots,n\]
where:
\begin{itemize*}
	\item \(\Par{W_j(t)}_t\) is a Brownian motion under the (extended) \(\mathbb{Q}^{T_j}\)-forward probability (see \cite{LyashenkoMercurio2019ModelingFramework} for details about the definition of the extended forward probability).
	\item the Brownian motions \(\Par{W_j}_{j=1,\dots,n}\) are correlated via \(dW_i(t) dW_j(t) = \rho_{ij} dt\)
	\item \(\sigma_j\) is an adapted process.
	\item \(g_j\) is a function such that \(g_j(t) = 1\) for \(t \leq T_{j-1}\), \(g_j(t)\) is monotonically decreasing between \(T_{j-1}\) and \(T_j\), and \(g_j(t)=0\) for \(t \geq T_j\).
\end{itemize*}

Following \cite{kohatsu2010jump}, we consider a \(d\)-dimensional L\'evy process \(Z\) with L\'evy measure \(\nu\) such that \(\int_{\Abs{z} > 1} \Abs{z} \nu(dz) < \infty\).
We assume (without loss of generality) \(Z\) to be drift-less and to be a pure jump L\'evy process; we refer to \cite{GrbacKriefTankov2016} for the case where \(Z\) is a general L\'evy process (with zero drift).\\
We propose to extend the FMM (written under its log-normal form) with the following dynamics under the final (extended) \(T_n\)-forward measure:
\[
dR_t = R_{t_-} \Par{b(t, T)dt + \Gamma(t) dZ_t}
\]
where \(R_t = \Par{R^1_t, \dots,R^n_t}^T\) and \(\Gamma(t)\) is a deterministic \(n\times d\) volatility matrix such that \(\Gamma := (\sigma_{i,j} g_i(t))\) where the function \(g_i\) is defined similarly as above and \(\sigma_{ij}\) are scalar values.\\
The drift term \(b(t, T)\) is determined by the no-arbitrage condition yielding to:
\[
\frac{dR_t^i}{R_{t-}^i} = \gamma^i(t) dZ_t - \int_{\Rd} \gamma^i(t) z \SquaredBracket{\prod_{j=i+1}^{n} \left(1 + \frac{\tau^j R^j_t \gamma^j(t) z}{1 + \tau^j R^j_t} \right) - 1} \nu(dz) dt
\]
where $Z$ is a $d$-dimensional pure jump L\'evy process with L\'evy measure $\nu$ and $\gamma^i(t), i=1,\dots,n$ are the $d$-dimensional deterministic volatility functions:
\(\gamma^i(t) = \Par{\Gamma_{i1}(t) \dots \Gamma_{id}(t)} \).
\remark{As in \cite{PapapantoleonSkovmand2010NumericalLevyLibor} we approximate the drift term, limiting ourselves to its first order expansion in \(\Abs{R_t}\) for simplicity.}

\subsubsection{MLMC applied to the pricing of swaptions}
As in \cite{kohatsu2010jump}, we let \(T_0, T_1,\dots,T_n\) be a set of times and consider the case of a receiver swaption with maturity \(T_1\), tenor \(T_n - T_1\) and strike \(K\), whose price, at time 0, is given by (see \cite{kohatsu2010jump}):
\[\frac{P(0, T_1)}{\prod_{i=1}^{n} (1 + \delta R^i_0)} \E{h\Par{R^1_{T_1},\dots,R^n_{T_1}}},\]
with \(h\) such as:
\[h\Par{R^1_{T_1},\dots,R^n_{T_1}} = \Par{\prod_{i=1}^{n} \Par{1 + \delta R^i_{T_1}} -1 -K \delta \sum_{i=1}^{n} \prod_{j=i+1}^{n}  \Par{1 + \delta R^i_{T_1}}}_+\]
We consider a two-dimensional L\'evy driver with margins given by two CGMY models (\hyperref[appendix:ModelParameters]{\cgmyO} and \hyperref[appendix:ModelParameters]{\cgmyI} with Blumenthal-Getoor indices respectively equal to 0.2 and 0.4) and dependence modeled by a Clayton copula \eqref{eq:ClaytonCopula} with \(\theta = 0.7\) and \(\eta = 0.3\). The MLMC is applied to a \(5Y\times5Y\) swaption (that is the option to enter into a 5Y swap in 5Y) with strike \(K=2\%\).\\
We choose:
\begin{itemize*}
	\item the term structure: \(T_0 =0, T_6 = 5\) and \(T_{i+1} - T_i =1Y\), \(i=1\dots,5\)
	\item flat forward rates \(R^i_0 = 2\%\) \(i=1\dots,6\) 
	\item and the matrix 
	\(
	\sigma := 
	{\small
	\begin{pmatrix}
		0.50 & 1.50\\
		0.80 & 1.25\\
		1.00 & 1.00\\
		1.25 & 0.80\\
		1.50 & 0.50
	\end{pmatrix}}
	\)
\end{itemize*}

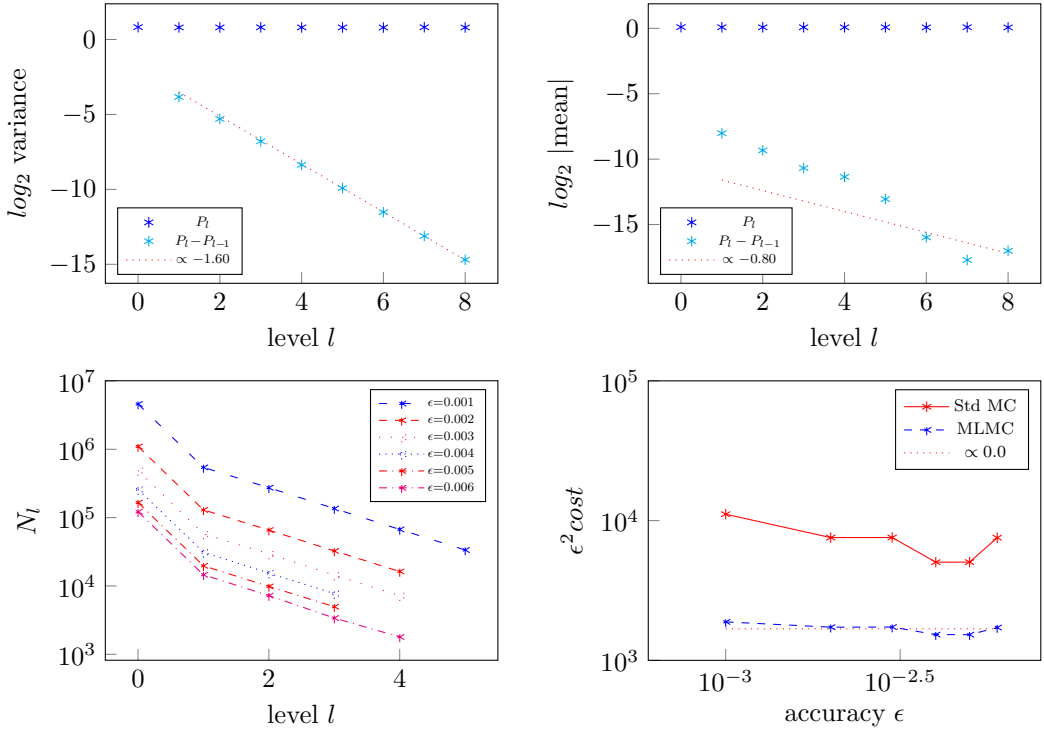
\def\mycaptionFour{MLMC applied to a the pricing of a swaption with a SDE driven by a L\'evy copula with margins \hyperref[appendix:ModelParameters]{\cgmyO} and \hyperref[appendix:ModelParameters]{\cgmyI}, \(\beta=0.4\)}
\newplotmlmc{pathconvergence=\pathgilesconvergence{sde}, pathapplied=\pathgilesapplied{sde}, modelname=cgmy02_cgmy04, bg=0.4, xmin=0.0006, xmax=0.008, ymin=10e2, ymax=10e4, thecaption=\mycaptionFour, label=fig:MLMCSde}{0.001,0.002,0.003,0.004,0.005,0.006}
\subsection{Comparison of the CTMC scheme against the series representation algorithm with a two-dimensional L\'evy copula}\label{NumApp:subsec:SeriesRepresentation}
In \cite{TankovSimulation2}, the author describes a methodology to simulate L\'evy copulas via a series representation truncated at a chosen level \(\tau\) (see \cite[Theorem 3]{TankovSimulation2}).
This algorithm is applied (in a simpler form) in \cite{TankovSimulation} to a L\'evy copula model with two Variance Gamma models and the Clayton copula function with coefficients: \(\theta=10\), \(\eta=0.75\) (strong tail dependence) and \(\theta=0.61\), \(\eta=0.99\) (weak tail dependence).
The Clayton copula is defined for any \(\theta >0\) and \(\eta \in [0,1]\) by:
\begin{equation}\label{eq:ClaytonCopula}
	F(u_1, \dots, u_d) = 2^{2 - d} \Par{\sum_{i=1}^{d} \Abs{u_i}^{-\theta}}^{\!-\frac{1}{\theta}} \Par{\eta \One{u_1 \dots u_d \geq 0} - (1-\eta)\One{u_1 \dots u_d < 0}},
\end{equation}
for \(u_1, \dots, u_d \in \overbar{\R}^d\)  (with \(\overbar{\R} := (-\inf, \inf]\)).\\
This setting is considered for two different option payoffs: an Asian call option (a call on the arithmetic average of the underlyings) and a Best-of call option (a call on the best of performances of the underlyings); the call options on the individual stocks are used as control variates.\\
We choose the same model parameters as in \cite{TankovSimulation}:
\begin{itemize*}
	\item \(S_1 = S_2= 1\),\; \(r=0.03\),\; \(d=0\),\; \(T=0.02\) (1 week option)
	\item \(VG_1\): \(\sigma_1=30\%,\; \nu_1=0.04,\; \theta_1=-0.2\)
	\item \(VG_2\): \(\sigma_2=20\%,\; \nu_2=0.04,\; \theta_2=-0.2\)
	\item the truncation parameter \(\tau\) is set to 2,000
\end{itemize*}
The spatial step \(h\) of the CTMC grid to \num{1e-6}, and the Monte-Carlo engine uses 100,000 paths in both cases.

Note that the confidence interval (``\(99\%\) CI") in the following Figures \ref{fig:seriesAsian} and \ref{fig:seriesBestOf} represents the addition of the two \(99\%\) confidence intervals of the CTMC and the series representation. The difference between the two cases comes from the Monte-Carlo error and the approximation error of both schemes.
The data is given in Appendix \ref{appendix:ModelParameters:Series} for completeness.

\remark{The series representation algorithm requires the inversion of a tail integral for each term of the (truncated) series. The resulting computing time can sometimes be a bottleneck in numerical applications as pointed out in the conclusion of the paper \cite{LiDelouxDieulle2016}. The CTMC \eqref{Scheme}, which does not suffer from such drawbacks,  provides a better performance in terms of speed; moreover, numerical testing showed that one can use a much less granular grid than a uniform grid with no significant deterioration in accuracy while reducing significantly the computing time.}

\plotBenchmarkingSeries{\DataForPaper/series}{asian}{Asian}{fig:seriesAsian}
\plotBenchmarkingSeries{\DataForPaper/series}{bestof}{Best-of}{fig:seriesBestOf}

\section{Monte-Carlo for functionals of a L\'evy process}\label{sec:mc}

\subsection{Preliminaries}\label{sec:prelims}

Given a L\'evy process \(Y\), we are interested in the computation of the quantity \(\E{f(Y)}\) where:
\begin{itemize*}
	\item \(f: D[0,T] \rightarrow \R\) is a Borel measurable function that is Lipschitz continuous with respect to the supremum norm on the space \(D[0,T]\) of c\`adl\`ag functions mapping \([0,T]\) to \(\Rd\).
	\item \(Y = (Y_t)_{0 \leq t \leq T}\) defines an \(\Rd\)-valued L\'evy process with L\'evy triplet \((\Sigma_Y, \lambda_Y, \mu_Y)_{\widetilde{c}}\) relative to the cut-off function \(\widetilde{c} := \mathds{1}_{[-V, V]^d}\) where \(V\) is either 1 or 0, the latter only if \(\int_{[-1,1]^d} |x| \lambda_Y(dx) < \infty\) (where \(\Abs{x}\) denotes the Euclidean norm).
\end{itemize*} 
For ease of presentation, we assume the Lipschitz constant of \(f\) is bounded above by 1, i.e. \(f \in Lip(1)\).

\subsection{Assumptions}\label{subsec:additionalAssumptions}

To perform the complexity analysis of the Monte-Carlo algorithm (section \ref{subsec:MC}), we require two additional assumptions and we assume that the L\'evy process \(Y\) satisfies (AS1) and (AS2) defined by:
\begin{align*}
	&\int_{\Rd \setminus [-1, 1]^d} |x|^2 \lambda_Y(dx) < \infty \label{def:AS1}\tag{AS1}\\
	&\max \left\{ h^2 \lambda_Y\left(\Rd \setminus [-h/2, h/2]^d \right), \int_{[-h,h]^d} |x|^2 \lambda_Y(ds) \right\} < \aleph \times h^{2 - \beta} \label{def:AS2}\tag{AS2}
\end{align*}
where \(\beta\) is the Blumenthal-Getoor index:
\begin{equation}\label{def:bg}
	\beta := \inf \left\{\eta>0:  \int_{[-1,1]^d} |x|^{\eta} \lambda_Y(dx) < \infty \right\}
\end{equation}
\remark{By definition of the L\'evy measure \(\int_{[-1,1]^d} |x|^2 \lambda_Y(dx) < \infty\), i.e. \(\beta \in [0,2]\).}

\remark{\
	\begin{enumerate}[label=(\roman*)]
		\item Under \eqref{def:AS1} the large jumps of \(Y\) (and hence the  marginals of \(Y\), see e.g. \cite[Ch. 5, Thm. 25.3]{sato2013levy}) have a finite second moment so that its law can be approximated effectively by a truncated L\'evy process \(X\) defined below.
		\item Assumption \eqref{def:AS2} gives the condition on the structure of the small jumps required for the convergence analysis. Note that \eqref{def:AS1} is satisfied under general conditions.
		\item We stress that neither the weak approximation for \(Y\) defined in \eqref{Scheme} nor the coupling of the Markov chain described in Section \ref{sec:coupling} require either \eqref{def:AS1} or \eqref{def:AS2}. Their role is to make the analysis of the Monte-Carlo algorithm based on \eqref{Scheme} feasible.
	\end{enumerate}
}

\subsection{The truncated L\'evy process \(X\)}\label{subsec:truncatedProcess}

We let \(X\) define a L\'evy process with the same characteristic triplet \(\Par{\Sigma, \lambda, \mu}_{\widetilde{c}}\) equal to \(Y\), i.e. \(\Sigma = \Sigma_Y\), \(\mu = \mu_Y\), except that its L\'evy measure is that of \(Y\) truncated at a large level \(R>0\).
\begin{equation}\label{eq:truncatedMeasure}
	\lambda(dx) := \One{[-R,R]}(x) \lambda_Y(dx)
\end{equation}
\remark{\
	\begin{enumerate}[label=(\roman*)]
		\item The main aim of this truncation is to reduce the problem of the simulation of the jumps in \eqref{Scheme} below to that of a random variable taking finitely many values.
		\item The precise value of \(R\) (as a function of the prescribed size of the mean-square error) will be determined later (see Sections \ref{subsec:MC} and \ref{subsec:errorAnalysis}). Note also that, if the jumps of \(Y\) have compact support, the truncation in \eqref{eq:truncatedMeasure} has no effect for \(R\) sufficiently large.
	\end{enumerate}	
}

The weak approximation of \(X\) in \eqref{Scheme} is based on the following representation into a sum of independent processes \(W=\Par{W_t}_{t \geq 0}\) and \(X^{\lambda} = \Par{X^{\lambda}_t}_{t \geq 0 }\).
\begin{eqnarray}\label{eq:decomposition}
	X_t = \mu t + \sigma W_t + X^{\lambda}_t
\end{eqnarray}
where \((W_t)_t\) is a standard \(d\)-dimensional Brownian motion, the matrix \(\sigma\) is a square root of \(\Sigma\) (i.e. \(\sigma \sigma^T = \Sigma\)) and \(X^{\lambda}\) is the pure-jump component of the process. \newline

The infinitesimal generator of \(X^{\lambda}\) is given by (see \cite[p. 208, Thm 31.5]{sato2013levy}):
\[
\forall f \in \mathcal{C}^2_0,\;\; \forall x \in \Rd, \;\; (L^{\lambda}\!f)(x) \!=\!\! \int_{\Rd} \!\left[\! f(x\!+\!y) \!-\! f(x)\! -\! \sum_{j=1}^d y_j  \partial_j f(x) \One{[-V, V]}(y) \right]\! \lambda(dy)
\]

\subsection{Continuous-Time Markov Chain Scheme}\label{subsec:CTMC}

We define a continuous-time Markov chain (CTMC) \(X^{h(\lambda)}\) approximating the pure-jump part of the process \(X^{\lambda}\) in decomposition \eqref{eq:decomposition}.\\
Let \(h \in (0,1)\) and consider the CTMC \((X^h_t)_{t \geq 0}\) having state space \(\Zd_h = h \Zd = \{hk : k \in \Zd\}\), initial state \(X^h_0 = 0\) and an infinitesimal generator \(L^h_{\lambda}\) given by a spatially homogeneous \(Q-\)matrix \(Q^h\) defined as (see \cite{MijatovicVidmarJacka12} for more details):
\begin{equation}\label{def:matrixQ}
	Q^h_{s s'} = \begin{cases}
		c^h_{s- s'} & \text{if } s- s' \ne 0.\\
		-\lambda(\Rd \setminus\! A^h_0), & \text{if } s - s' = 0
	\end{cases}
\end{equation}
where we have the following notations:
\begin{itemize*}
	\item
	for any \(s \in \Zd_h\),
	\begin{equation}\label{def:Ah}
		A^h_s := \prod_{j=1}^{d} I^h_{s_j}, \text{where, for any } r \in \Z_h, I_r^h = 
		\begin{cases} 
			\left[r-h/2, r+h/2\right)  & \text{if } r < 0.\\
			\left[-h/2, h/2\right]     & \text{if } r = 0.\\
			\left(r-h/2, r+h/2\right]  & \text{if } r > 0.
		\end{cases}
	\end{equation}
	\item for \(s \in \Zd_h \setminus\! \{0\}\), we set \(c^h_s := \lambda(A^h_s)\)
\end{itemize*}

The approximation process \((X^h_t)_t\) of \((X_t)_t\) is defined by the following scheme:
\begin{equation}\label{Scheme}\tag{\textit{Scheme}}
	X_t^h := (\mu + \widetilde{\mu})t + \sigma W_t + c^h B_t + X^{h(\lambda)}_t - \mu^{h(\lambda)}  t 
\end{equation}
where \(W\) and \(B\) are two independent \(d\)-dimensional Brownian motions, independent of the CTMC \(X^{h(\lambda)}\) constructed as in Section \ref{subsec:CTMC} and \(c^h\) is a square-root of the non-negative definite matrix \(C^h\) which components are given by:
\begin{equation}\label{def:matrixC}
	C^h_{i,j} := \int_{A^h_0} x_i x_j \One{[-V, V]^d}(x) \lambda(dx) 	
\end{equation}
This makes \(C^h \in \R^{d \times d}\) a symmetric, non-negative definite matrix. Let \(c^h \in \R^{d \times d}\) be \(c^h := U \Lambda^{1/2}\) where \(C^h = U \Lambda U^T\) is the eigen decomposition of \(C^h\) with \(\Lambda\) a diagonal matrix with non-negative elements and \(U\) a real orthogonal matrix. Note that \(C^h = c^h c^{h T}\) and \(\Sigma = \sigma \sigma^T\).

\remark{\label{rem:scheme}\
	The Brownian motions \((W_t)_{t \geq 0}\) and \((B_t)_{t \geq 0}\) are independent, hence \((\sigma W_t + c^h B_t)_{t \geq 0}\) is equal (in law) to \((\phi_h \widetilde{W}_t)_{t \geq 0}\) where \((\widetilde{W}_t)_{t \geq 0}\) is a Brownian motion and \(\phi_h\) is a square-root of \(\Sigma + C_h\), i.e. \(\phi_h^T \phi_h = \Sigma + C_h\).

The coordinates of the vector \(\widetilde{\mu}\) are given by:
\[\widetilde{\mu}_j = \int_{\Rd \setminus [-V, V]^d} y_j \lambda(dy), \quad j \in \Bracket{1,\dots,d}.\]
Furthermore, we have a well-defined drift vector of the compound Poisson process \(X^{h(\lambda)}\) given by
\begin{equation}\label{muh}
	\mu^{h(\lambda)} := \sum_{s \in \Zd \setminus \{0\}} s \lambda(A^h_s).
\end{equation}
Note that, due to the truncation in \eqref{eq:truncatedMeasure}, the components of \(\tilde{\mu}\) are finite and the sum in \eqref{muh} possesses only finitely many summands (and hence is finite).
\remark{The drift correction \(\tilde{\mu}\) appears in \eqref{Scheme} due to the representation of the L\'evy triplet with respect to the cut-off function \(\tilde{c} = \One{\SquaredBracket{-V, V}^d}\). The characteristics triplet of \(X\), \(\Par{\Sigma, \lambda, \mu}_{\tilde{c}}\), can be expressed as \(\Par{\Sigma, \lambda, \mu + \tilde{\mu}}_{\R}\). Hence, the drift in our scheme is nothing other than what is known as the ``center" drift\footnote{A straightforward calculation yields \(\gamma + \tilde{\mu} = \bar{\gamma} + \int_{\R^d \backslash [-V, V]^d} x \lambda(dx) := \gamma_c \) where \(\bar{\gamma}\) is the drift corresponding to the cut-off function \(c(x) = \One{\Abs{x} \leq 1} \) and \(\gamma_c\) is the \textit{center} drift corresponding to the cut-off function \(c(x) = \One{\R^d}(x)\).} of the process \(X\).}

\remark{Note that, for each coordinate of the vector \(\mu^{h(\lambda)}\), the formula \eqref{muh} simplifies to the sum over the coordinates \(s^i\) (of the \(i\)-th axis) of the product of \(s^i\) and the \(i\)-th marginal tail integral of the (truncated) L\'evy measure \(\lambda\) on the interval \(\SquaredBracket{s^i-\frac{h}{2}, s^i + \frac{h}{2}}\). 
	Consequently, it suffices to compute \(n\times d\) terms instead of \(n^d\) terms as in formula \eqref{muh}.}

\subsection{Monte-Carlo algorithm}\label{subsec:MC}
The expectation \(\E{f(Y)}\) is approximated by the estimator:
\begin{equation}\label{def:mcestimator}
	\widehat{S}_{n,h} = \frac{1}{n} \sum_{i=1}^n f\left ( X^{h, (k)} \right)
\end{equation}
where \(X^{h, (1)}, X^{h, (2)}, \dots, X^{h, (n)}\) are \(n\) independent copies of \(X^h\) given by \eqref{Scheme}.\\
The simulation of the path of \(X^h\) follows a two-step procedure:

\begin{algorithm}[H]
	\caption{Simulation of \( X^{h, (k)} \)}
	\label{algo:MCAlgorithm}
	\begin{algorithmic}[1]		
		\STATE
		\begin{enumerate}[label=(\theenumi\alph*)]
			\item Draw a Poisson random variable \(\mathcal{N} \sim \mathcal{P}\!\left( \nu \right) \) with  \( \nu\!=\!T\! \times\! \lambda( \Rd \setminus\! A^h_0) \)
			\item Draw \( \mathcal{N} \) IID jump sizes from the probability mass function on \( \Zd_h \) proportional to:
			\begin{equation}\label{algo:massFunction}
				s \rightarrow \lambda(A^h_s), s \in \Zd_h
			\end{equation}
		\end{enumerate}
		\STATE Simulate the required Brownian increments
	\end{algorithmic}
\end{algorithm}
\clearpage
\remark{\
	\begin{enumerate}[label=(\roman*)]
		\item  The first step of the algorithm above corresponds to the jump chain-holding time construction of CTMCs. The key point is that the jumps of the CTMC \(X^{\lambda(h)}\) take values in the finite set \([-R, R] \cap \Zd_h\) and can be simulated efficiently in the multidimensional context: the jumps are IID random variables taking at most a constant multiple of \((R/h)^d\) values and can hence be simulated at most
		\[ \bigO{d\Par{\log R + \log 1/h}} \] steps. The simplest such simulation algorithm is based on the complete binary tree method but further improvements (like Huffman's tree, the Alias and the Table Look-up methods) are possible, see the monograph \cite{Devroye} for possible algorithms for the simulation of finite-valued random variables. Furthermore, it follows from \cite{Devroye} that it is not always necessary to compute the normalizing constant of the distribution \(s \rightarrow \lambda\Par{A_s^h}\) in order to carry out the simulation.
		\item The simulation of the sample path of the CTMC \(X^{\lambda(h)}\) might require pre-computations, most notably of the mass function proportional to \eqref{algo:massFunction}. These point masses are linear combinations of tail integrals of the L\'evy measure, which are directly available in the case the multidimensional L\'evy measure \(\lambda_Y\) (and hence its truncation \(\lambda\) given in \eqref{eq:truncatedMeasure}) is specified via a L\'evy copula (see the seminal paper \cite{TankovKallsen} for details). As shown in \cite[Thm 3.6]{TankovKallsen}, L\'evy copulas provide a natural and completely general way of specifying multidimensional L\'evy measures. Hence the Monte-Carlo algorithm in \eqref{algo:MCAlgorithm} can be used to simulate \textbf{any} L\'evy process, as long as its L\'evy measure can be expressed in terms of a L\'evy copula in a tractable way.
		\item In the Monte-Carlo algorithm \ref{algo:MCAlgorithm} we assume that we can evaluate the path-functional \(f\) on the simulated path of \(X^h\). This is true if for example \(f\) depends on the values of the process at an arbitrary but finite set of times. If this is not the case (e.g. \(f\) is the payoff of an Asian option), a further discretization is needed. This straightforward extension appears naturally in the Euler scheme for L\'evy-driven SDEs and is hence incorporated in the analysis in Section \ref{subsec:mlmcAlgorithm:LevyDrivenSDE} below. For the sake of the exposition, in this Section we assume that the expected number of steps required to evaluate \(f(X^h)\) is \(\bigO{\lambda_Y \Par{\Rd \setminus\! A_0^h}}\) as \(h \rightarrow 0\), i.e. a constant factor times the expected number of jumps of \(X^h\).	
	\end{enumerate}\label{rem:algo}
}
\remark{\label{rem:jumptimes}
	Note that they are two ways to simulate the jumps: either simulate the arrival times and then the corresponding jump values or simulate the number of jumps and then the jump values. The latter case is simpler and can be used if the functional \(f\) does not depend on the arrival times, e.g. when \(f\) is the payoff of a call option, it only depends on the final value of the spot underlying, that is only on the sum of all the jump increments. If \(f\) depends on the structure of the path, e.g. \(f\) is the payoff of a CDS product (whose time to default can be modeled as the first time of a large negative instantaneous jump, as in Section \ref{NumApp:subsec:GarreauKercheval} below), then the simulation of the arrival times is needed.}

\subsection{Error analysis of the Monte-Carlo algorithm \ref{algo:MCAlgorithm}}\label{subsec:errorAnalysis}
The mean-square error of the estimator \eqref{def:mcestimator} is defined as the \(L^2\)-distance between the constant \(\E{f(Y)}\) and the random variable \(\hat{S}_{n, h}\).
\begin{equation}\label{def:mse}
	MSE(\widehat{S}_{n,h}) := \E{\left(\E{f(Y)} - \widehat{S}_{n,h} \right)^2} = Bias(h)^2 + \frac{1}{n} \Var\left(f(X^h)\right)
\end{equation}
where the bias, given by:
\begin{equation}\label{def:bias}
	Bias(h) := \left| \E{f(Y)} - \E{f(X^h)} \right|	
\end{equation}
depends only on \(h\) while the variance depends (asymptotically) only on the number of simulations \(n\) (see Section \ref{subsec:MCComplexity}). The task is now to control the two terms separately.

Since \(f \in Lip(1)\), it holds that:
\begin{equation}\label{eq:bias}
	Bias(h) \leq \E{\sup_{t \in [0,T]} ||Y_t - X_t ||_{\infty}} + \underbrace{\left|\E{f(X)} - \E{f(X^h)} \right|}_{:= Bias_X(h)}	
\end{equation}

The following lemma and proposition provide a bound for the first and second terms on the right hand side of this inequality (their proof is in  Appendix \ref{proof:mcErrorAnalysis}).\\

{\lemma{\label{lemma:MeanBound}
	Recall that \(Y\) satisfies condition \eqref{def:AS1} and \(X\) is defined as in Section \ref{subsec:truncatedProcess}. The following inequality holds:
	\[	\E{\sup_{t \in [0,T]} ||Y_t - X_t ||_{\infty}} \leq T \kappa(R), \quad \text{where  } \kappa(R) := \int_{\Rd \setminus A^R_0} ||x||_{\infty} \lambda_Y(dx) \]
	Furthermore, it holds that \(\kappa(R) = o(1)/R\) as \(R \rightarrow \infty\).
}}\\
\(||.||_{\infty}\) denotes the standard supremum norm: \(||x||_{\infty} := \sup\Par{\Abs{x_i}, i=1,\dots,d}\).
\remark{The function \(\kappa(R)\) may, in specific examples, decay faster than stated; e.g. if \(\lambda_Y\) has exponential tails, \(\kappa(R)\) will decay exponentially (see \eqref{def:Ah} for the definition of \(A_0^R\)).}

According to \eqref{def:bias} and Lemma \ref{lemma:MeanBound}, controlling the bias of the estimator \eqref{def:mcestimator} amounts to establishing a bound on the difference:
\[Bias_X(h) = \left| \E{f(X)} - \E{f(X^h)} \right|\]
for the L\'evy process \(X\) defined in Section \ref{subsec:truncatedProcess} and \(X^h\) given by \eqref{Scheme}. This is achieved in Proposition \ref{prop:MeanBound} below whose proof is deferred to Appendix \ref{proof:prop:MeanBound}.

\proposition{\label{prop:MeanBound}\
	\begin{itemize*}
		\item[(I)] Assume that for some \(\theta \geq 1\), we have:
		\[
		\int_{A^h_0} \langle y', x \rangle^2 \lambda_Y(dx) \leq \theta \int_{A^h_0} \langle y, x \rangle^2 \lambda_Y(dx)
		\]
		for any \(y, y' \in \R\) with \(|y| = |y'| = 1\).\\ Then the following inequality holds for any time horizon \(T > 0\) and all small \(h > 0\)
		\[
		Bias_X(h)^2 \leq 8Tdh^2 \lambda_Y(\Rd \setminus\! A^h_0) + 2 \frac{h^2 \theta}{c_1} \left(\! 2 + c_2 \log \max \left\{\!e, \frac{T}{h^2}\! \int_{A^h_0}\!\! |x|^2 \lambda_Y(dx) \right\} \right)^2
		\]
		where constant \(c_1\) and \(c_2\) depend only on the dimension \(d\) of the L\'evy process \(Y\).
		\item[(II)] If in addition \(Y\) satisfies \eqref{def:AS2} and the Blumenthal-Getoor index \(\beta\) is in \([0,2)\), then for any \(h < \exp(\beta)\)
		\[
		Bias_X(h)^2 \leq h^{2-\beta} \left[8T\aleph + 8h^{\beta} \frac{\theta}{c_1} \Par{2 + c_2 \log^2(\aleph T) + c_2^2 \beta^2 \log^2(1/h)} \right]
		\]		 
	\end{itemize*}	
}

\remark{\label{Remark:MCBias}\
	\begin{enumerate}[label=(\roman*)]
		\item Under assumption \eqref{def:AS2}, the following limits hold: \(\lim_{h \downarrow 0} h^2 \lambda_Y \Par{\Rd \setminus\! A_0^h} = 0\) and \\
		\(\lim_{h \downarrow 0} \int_{A^h_0} \Abs{x}^{\beta} \lambda_Y \Par{dx} = 0\). The same holds for the truncated L\'evy measure \(\lambda\) in \eqref{eq:truncatedMeasure}.
		\item The condition \(\int_{\SquaredBracket{-1, 1}} \Abs{x}^{\beta} \lambda_Y \Par{dx} < \infty\), which is satisfied e.g. if \(\lambda_Y\) has  density in some neighborhood of the origin in \(\Rd\) asymptotically equivalent to \(x \to \Abs{x}^{\delta}\) for some power \(\delta\), implied the assumption in \eqref{def:AS2}.
	\end{enumerate}
	\label{rem:meanMCBound}
}%
The claims in Remark \ref{rem:meanMCBound} are proven in Appendix \ref{proof:prop:MeanBound}.\\
The next proposition provides a upper bound for the variance of the estimator \eqref{def:mcestimator} (the proof is given in Appendix \ref{proof:mcErrorAnalysis}):
\proposition{\label{prop:VarianceBound}
	Under \eqref{def:AS1} and \eqref{def:AS2}, the following inequality holds for all \(h > 0\) and any \(f \in Lip(1)\)
	\[
	\Var \left(f(X^h) \right) \leq 3 \left( 2T^2 (|\mu|^2 +\! M_{\mu}^2) + 8Td ||\sigma||_{\infty}^2 +4T\! \int_{\Rd}\! |x|^2 \lambda_Y(dx) + 12dT \aleph h^{2-\beta}  \right)
	\]	
	where \( M_{\mu} = \int_{\Rd \setminus [-V,V]^d} ||y||_1 \lambda_Y(dx)\)
}
\remark{Recall that by definition \(V\) is either 1 or 0, the latter only if \(\int_{\SquaredBracket{-1, 1}} \Abs{x} \lambda_Y \Par{dx} < \infty\).}

\subsection{Complexity analysis of the Monte-Carlo algorithm \ref{algo:MCAlgorithm}}\label{subsec:MCComplexity}
Under \eqref{Scheme}, the form of the mean-square error in Lemma \ref{lemma:MeanBound}, Proposition \ref{prop:MeanBound} and Proposition \ref{prop:VarianceBound} imply that:
\begin{equation}\label{eq:mse}
	MSE(\widehat{S}_{n,h}) \leq \frac{D_V}{n} + D_B h^{2-\beta} + \frac{D_T}{R^2}	
\end{equation}
for the constants \(D_V = (T \lor 1)^2 d^3 D_0\), \(D_B = (T \lor 1) d D_0\), \(D_T = T D_0\), where \(D_0\) depends neither on the time horizon \(T\) nor on the dimension \(d\). Given a root mean-square error \(\epsilon\), we can distribute it equally among the summands leading to:
\begin{equation}\label{def:hRn}
	h = (3D_B)^{-1/(2-\beta)}\epsilon^{2/(2-\beta)}, \quad R = \sqrt{3D_T} \epsilon^{-1}, \quad n = \lceil 3 D_V \epsilon^{-2} \rceil
\end{equation}
By Remark \ref{rem:algo} (i) and (iii) and the definition in \eqref{Scheme}, the expected cost \(\mathcal{C}_h\) of generating a value of \(f(X^h)\) satisfies:
\begin{equation}\label{mc:cost}
	\mathcal{C}_h \leq D_1 T \lambda(\Rd \setminus\! A^h_0) d[1 + \log(R/h)] \leq D_1 \aleph T h^{-\beta} d [1 + \log(R/h)]
\end{equation}
where the constant \(D_1 > 0\) depends neither on the process \(Y\) nor on \(T, d\), and the second inequality comes from \eqref{def:AS2}. In this estimate, we assume that the generation of a Poisson random variable with intensity \(T \lambda \Par{\Rd \setminus\! A_0^h}\) can be carried out in \(\bigO{1}\) steps for any \(h > 0\) (see e.g. \cite{Devroye}) and that the Brownian increments and the functional evaluation of \(f\) require at most \(\bigO{dT \lambda \Par{\Rd \setminus\! A_0^h}}\) operations. Hence the computational complexity \(\mathcal{C}(\epsilon) = n \mathcal{C}_h\) of the Monte-Carlo algorithm in \eqref{algo:MCAlgorithm}, i.e. the expected cost required to ensure \(MSE\Par{\hat{S}_{n,h}} \leq \epsilon^2\), is by \eqref{eq:mse}---\eqref{mc:cost} bounded above by:
\begin{equation}\label{eq:complexityCost}
	\mathcal{C}(\epsilon) \leq D_2 \epsilon^{-2 - 2 \beta / (2 - \beta)} \SquaredBracket{\log \Par{1/\epsilon} + D_3}
\end{equation}
since by \eqref{def:hRn} it holds that \(1 + \log (R/h) = \SquaredBracket{\log(1/\epsilon) + D_3}(4-\beta)/(2 - \beta)\), where the constants equal
\[
D_2 := D_1 T  \aleph \Par{3D_B}^{\beta/(2-\beta)}\Par{1 + 3D_V} \frac{4 - \beta}{2- \beta}, \quad D_3 := \Par{1 + \log \Par{\sqrt{3D_T} \Par{3D_B}^{1/(2-\beta)}}}\frac{2- \beta}{4 - \beta}
\]
Note that all the constants that appear in \eqref{eq:complexityCost} depend explicitly on the characteristics of the process \(Y\) and the parameters \(T\) and \(d\).

\section{Multilevel Monte-Carlo algorithm for L\'evy-driven SDEs}\label{sec:mlmcAlgorithm}

\subsection{The need for coupling}\label{subsec:mlmcAlgorithm:Background}
In Section \ref{sec:mc} we developed a Monte-Carlo algorithm for computing expectations of the path-functionals of L\'evy processes and found its complexity \eqref{eq:complexityCost}. In order to reduce it, it is natural to attempt to apply ideas from the multilevel Monte-Carlo to the estimator in \ref{def:mcestimator}.\\
The first step in MLMC towards computing \(\E{f(Y)}\) is to approximate it by \(\E{f(X^{h_L})}\) and use the telescopic sum \(\E{f(X^{h_L})} = \E{f(X^{h_0})} + \sum_{l=1}^{L} \E{f(X^{h_l}) - f(X^{h_{l-1}})}\), where \(X^{h}\) is a stochastic approximation of \(Y\), which converges weakly to \(Y\) as \(h \to 0\). Here \(L \in \N^*\) is a fixed number of levels, \(h_l = h_0 2^{-l}, l=0, 1, \dots, L\), for some \(h_0 > 0\) and the MLMC estimator \(\hat{S}\) is the unbiased estimator of the right-hand side (cf. Section \ref{subsec:mlmcAlgorithm:Algo} below) with the mean-square error:
\[
MSE(\widehat{S}) = \Abs{\E{f\Par{Y}} - \E{f\Par{X^{h_L}}}}^2 + \frac{1}{N_0}\Var\SquaredBracket{f\Par{X^{h_0}}} + \sum_{l=1}^{L} \frac{1}{N_l}\Var\SquaredBracket{f\Par{X^{h_l}} - f\Par{X^{h_{l-1}}}} 
\]
Note in particular that this representation of \(MSE(\widehat{S})\) assumes that (1) the random elements \(X^{h_0}\), \(\Par{X^{h_l}, X^{h_{l-1}}}, l=1,\dots,L\), are independent, which is easy to satisfy, and, more importantly, that (2) each pair \(\Par{X^{h_l}, X^{h_{l-1}}}\) is defined jointly on some probability space.
To apply Giles' Complexity Theorem (see \cite{Giles15}), the following type of bounds are sought:
\begin{equation}\label{Giles:condition}
	 \Var\SquaredBracket{f\Par{X_T^{h_l}} - f\Par{X_T^{h_{l-1}}}} \leq C_2 h_l^B, \quad \Abs{\E{f\Par{X_T^{h_l}} - f\Par{Y_T}}} \leq C_1 h_l^A,\quad \mathcal{C}_h \leq C_3 h^{-G}
\end{equation}
where \(\mathcal{C}_h\) denotes the expected computational cost of simulating one realization of \(X_T^h\). The computational complexity \(\mathcal{C}(\epsilon)\) of the multilevel estimator needed to achieve the accuracy \(\epsilon\) (in terms of the root mean-square error) is proportional to \(\epsilon^{-2} \Par{\log \epsilon}^2\) if \(B=G\) and \(\epsilon^{-2 - (G-B)/{A}}\) if \(B < G\).

\remark{The standard way of checking the level variance satisfies assumption \eqref{Giles:condition} is to prove that the underlying approximation scheme has weak convergence of order \(A\) and strong convergence of order \(B/2\). Indeed, in this case it holds that:
\[ \Var\SquaredBracket{ f\Par{ X_T^{h_l} } - f\Par{ X_T^{h_{l-1}} } } \leq \E{ \Abs{X_T^{h_l} - Y_T}^2 } + \E{ \Abs{X_T^{h_{l-1}} - Y_T}^2 } \leq 2 C_0 h^B_{l-1}\] for any \(f \in Lip(1)\) and some constant \(C_0 > 0\). However, as noted before, our approximation \eqref{Scheme} from Section \ref{sec:mc} is weak. In order to circumvent the issue of its strong convergence to the limiting process \(Y\), we introduce a coupling between consecutive approximation levels of \eqref{Scheme}, which (a) ensures that the two processes are strongly dependent, thus enabling us to prove directly that the level variance decays sufficiently fast without having to refer to the strong limit and (b) allows us to jointly simulate the pair of processes from the neighboring levels. In the next section, we give an intuitive description of this coupling and its basic properties.}

\subsection{Intuitive description of the coupling of \(X^h\) and \(X^{2h}\) and its fundamental properties}\label{subsec:mlmcAlgorithm:Intuitive}
The aim is to define the processes \(X^h\) and \(\widetilde{X}^{2h}\), marginally given by \eqref{Scheme} for \(h\) and \(2h\), on the same probability space so that their respective trajectories stay close to each other with as high probability as possible. It is intuitively clear that such a property would reduce the variance of the difference \(f\Par{X^{h_l}} - f\Par{X^{h_{l-1}}}\).\\ Coupling by:
\begin{align*}
	X^h_t &= (\mu + \widetilde{\mu})t + \sigma W_t + c^h B_t + X^{h(\lambda)}_t - \mu^{h(\lambda)}t  \\
	\widetilde{X}^{2h}_t &= (\mu + \widetilde{\mu})t + \sigma W_t + c^{2h} B_t + \widetilde{X}^{{2h}(\lambda)}_t - \mu^{{2h}(\lambda)} t
\end{align*} based on the same Brownian motions \(W\) and \(B\) is a natural choice.  
The key construction, given in Section \ref{sec:coupling} below, defines a coupling of the compound Poisson processes \(X^{h(\lambda)}\) and \(\widetilde{X}^{2h(\lambda)}\).\\
The basic idea is to define the paths of \(\widetilde{X}^{2h(\lambda)}\) using those of \(X^{h(\lambda)}\) and some ``additional randomness" independent of the process \(X^{h(\lambda)}\). Put differently, as we shall see in Section \ref{sec:coupling}, we both thin and transform the Poisson point process of jumps of \(X^{h(\lambda)}\) to obtain a trajectory of \(\widetilde{X}^{2h(\lambda)}\). More precisely, if a jump size \(s\) of \(X^{h(\lambda)}\) is in \(\Zd_{2h}\) (it is also in \(\Zd_h\)), then we take that jump also to be a jump of \(\widetilde{X}^{2h(\lambda)}\). If not (i.e. \(s \in \Zd_h \setminus\! \Zd_{2h}\)), then we use an independent random variable to determine a point \(\widetilde{s}\) in \(\Zd_{2h}\), which is close to \(s\), and define the jump of \(\widetilde{X}^{2h(\lambda)}\) to be \(\widetilde{s}\). Our construction of the coupling allows us to control the difference in jump sizes of the processes \(X^{h(\lambda)}\) and \(\widetilde{X}^{2h(\lambda)}\). In fact this difference is bounded above by \(h\). In particular, thinning of the Poisson point process of jumps of \(X^{h(\lambda)}\) may occur with positive probability at a jump time of \(X^{h(\lambda)}\) only if that jump of \(X^{h(\lambda)}\) is of the size \(h\) according to the norm \(\NormSup{\cdot}\) (see Remark \ref{rem:couplingThm} below for a precise description).\\
In order to base a successful MLMC scheme on the coupling we have just described, we need to be able to (1) efficiently simulate the process  \(\Par{X^h, \widetilde{X}^{2h}}\) and (2) control the level variances:
\begin{enumerate}[label={(\arabic*)}]
	\item First simulate the trajectories of \(X^h\). At every jump time \(t\) of \(X^h\), conditional on the jump size \(\Delta X^h_t = s \in \Zd_h \setminus\! \Zd_{2h}\), the difference in jump sizes \(\Delta \widetilde{X}^{2h}_t - \Delta X^h_t\) is distributed according to the mass function \(p^{\lambda}(s, m) = \frac{\lambda\left(A^{2h}_{s+m} \bigcap A^h_s \right)}{\lambda(A^h_s)}\) where \(m \in \Zd_h\), such that \(||m||_{\infty} = h\) and \(s+m \in \Zd_{2h}\), and the boxes of the form \(A_s^h\) are defined in \eqref{def:Ah}. A pictorial description of this coupling of jump sizes is presented in Figure \eqref{fig:gridExample}. It is important to note that the mass function \(p(s,\cdot)\) is explicit in the L\'evy measure of \(X\) and non-zero only on a finite set. Furthermore, \(p(s,\cdot)\) is easily computable if the jumps of \(X\) are described by a L\'evy copula making the coupling easy to simulate. Also note that, as stated in Remark \ref{rem:jumptimes}, the precise knowledge of the jump times might not be required, and that only the jump sizes play a part in the coupling.
	\item If the process \(Y\) satisfies \eqref{def:AS1} and  \eqref{def:AS2}, and its Blumenthal-Getoor index \(\beta\), defined in \eqref{def:bg}, is in the interval \([0,2)\), for all \(h > 0\), it holds:
	\begin{equation}\label{eq:couplingVariance}
		\Var \Par{ f\Par{X^h} - f\Par{\widetilde{X}^{2h}} } \leq D_V^{ML} h^{2 - \beta}, \quad \text{with: } D_V^{ML} := 80 dT \aleph
	\end{equation}
	making the MLMC scheme feasible. A more general result on the asymptotic behavior of the level variance is given in Proposition \ref{prop:MLVariance} below.
\end{enumerate}

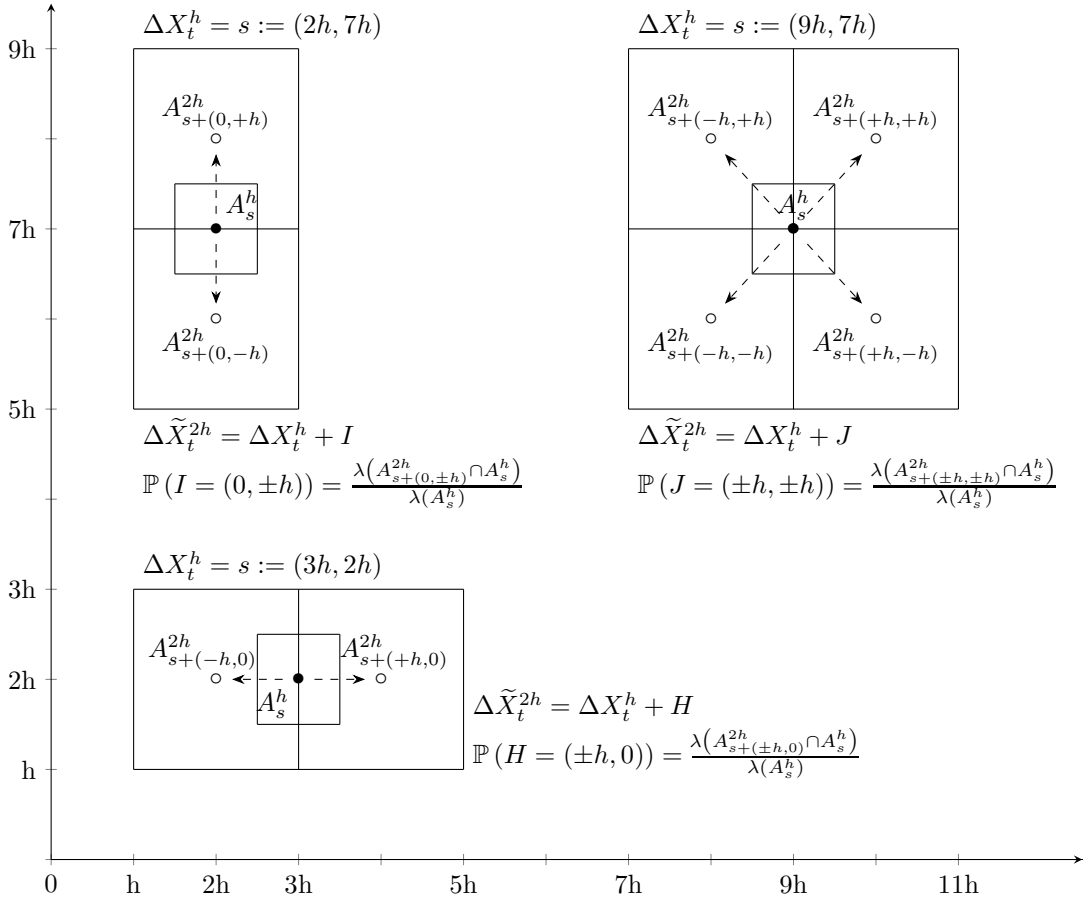
\begin{figure}
	\centering
	\begin{tikzpicture}
		\begin{axis}[
			scale       = 2.0,
			axis x line	= left,
			axis y line	= left,
			xmin = 0, xmax = 12.5,
			ymin = 0, ymax = 9.5,
			xtick 		= {0,...,11},
			ytick 		= {0,...,9},
			xticklabels = {0, h, 2h, 3h,, 5h,, 7h,, 9h,, 11h},
			yticklabels = {,h, 2h, 3h,, 5h,, 7h,, 9h},
			]
			\draw[->]  (-0.3, 0) -- (22.2, 0);
			\draw[->]  (-0.3, 0) -- (0, 22.2);
			
			\node (X) at (2, 7) {\textbullet};	\node[above right] at (X) {\(A^h_s\)};
			\node (T) at (2, 8) {\(\circ\)};		\node[above]       at (T) {\(A^{2h}_{s + (0,+h)}\)};
			\node (B) at (2, 6) {\(\circ\)};		\node[below]       at (B) {\(A^{2h}_{s + (0,-h)}\)};
			\node[above right] at (1,9) {\(\Delta X^h_t = s := (2h, 7h)\)};
			\node[below right] at (1,5) {\(\Delta \widetilde{X}^{2h}_t = \Delta X^h_t + I\)};
			\node[above right] at (1,3.8) {\(\Prob{I = (0, \pm h)} = \frac{\lambda \Par{A^{2h}_{s + (0,\pm h)} \cap A^h_s}}{\lambda \Par{A^h_s}}\)};
			
			\draw[dashed, ->, >=Stealth] (X) -- (T); 
			\draw[dashed, ->, >=Stealth] (X) -- (B);
			\draw[-] (1, 7) -- (3, 7);
			\draw[-] (1, 9) -- (3, 9);	\draw[-] (1, 5) -- (3, 5);	\draw[-] (1, 5) -- (1, 9);	\draw[-] (3, 5) -- (3, 9);
			\draw[-] (1.5, 7.5) -- (2.5, 7.5);	\draw[-] (1.5, 6.5) -- (2.5, 6.5);	\draw[-] (1.5, 6.5) -- (1.5, 7.5);	\draw[-] (2.5, 6.5) -- (2.5, 7.5);

			\node (Y) at (3, 2) {\textbullet};	\node[below left] at (Y) {\(A^h_s\)};
			\node (L) at (2, 2) {\(\circ\)};		\node[above] at (L) {\(A^{2h}_{s + (-h,0)}\quad \)};
			\node (R) at (4, 2) {\(\circ\)};		\node[above] at (R) {\(\quad A^{2h}_{s + (+h,0)}\)};
			\node[above right] at (1,3) {\(\Delta X^h_t = s := (3h, 2h)\)};
			\node[below right] at (5,2) {\(\Delta \widetilde{X}^{2h}_t = \Delta X^h_t + H\)};
			\node[above right] at (5,0.8) {\(\Prob{H = (\pm h, 0)} = \frac{\!\lambda \Par{A^{2h}_{s + (\pm h, 0)} \cap A^h_s}}{\lambda \Par{A^h_s}}\)};

			\draw[dashed, ->, >=Stealth] (Y) -- (L); 
			\draw[dashed, ->, >=Stealth] (Y) -- (R);
			\draw[-] (3, 1) -- (3, 3);
			\draw[-] (1, 1) -- (1, 3);	\draw[-] (5, 1) -- (5, 3);	\draw[-] (1, 1) -- (5, 1);	\draw[-] (1, 3) -- (5, 3);
			\draw[-] (2.5, 1.5) -- (2.5, 2.5);	\draw[-] (3.5, 1.5) -- (3.5, 2.5);	\draw[-] (2.5, 1.5) -- (3.5, 1.5);	\draw[-] (2.5, 2.5) -- (3.5, 2.5);

			\node at (9, 7) {\textbullet};
			\node (Z)  at ( 9, 7) {\textbullet};\node[above] at (Z) {\(A^h_s\)};
			\node (TL) at ( 8, 8) {\(\circ\)};		\node[above] at (TL) {\(A^{2h}_{s + (-h,+h)}\)};
			\node (TR) at (10, 8) {\(\circ\)};		\node[above] at (TR) {\(A^{2h}_{s + (+h,+h)}\)};
			\node (BL) at ( 8, 6) {\(\circ\)};		\node[below] at (BL) {\(A^{2h}_{s + (-h,-h)}\)};
			\node (BR) at (10, 6) {\(\circ\)};		\node[below] at (BR) {\(A^{2h}_{s + (+h,-h)}\)};
			\node[above right] at (7,9) {\(\Delta X^h_t = s := (9h, 7h)\)};
			\node[below right] at (7,5) {\(\Delta \widetilde{X}^{2h}_t = \Delta X^h_t + J\)};
			\node[above right] at (7,3.8) {\(\Prob{J = (\pm h, \pm h)} = \frac{\!\lambda \Par{A^{2h}_{s + (\pm h,\pm h)} \cap A^h_s}}{\lambda \Par{A^h_s}}\)};
			
			\draw[dashed, ->, >=Stealth] (Z) -- (TL); 
			\draw[dashed, ->, >=Stealth] (Z) -- (TR); 
			\draw[dashed, ->, >=Stealth] (Z) -- (BL);
			\draw[dashed, ->, >=Stealth] (Z) -- (BR);
			\draw[-] (7, 7) -- (11, 7); \draw[-] (9, 5) -- (9, 9);
			\draw[-] (7, 5) -- (7, 9);	\draw[-] (11, 5) -- (11, 9);	\draw[-] (7, 5) -- (11, 5);	\draw[-] (7, 9) -- (11, 9);
			\draw[-] (8.5, 6.5) -- (8.5, 7.5);	\draw[-] (9.5, 6.5) -- (9.5, 7.5);	\draw[-] (8.5, 6.5) -- (9.5, 6.5);	\draw[-] (8.5, 7.5) -- (9.5, 7.5);
		\end{axis}
	\end{tikzpicture}	
	\caption{This figure gives a graphical description of the coupling \(\Par{X^h, \widetilde{X}^{2h}}\). More precisely it depicts the coupling of the jumps of the two chains in the case \(\Delta X^h_t\) is in the set \(\Zd_{h} \setminus \Zd_{2h}\) (here \(d=2\) and recall that box \(A^h_s\) is defined in \eqref{def:Ah}). If \(\Delta X^h_t \in \Zd_{2h}\), then the jumps of the coupled chains coincide. The random variables \(I, J, H\) are independent of the jump \(\Delta X^h_t\) with distributions given in the figure, making the coupling very easy to simulate.}
	\label{fig:gridExample}
\end{figure}

\subsection{The multilevel Monte-Carlo algorithm}\label{subsec:mlmcAlgorithm:Algo}
In this Section, we develop a multilevel version of the Monte-Carlo algorithm described in Section \ref{sec:mc}, based on the coupling from Section \ref{sec:coupling}.\\
Given the Monte-Carlo algorithm \ref{algo:MCAlgorithm}, to compute \(\E{f(Y)}\) at a lower computational cost, it is natural to define a multilevel scheme in space (rather than time) as follows: for \(L \in \N^*\), fix a sequence \(h_l := h_0 2^{-l}, l=0,1,\dots,L\), for some \(h_0 \in (0,1]\). Let \(Y\) satisfies \eqref{def:AS1} and then define the multilevel estimator \(\widehat{S}\) by:
\begin{equation}\label{def:mlmcEstimator}
	\widehat{S} := \frac{1}{N_0} \sum_{k=1}^{N_0} f\left(X^{h_0, (k)} \right) + \sum_{l=1}^{L} \frac{1}{N_l} \sum_{k=1}^{N_l} \left[f\left(X^{h_l, (k)} \right) - f\left(\widetilde{X}^{h_{l-1}, (k)} \right) \right]
\end{equation}
In \eqref{def:mlmcEstimator} we simulate \(N_0\) independent copies \(X^{h_0, (1)}, \dots, X^{h_0, (N_0)}\) of \(X^{h0}\) given in \eqref{Scheme} and \(N_l\) independent copies \(\Par{ X^{h_l, (k)}, \widetilde{X}^{h_{l-1}, (k)} }\) of the process \(\Par{X^{h_l}, \widetilde{X}^{h_{l-1}}}\) defined in \eqref{def:couplingScheme}. The simulation of the coupled compound Poisson processes in \eqref{def:couplingScheme} is carried out as described in Theorem \ref{thm:coupling}. The simulated processes \(X^{h_0}, \Par{X^{h_1}, \widetilde{X}^{h_0}}, \dots, \Par{X^{h_l}, \widetilde{X}^{h_{l-1}}}\) are assumed to be independent. As in the Monte-Carlo algorithm \ref{algo:MCAlgorithm}, we assume that we can evaluate the path-functional \(f\) on any simulated trajectory of the process \(X^{h_l}\).\\
\begin{algorithm}[H]
	\caption{Simulation of \((X^{h_l}, \widetilde{X}^{h_{l-1}})\)}
	\label{algo:MLMCAlgorithm}\
	\begin{flushleft}for each level \(l\) in \(\SquaredBracket{0,L}\)\end{flushleft}
	\begin{algorithmic}[1]
		\STATE Simulate the jumps of \(X^{h_l}\) as in Algorithm \ref{algo:MCAlgorithm}
		\STATE For each jump size of \(X^{h_l}\), use the coupling described in \eqref{def:couplingScheme} to simulate the jump size of \(\widetilde{X}^{h_{l-1}}\)
		\STATE Simulate the Brownian motion increments
	\end{algorithmic}
\end{algorithm}
It follows from the coupling in Theorem \ref{thm:coupling} that the laws of \(\widetilde{X}^{l-1}\) and \(X^{l-1}\) coincide and hence \(\E{\hat{S}} = \E{f\Par{X^{h_0}}} + \sum_{l=1}^{L} \E{f(X^{h_l}) - f(\widetilde{X}^{h_{l-1}})} = \E{f(X^{h_L})}\). Therefore the mean-square error \(MSE(\hat{S})= \E{\Par{\E{\hat{S}} - \hat{S}}^2}\) takes the form:
\begin{equation}\label{eq:mlmcMse}
	MSE(\hat{S}) = Bias(h_L)^2 + \frac{V_0}{N_0} + \sum_{l=1}^{L} \frac{V_l}{N_l}, \quad \text{where  } V_l := \Var \SquaredBracket{f\Par{X^{h_l}} - f\Par{\widetilde{X}^{h_{l-1}}}}
\end{equation}
for \(l=1,\dots,L\), \(Bias(h_l)\) is defined in \eqref{def:bias} and \(V_0 := \Var \SquaredBracket{f(X^{h_0})}\).\\
In order to perform the complexity analysis of the multilevel estimator in \eqref{def:mlmcEstimator}, we need to bound the expected costs \(\mathcal{C}_l, l=1,\dots,L\) (resp. \(\mathcal{C}_0\)) for computing a sample path of the process \(\Par{X^{h_l}, \widetilde{X}^{h_{l-1}}}, l=1,\dots,L\) (resp. \(X^{h_0}\)) defined in \eqref{def:couplingScheme} (resp. \eqref{Scheme}). Under \eqref{Scheme}, the cost \(\mathcal{C}_0\) is equal to \(\mathcal{C}_{h_0}\) and has been estimated in \eqref{mc:cost} in terms of \(h_0\) and the truncation level \(R\). Furthermore, it follows from Theorem \ref{thm:coupling}, that the expected computational cost \(C_l\) of simulating \(f\!\Par{X^{h_l}} - f\!\Par{\widetilde{X}^{h_{l-1}}}\), where the coupling \(\Par{X^{h_l}, \widetilde{X}^{h_{l-1}}}\) is given in \eqref{def:couplingScheme}, satisfies:
\begin{equation}\label{eq:mlmcCost}
	C_l \leq T \lambda(\Rd \setminus\! A^{h_l}_0) D_1^{ML} d (1 + \log(R/h_l)) \leq d T \aleph D_1^{ML}h_l^{-\beta} (1 + \log(R/h_l))  
\end{equation}
for any \(l=1, \dots, L\) and some constant \(D_1^{ML} \geq D_1 > 0\) that, as in \eqref{mc:cost}, depends neither on the process \(Y\) nor on \(T\), \(d\). The second inequality in \eqref{eq:mlmcCost} follows from \eqref{def:AS2}. Note that by \eqref{mc:cost} the bounds in \eqref{eq:mlmcCost} holds also for \(l=0\).\\
We now derive a bound for the minimal total expected cost \(\mathcal{C}^{ML}(\epsilon) = \sum_{l=0}^{L} N_l \mathcal{C}_l\), under the constraint \(MSE(\hat{S}) \leq \epsilon^2\). Stipulating that \(Bias(h_L)^2 \leq \epsilon^2/2\), we can, as shown by Giles in \cite{Giles15}, treat \(N_l\) as continuous variables and use the Lagrange multipliers, yielding to:
\begin{equation}\label{eq:mlmcTotalCost}
	\mathcal{C}^{ML}(\epsilon) = \epsilon^2 \left(\sum_{l=0}^{L} \sqrt{V_l \mathcal{C}_l}\right)
\end{equation}
Hence it is key to determine whether a bound on the function \(l \to V_l \mathcal{C}_l\) is increasing, decreasing, or constant in \(l\). Recall that the constant \(D_V\) (resp. \(D_V^{ML}\)) is given in \eqref{eq:mse} (resp. \eqref{eq:mlmcCost}). For any \(l=0,1,\dots,L\), the following inequality holds:
\begin{equation}\label{eq:vlcl}
	V_l \mathcal{C}_l \leq \bar{D} \SquaredBracket{\log \Par{1/\epsilon} + D_3}2^{-(2 - 2 \beta)/l}, \quad \text{where: } \bar{D} := \Par{D_V^{ML} + D_V} d T \aleph D_1^{ML} \frac{4 - \beta}{2- \beta} h_0^{-\beta}
\end{equation}
by the representation for \(1 + \log \Par{R/h_l}\) following \eqref{eq:complexityCost}. Analogous to \eqref{def:hRn}, the control on the bias of the MLMC scheme in \eqref{def:mlmcEstimator} requires \(R = \sqrt{3D_T}\epsilon^{-1}\) and \(h_l = \Par{3D_B}^{-1/(2-\beta)} \epsilon^{2/(2-\beta)}\). As \(h_L = h_0 2^{-L}\), we hence find that the number of levels \(L\) in our MLMC scheme is bounded above by:
\begin{equation}\label{eq:mlmcMaxLevel}
L \leq \left\lfloor \frac{\log{3D_B h_0^{2-\beta}} - 2 \log{\epsilon}}{(2- \beta) \log{2}} \right\rfloor 
\end{equation}
where \( \lfloor x \rfloor\) denotes the smallest integer \(n\) such that \(n \leq x < n + 1\).\\
In the case of our scheme \eqref{def:mlmcEstimator}, the representation of \(\mathcal{C}^{ML}(\epsilon)\) in \eqref{eq:mlmcCost} can be bounded above by:
\begin{equation}\label{eq:mlmcCostBounds}
	\mathcal{C}^{ML}(\epsilon) \leq \bar{D} \epsilon^{-2} (\log(1/\epsilon) + D_3) \times	
	\begin{cases}
		D_5, 									& 0 \leq \beta < 1\\
		\log^2(1/\epsilon), 					& \beta = 1\\
		D_6 \epsilon^{-4(\beta-1)/(2-\beta)},	& 1 < \beta < 2
	\end{cases}
\end{equation}
where \(D_5=(1 - 2^{-(1-\beta)})^{-2}\) and \(D_6 = D_5 (h_0(3D_B)^{1/(2-\beta)})^{2(\beta-1)}\).

\subsection{MLMC for L\'evy-driven SDE}\label{subsec:mlmcAlgorithm:LevyDrivenSDE}
Let \(Y\) be a \(\R^d\)-valued L\'evy process from Section \ref{subsec:additionalAssumptions} with the L\'evy triplet \(\Par{\Sigma_Y, \lambda_Y, \mu_Y}_{\widetilde{c}}\). Consider the L\'evy-driven stochastic equation
\begin{equation}\label{eq:sde}
	Z_t = z_0 + \int_0^ta(Z_{s^-}) dY_s, \quad z_0 \in \R^m
\end{equation}
Following \cite{Dereich2011} we assume the coefficient \(a: \R^m \to \R^{m \times d}\) and the driving L\'evy process \(Y\) satisfy
\begin{equation}\label{eq:sdeConditions}
	\Abs{a(z) - a(z')} \leq K \Abs{z - z'}, \quad \max \Bracket{\Abs{a(z_0)}, \Abs{\mu_Y}, \Abs{\Sigma_Y}} \leq K, \quad \int_{\R^d} \Abs{x}^2 \lambda(dx) \leq K^2
\end{equation}
for all \(z, z' \in \R^m\) and some constant \(K \in (0, \infty)\) (\(\Abs{M}\) denoting the Frobenius norm for a matrix \(M\)). Under these assumptions, the SDE \eqref{eq:sde} has a unique strong solution \(Z = \Par{Z_t}_{t \in [0, \infty)}\) (see e.g \cite{Applebaum2009}).

\subsubsection{Monte-Carlo algorithm}
The aim is to compute the expectation \(\E{f\Par{Z}}\) via a Monte-Carlo approximation that combines the algorithm in \cite{Dereich2011} with \eqref{Scheme} for the driving L\'evy process \(Y\). More precisely, fix \(h, \epsilon >0\) and define random times
\(T^h_j\), \(j \in \N\), recursively as follows:
\begin{equation}\label{eq:sdeRecursion}
	T^h_0 :=0 \text{  and  } T^h_j := \inf \Bracket{t > T^h_{j-1}: \Delta X^h_t \neq 0 \text{ or } t = T^h_{j-1} + \epsilon}, j \in \N^*
\end{equation}
where \(X^h\) is defined in \eqref{Scheme}. The Euler Scheme approximation \(Z^h\) to the solution \(Z\) of SDE \eqref{eq:sde} is defined recursively by
\begin{equation}\label{eq:sdeEulerScheme}
	Z^h_{T^h_{j+1}} = 	Z^h_{T^h_{j}} + a\Par{Z^h_{T^h_{j}}} \Par{X^h_{T^h_{j+1}} - X^h_{T^h_j}}, \text{  for any } j \in \N
\end{equation}
and can hence be simulated if the increments of \(X^h\) can be simulated. It is natural to define \(\epsilon:=h^{\beta}\) (see Remark \ref{rem:sdeEpsilon} below), where \(\beta\) is the Blumenthal-Getoor index of the L\'evy driver \(Y\). In order to approximate the expectation \(\E{f\Par{Z}}\), we simulate \(n \in \N^*\) IID copies \(X^{h, (1)}, \dots, X^{h, (n)}\) of \(X^h\) from \eqref{Scheme} by applying the algorithm in Section \ref{subsec:MC} over each time interval \([T^h_j, T^h_{j+1})\) with \(T^h_j \leq T\). Compute the corresponding paths \(Z^{h, (1)}, \dots, Z^{h, (n)}\) of the Euler scheme \(Z^h\) via \eqref{eq:sdeEulerScheme}, and define the estimator
\begin{equation}\label{def:sdeEstimator}
	\hat{S}_{n, h} := \frac{1}{n} \sum_{k=1}^n f \Par{Z^{h, (k)}}.
\end{equation}
Since \(f\) acts on c\`adl\`ag paths, in formula \eqref{def:sdeEstimator} we regard \(Z^{h, (k)}\) as a piecewise constant continuous-time path on the interval \([0,T]\).

\remark{\label{rem:sdeEpsilon}\
	It is natural to choose \(\epsilon\) such that, asymptotically, there is a bounded number of jumps of \(X^h\) in any interval of length \(\epsilon\), i.e. \(\epsilon \lambda \Par{\R^d \setminus\! A^h_0} = \bigO{1}\) as \(h \downarrow 0\). Hence \eqref{def:AS2} leads to the choice \(\epsilon = h^{\beta}\).
}

\proposition{\label{prop:sdeBias}\
	Let the driving L\'evy process \(Y\) in \eqref{eq:sde} satisfy the non-degenerative assumption on its L\'evy measure \(\lambda_Y\) in Proposition \ref{prop:MeanBound} (I), as well as \eqref{def:AS1} and \eqref{def:AS2} with the Blumenthal-Getoor index
	\(\beta \in (1, 2)\). Then there exists constants \(k_1, k_2 > 0\), which depend only on the dimension \(d\), time-horizon \(T\) and constant \(K\) in \eqref{eq:sdeConditions}, such that the following inequality holds:
	\[
	Bias(h)^2 = \Abs{\E{f\Par{Z}} - \E{f\Par{Z^h}}}^2 \leq k_1 \Par{h^{2-\beta} + R^{-2}} \text{  and } \Var \Par{f\Par{Z^h}} \leq k_2
	\]
	for small \(h>0\), any sufficient \(R>0\) and any \(f \in Lip(1)\).
}

\remark{\label{rem:sdeBias} Recall that \(R\) is the truncation level of the very big jumps of \(Y\) (see Section \ref{subsec:truncatedProcess} above). The proof of Proposition \ref{prop:sdeBias}, given in Appendix \ref{proof:sdeLevyDrivenProcess}, is based on estimating the ``distance" between the Euler scheme in \eqref{eq:sdeEulerScheme} and the simulation scheme in \cite{Dereich2011}. This is achieved by a natural coupling and a standard argument based on Gr\"onwall's lemma. The main result for the bias in \cite{Dereich2011} is then applied to establish Proposition \ref{prop:sdeBias}. The assumption \(\beta \in (1,2)\) is not essential in this argument. It however encompasses the most interesting case and shorten the proof. We also stress that the pure-jump case, i.e. \(\Abs{\Sigma_Y} = 0\), is covered by the proposition.}

In order to determine the computational complexity \(\mathcal{C}(\epsilon)\) of the Euler scheme satisfying \(MSE \Par{\hat{S}_{n, h}} \leq \epsilon^2\), where \(\hat{S}_{n, h}\) is given by \eqref{def:sdeEstimator} and \(MSE \Par{\hat{S}_{n, h}}\) is defined by the analogue of \eqref{def:mse}, we note that the expected number of breakpoints of each simulated trajectory of \(Z^h\) is bounded above two times the expected number of jumps plus one. As in \cite{Dereich2011}, we assume that \(a(z)\) can be evaluated in \(\bigO{1}\) steps for any \(z \in \R^m\) and that the evaluation of \(f\) on piecewise constant paths (in time) is less than a constant multiple of the breakpoints of the path. Hence by \eqref{def:AS2} we have, the expected cost \(\mathcal{C}_h\) of computing \(f\Par{Z^h}\) is bounded above as in \eqref{mc:cost}, with the constant \(D_1\) replaced by possibly a larger constant \(D_1'\). Since \(\mathcal{C}(\epsilon) = n \mathcal{C}_h\), Proposition \ref{prop:sdeBias} and an analogous argument to the one in Section \ref{subsec:MCComplexity} above imply that \(\mathcal{C}(\epsilon)\) can be controlled as in \eqref{eq:complexityCost} with possibly enlarged constants \(D_2\) and \(D_3\).

\subsubsection{Multilevel Monte-Carlo algorithm for SDE \eqref{eq:sde}}
\proposition{\label{prop:sdeMLMCCost}\
	Let the driving L\'evy process \(Y\) in \eqref{eq:sde} satisfy \eqref{def:AS1} and \eqref{def:AS2} and the components of the process \(\Par{Z^h, \widetilde{Z}^{2h}}\) be each given by the recursion in \eqref{eq:sdeEulerScheme}, driven by the components of the coupling \(\Par{X^h, \widetilde{X}^{2h}}\) from \eqref{def:couplingScheme} respectively. Then there exists a constant \(k_3 > 0\), which depends only on the dimension \(d\), the time-horizon \(T\) and constant \(K\) in \eqref{eq:sdeConditions}, such that the inequality
	\[
	\Var \Par{f\Par{Z^h} - f\Par{\widetilde{Z}^{2h}}} \leq k_3 h^{2 - \beta}
	\]
	holds for sufficiently small \(h > 0\) and any \(f \in Lip(1)\).
}

\remark{The proof of Proposition \ref{prop:sdeMLMCCost}, which follows from Gr\"onwall's inequality, is given in Appendix \ref{proof:sdeLevyDrivenProcess}.}

\subsection{Coupling of CTMCs \(X^{h(\lambda)}\) and \(X^{2h(\lambda)}\) and the level variances}\label{sec:coupling}

The convergence of \(X^h\) to \(X\) as \(h \downarrow 0\), studied in \cite{MijatovicVidmarJacka12} is weak. Put differently, the processes \(X^h\) and \(X\) are not defined on the same probability space. In fact, if \(X\) is a Brownian motion on some filtered probability space and \(X^h\) is a CTMC with respect to that filtration, then by e.g. \cite[Lem. 2.3]{JackMijatovic12} \(X\) and \(X^h\) are necessarily independent for all \(h > 0\), making a stronger path-wise convergence impossible. However, for any \(h > 0\) it is possible to define a co-adapted coupling of \(X^h\) and \(X^{2h}\) (i.e. an adapted process \(\Par{X^h, X^{2h}}\) on a filtered probability space with state space \(\Rd \times \Rd\)) such that:
\begin{enumerate}[label={(\alph*)}]
	\item the laws of the components \(X^h\) and \(X^{2h}\) are respectively given by the \(Q\)-matrices \(Q^h\) and \(Q^{2h}\) in \eqref{def:matrixQ} and
	\item the dependence between them is strong enough to reduce the complexity of the naive Monte-Carlo algorithm via a multilevel approach.
\end{enumerate}
We now define this coupling and prove that the constructed components satisfy the requirements in (a) and (b).

In this Section we apply the coupling construction described in Appendix \ref{NumApp:subsec:couplingOfJumps} to bound the level variances of the MLMC algorithm based on \eqref{Scheme}. Our aim is to control the variance:
\begin{equation}
	\Var \Par{f\Par{X^h} - f\Par{\widetilde{X}^{2h}}} \leq \E{\Par{f\Par{X^h} - f\Par{\widetilde{X}^{2h}}}^2} \leq \E{ \sup_{s \in [0,T]} \NormSup{X^h_s - \widetilde{X}^{2h}_s}^2}
\end{equation}
where \(X^h\) and \(\widetilde{X}^{2h}\), defined on the same probability space, are in \eqref{Scheme}:
\begin{align}\label{def:couplingScheme}
	X^h_t &= (\mu + \widetilde{\mu})t + \sigma W_t + c^h B_t + X^{h(\lambda)}_t - \mu^{h(\lambda)}t \nonumber \\
	\widetilde{X}^{2h}_t &= (\mu + \widetilde{\mu})t + \sigma W_t + c^{2h} B_t + \widetilde{X}^{{2h}(\lambda)}_t - \mu^{{2h}(\lambda)}t 
\end{align}\\
with the same Brownian motions \(W\) and \(B\) and the compound Poisson processes \(X^{h(\lambda)}\) and \(\widetilde{X}^{2h(\lambda)}\) coupled as in Theorem \ref{thm:coupling}. As it will become clear in the proof of Proposition \ref{prop:MLVariance} (deferred to Appendix \ref{proof:prop:MLVariance}), the level variances are controlled both by the coupling in Theorem \ref{thm:coupling} and the bound on the covariance matrix \(C_h\) of the Brownian component of the scheme, arising from the small jumps (see \eqref{def:matrixC}).

\proposition{\label{prop:MLVariance}\
	Let \( \Par{X^h, \widetilde{X}^{2h}} \) be as in \eqref{def:couplingScheme} and assume \(Y\) satisfies \eqref{def:AS1}.
	\begin{enumerate}[label=(\Roman*)]
		\item The following inequalities hold for any time horizon \(T>0\) and all \(h>0\).
		\[ \Var \Par{f\Par{X^h} - f\Par{\widetilde{X}^{2h}}} \leq 16 d T h^2 \lambda_Y \Par{\Rd \setminus\! A^h_0} + 64 d T \int_{A^{2h}_0} \Abs{x}^2 \lambda_Y (dx) \]
		\item If in addition the Blumenthal-Getoor index \(\beta\) of \(Y\) satisfies \(\beta \in [0,2)\) and \eqref{def:AS2} holds, for all \(h>0\) we have:
		\[ \Var \Par{f\Par{X^h} - f\Par{\widetilde{X}^{2h}}} \leq 80dT \aleph h^{2-\beta} \]
	\end{enumerate}%
}

\remark{\
	\begin{enumerate}
		\item Under \eqref{def:AS1}, Remark \ref{rem:meanMCBound} implies that the right-hand side of the inequality in Proposition \ref{prop:MLVariance} (I) tends to \(0\) as \(h \downarrow 0\).
		\item The first term on the right-hand side of the inequality in Proposition \ref{prop:MLVariance} (I) bounds the difference between the levels due to the large jumps of the L\'evy process \(X\). The second term control the difference in the levels which is due to the small jumps of \(X\).
	\end{enumerate}
}

\appendix
\section{Proofs}\label{appendix:proofs}

\subsection{Blumenthal-Getoor index of a multidimensional L\'evy process}\label{proof:bg}
The Blumenthal-Getoor index of a L\'evy copula is equal to $\max_i \beta_i$ where $\beta_i$ is the Blumenthal-Getoor index of the \(i\)-th margin.
\begin{proof}
	Let \(X = (X^1, X^2, \dots, X^d)\) be a \(\mathbb{R}^d\)-valued L\'evy process with L\'evy measure \(\nu\),  Blumenthal-Getoor index \(\beta\) and i-th marginal L\'evy measure \(\nu_i\) with respective Blumenthal-Getoor index \(\beta_i\).  
	\begin{itemize}
		\item[1)] \(\forall x  \in \Rd, \Abs{x} \leq \sqrt{d} \NormSup{x}\), therefore:
		\begin{align*}
			\forall \eta \geq 0, \quad & \Abs{x}^{\eta} \leq {d}^{\eta/2} \NormSup{x}^{\eta} \leq {d}^{\eta/2} \sum_{i=1}^{d} \Abs{x_i}^{\eta} \\
			\implies & \int_{\Abs{x} < 1} \Abs{x}^{\eta} \nu(dx) \leq d^{\eta/2} \sum_{i=1}^{d} \int_{\Abs{x} < 1} \Abs{x_i}^{\eta} \nu(dx) \leq d^{\eta/2} \sum_{i=1}^{d} \int_{\Bracket{x \in \Rd,  \Abs{x_i} < 1}} \Abs{x_i}^{\eta} \nu(dx)  \\
			\implies & \int_{\Abs{x} < 1} \Abs{x}^{\eta} \nu(dx) \leq d^{\eta/2} \sum_{i=1}^{d} \int_{\Abs{x_i} < 1} \Abs{x_i}^{\eta} \nu_i(dx_i)
		\end{align*}
		Hence if \(\eta > \max_i \beta_i\), \(\sum_{i=1}^{d} \int_{\Abs{x_i} < 1} \Abs{x_i}^{\eta} \nu_i(dx) < \infty \), which yields \(\eta \geq \beta\), and therefore \(\max_i \beta_i \geq \beta\).\\
		\item[2)] \(\forall i, 1\leq i \leq d\), \(\Abs{x_i} \leq \Abs{x}\), yielding:
		\begin{align*}
			\forall \eta \geq 0, \quad & \Abs{x_i}^{\eta} \leq \Abs{x}^{\eta} \\
			\implies & \int_{\Abs{x_i} < 1} \Abs{x_i}^{\eta} \nu_i(dx_i) \leq \int_{\Abs{x_i}<1} \Par{1 \land \Abs{x_i}^{\eta}} \nu_i(dx_i) \leq \int_{\Bracket{x \in \Rd, \Abs{x_i}<1}} \Par{1 \land \Abs{x}^{\eta}} \nu(dx)\\
			\implies & \int_{\Abs{x_i} < 1} \Abs{x_i}^{\eta} \nu_i(dx_i) \leq \int_{\Rd} \Par{1 \land \Abs{x}^{\eta}} \nu(dx)
		\end{align*}
		and as: \(\int_{\Rd} \Par{1 \land \Abs{x}^{\eta}} \nu(dx) = \int_{\Abs{x} < 1} \Abs{x}^{\eta} \nu(dx) + \underbrace{\int_{\Abs{x} \geq 1} \nu(dx)}_{< \infty}\), then for all \(\eta > \beta\), \(\int_{\Abs{x} < 1} \Abs{x}^{\eta} \nu(dx) < \infty\), which yields \(\int_{\Abs{x_i} < 1} \Abs{x_i}^{\eta} \nu_i(dx_i) < \infty\) and \(\eta \geq \beta_i\) for all \(i\), that is \(\eta \geq \max_i \beta_i\) and \(\beta \geq \max_i \beta_i\)
	\end{itemize}
	From 1) and 2), we conclude that \(\beta = \max_i \beta_i\)
\end{proof}

\remark{
	As in \cite[section 4.4]{TankovThesis}, we have used the fact that the measure \(\nu_i^{'}\) defined by:
	\(
	\nu_i^{'}(A) := \int_{\Bracket{x \in \Rd, x_i \in A}} \nu(dx), \text{ for } A \in  \Borel
	\)
	coincide with the measure \(\nu_i\).}	

\subsection{Error analysis of the Monte-Carlo algorithm \eqref{def:mcestimator}}\label{proof:mcErrorAnalysis}

\subsubsection{Proof of Lemma \ref{lemma:MeanBound}}

\proof{
	Let \(Z := Y - X\), \(Z\) is a compound Poisson process corresponding of the jumps of \(Y\) greater than \(R\), i.e.
	\(
	Z_t = \sum_{s \leq t} \Delta Y_s \One{\{|\Delta Y_s| \geq R\}}.
	\)
	where \(\Delta Y_t\) corresponds to the jump of \(Y\) at time \(t\).\\ 
	It follows:%
	\[
	\E{\sup_{t \in [0,T]} \NormSup{Z_t}} \leq T \int_{\Rd \setminus\! A^R_0} ||x||_{\infty} \lambda_Y(dx)
	\]
	Furthermore, for all \(x \in \Rd \setminus\! A^R_0\), \(||x||_{\infty}/R \leq (||x||_{\infty}/R)^2 \leq (|x|/R)^2\) and therefore
	\(\kappa(R) \leq \frac{1}{R} \int_{\Rd \setminus\! A^R_0} |x|^2 \lambda_Y(dx)\) and \(\kappa(R) = o(1)/R\) as \(R \rightarrow \infty\)	
}

\subsubsection{Controlling the bias of the Monte-Carlo algorithm \eqref{def:mcestimator}}\label{proof:subsec:biasMC}

\begin{proof}[Proof of Proposition \ref{prop:MeanBound}]\label{proof:prop:MeanBound}
	We write the process \(X^{\lambda}\) from the decomposition in equation \eqref{eq:decomposition} as: \(X^{\lambda} = L^{'} + L^{''}\) where \(L^{''}\) is the process of the jumps less than \(h\) of \(X^{\lambda}\) with L\'evy triplet \((0, \One{A^h_0} \lambda, 0)_{\widetilde{c}}\).\\
	\(L^{'} = X^{\lambda} - L^{'}\) is a compound Poisson process with intensity \(\lambda\left(\Rd \setminus\! A^h_0 \right)\)  and compensator \(\mu_h\):
	\[
	\mu_h = \int_{\Rd \setminus A^h_0 } x \lambda(dx)
	\]
	We now define a coupling between \(L^{'}\) and \(X^{h(\lambda)}\) via the process \(\widetilde{X}^{h(\lambda)}\).
	\begin{equation}\label{eq:BigJumpsCoupling}
		\forall t \geq 0, \quad \widetilde{X}^{h(\lambda)}_t := \sum_{s \leq t} \eta_h\left(\Delta L^{'}_s \right)
	\end{equation}
	where the function \(\eta : \Rd \rightarrow \Zd_h\) is defined by:
	\[
	\eta_h(x) = s \quad \text{if  } x \in A^h_s, s \in \Zd_h
	\]
	Note that \(\widetilde{X}^{h(\lambda)}\) has the same law as \(X^{h(\lambda)}\) and, furthermore, the difference \(L^{'} - \widetilde{X}^{h(\lambda)}\) is a compound Poisson process with intensity \(\lambda\left(\Rd \setminus\! A^h_0 \right)\) and jump sizes at most \(h\).\\
	We now define a copy \(\widetilde{X}^h\) of the approximation \(X^h\) from \eqref{Scheme} and represent the limit \(X\) as follow:
	\begin{align*}
		\widetilde{X}^h_t &:= (\mu +\widetilde{\mu})t + \sigma W_t + c^h B_t +\widetilde{X}^{h(\lambda)}_t - t\mu^{h(\lambda)}\\
		X_t 			  &= (\mu +\widetilde{\mu})t + \sigma W_t + L^{''} + L^{'} - t\mu_h
	\end{align*}
	where \(B\) and \(W\) are standard \(d\)-dimensional Brownian motions. All the processes on the right-hand side are independent, except \(c^h B\) and \(L^{''}\) which are coupled via the Koml\'os-Major-Tusn\`ady coupling (see Theorem 6.1 in \cite{Dereich2011}) and "large jump" compound Poisson processes \(L^{'}\) and \(\widetilde{X}^{h(\lambda)}\) which are coupled as in equation \eqref{eq:BigJumpsCoupling}.
	
	Note that the process \(M = (M_t)_{t \geq 0}\), given by \(M_t := L^{'}_t - t \mu_h - \left(\widetilde{X}^{h(\lambda)}_t - t\mu^{h(\lambda)} \right)\), is a martingale.\\
	From the triangle inequality and given that \(f \in Lip(1)\),
	\begin{align*}
		Bias_X(h)^2 
		&\leq \E{\sup_{t \in [0,T]} \NormSup{L^{''}_t - c^h B_t} + \sup_{t \in [0,T]} \NormSup{M_s}}^2 \\
		& \leq 2 \E{\sup_{t \in [0,T]} \NormSup{L^{''}_t - c^h B_t}^2} + 2 \E{\sup_{t \in [0,T]} \NormSup{M_s}^2}
	\end{align*}
	Each component \(M^{(i)}\), \(i=1,\dots, d\),  of the process \(M\) has zero mean since it is a martingale. Hence:
	\begin{equation}\label{eq:BiasX}
		\E{\left(M^{(i)}_T \right)^2} = \Var(M^{(i)}_T) = \Var\left( \left(L^{'}_T - \widetilde{X}^{h(\lambda)}_T \right)^{(i)} \right) \leq h^2 \E{N^h_T} = Th^2 \lambda\left(\Rd \setminus\! A^h_0 \right)
	\end{equation}
	where \(N^h\) is a Poisson process with intensity \(\lambda\left(\Rd \setminus\! A^h_0 \right)\). The inequality in equation \eqref{eq:BiasX} holds as \(L^{'} - \widetilde{X}^{h(\lambda)}\) is a compound Poisson process with intensity \(\lambda\left(\Rd \setminus\! A^h_0 \right)\) and jump sizes at most \(h\).\\ Furthermore Doob's inequality applied to the non negative submartingale \(|M^{(i)}|\) gives \(\E{\sup_{t \in [0,T]} \left( M^{(i)}_t\right)^2} \leq 4 \E{\left(M^{(i)}_T\right)^2}\), leading to:
	\begin{equation}\label{eq:BoundMartingaleM}
		\E{\sup_{t \in [0,T]} \NormSup{M_s}^2} \leq \sum_{i=1}^{d} \E{\sup_{t \in [0,T]} \left(M^{(i)}_s\right)^2} \leq 4 \sum_{i=1}^{d} \E{\left(M^{(i)}_s\right)^2} \leq 4dTh^2 \lambda(\Rd \setminus\! A^h_0)
	\end{equation}
	Applying Corollary 6.2 in \cite{Dereich2011}, there exist strictly positive constants \(c_1\) and \(c_2\) depending only on \(d\) such that:
	\[
	\E{\sup_{t \in [0,T]} \NormSup{L^{''}_t - c^h B_t}^2}^{\frac{1}{2}} \leq \frac{\sqrt{\theta} h}{\sqrt{c_1}} \left(2 + c_2 \log\max\left(\frac{T}{h^2}\int_{A^h_0 } |x|^2 \lambda(dx), e \right) \right)
	\]
	which, coupled with the previous inequality, leads to the bound in part (I) of the proposition.\\
	The inequality in (II) follows from (I), assumption \eqref{def:AS2} and elementary estimates.
\end{proof}

\subsubsection{Bounding the variance of the Monte-Carlo algorithm \eqref{def:mcestimator}}

Note that the following proof uses notation and results of the previous Section \ref{proof:prop:MeanBound}.
\begin{proof}[Proof of Proposition \ref{prop:VarianceBound}]
	Since \(f \in Lip(1)\), the following inequality holds:
	\[
	\Var\Par{f\Par{X^h}} = \Var\Par{f\Par{X^h} - f(0)} \leq \E{\Par{f\Par{X^h} - f(0)}^2} \leq \E{\sup_{t \in [0,T]} \NormSup{X^h_t}^2}
	\]
	It follow from \eqref{Scheme} that the inequality:
	\[
	\NormSup{X^h_t} \leq t(|\mu| + |\widetilde{\mu}|) + \Par{\NormSup{\sigma} + \NormSup{c^h}} |W_t| + \NormSup{X^{h(\lambda)}_t - \mu^{h(\lambda)}t}
	\]
	holds for all \(t \in [0, T]\) and hence:
	\begin{equation}\label{eq:VarXh}
		\NormSup{X^h_t}^2 \leq 3 \left[ 2T^2(|\mu|^2 + |\widetilde{\mu}|^2) + 2\Par{\NormSup{\sigma}^2 + \NormSup{c^h}^2} |W_t|^2 + \NormSup{X^{h(\lambda)}_t - \mu^{h(\lambda)}t}^2 \right]
	\end{equation}
	The inequality:
	\[
	\E{\sup_{t \in [0,T]} \NormSup{X^{h(\lambda)}_t - \mu^{h(\lambda)}t}} \leq 4dT \underbrace{h^2 \lambda(\Rd \setminus\! A^h_0)}_{\leq \aleph h^{2-\beta} \text{ by \eqref{def:AS2}}} + \E{\sup_{t \in [0,T]} \NormSup{L^{''}_t - \mu_h t}^2}
	\]
	follows by the bound in equation \eqref{eq:BoundMartingaleM}.\\
	Since \((L^{''}_t - \mu_h t)_{t \in [0,T]}\) is a martingale, applying the same arguments as for equation \eqref{eq:BoundMartingaleM} yields:
	\begin{align*}
		\E{\sup_{t \in [0,T]} \NormSup{L^{''}_t - \mu_h t}^2} \leq 4 \underbrace{\sum_{i=1}^d \E{\Par{L^{'' (i)}_T - T \int_{\Rd \setminus\! A^h_0 } x_i \lambda(dx)}^2}}_{ \xrightarrow{h \downarrow 0} T \sum_{i=1}^d\int_{\Rd} x^2_i \lambda(dx) = T \int_{\Rd} |x|^2 \lambda(dx) \leq T \int_{\Rd} |x|^2 \lambda_Y(dx)}
	\end{align*}
	by Campbell's formula (see Theorem 2.17 in \cite{kyprianou2006introductory} and  Proposition 11.10 in \cite{sato2013levy}), since each component \(L^{''(i)}\) is a one-dimensional L\'evy process with L\'evy measure given by the \(i\)-th marginal of \(\lambda\).
	
	In order to control the Brownian motion term in equation \eqref{eq:VarXh}, note that Doob's submartingale inequality yields \(\E{\sup_{t \in [0,T]} |W_t|^2} \leq 4dT\) and we will show that \(\NormSup{c^h}^2 \leq \aleph h^{2-\beta}\) and hence:
	\[
	\E{\sup_{t \in [0,T]} \NormSup{X^h_t}^2} \leq 3 \left[ 2T^2 (|\mu|^2 + |\widetilde{\mu}|^2) + 8dT \NormSup{\sigma}^2 + 12dT\aleph h^{2-\beta} + 4T \int_{\Rd} |x|^2 \lambda_Y(dx)\right]
	\]
	To finish the proof, we need to show that \(\NormSup{c^h}^2 \leq h^{2-\beta} \aleph\).
	Let \(Tr(M)\) denote the trace of the matrix \(M\).\\ 
	Note that $\NormSup{c^h} \leq \Abs{c^h}$ and $\Abs{c^h}^2 = Tr\Par{c^h c^{hT}} = Tr\Par{C^h}$ with: 
	\begin{equation}\label{eq:inequality_ch}
		Tr\Par{C^h} = \int_{A^h_0} \Abs{x}^2 \One{\SquaredBracket{-V, V}}(x) \lambda(dx) \leq \aleph h^{2 - \beta}
	\end{equation}
	where the last inequality follows by assumption \eqref{def:AS2}.\\
\end{proof}

\subsection{Proofs for L\'evy-driven SDE: Proposition \ref{prop:sdeBias} and \ref{prop:sdeMLMCCost}}\label{proof:sdeLevyDrivenProcess}

As our convergence proofs rely heavily both on the results and techniques in \cite{Dereich2011}, in this section, as in \cite{Dereich2011}, it is assumed the time horizon \(T=1\) for convenience. Hence only the portion of the path of \(Z^h\) on the time interval \([0,1]\) in \eqref{def:sdeEstimator} is considered.

\begin{proof}[Proof of Proposition \ref{prop:sdeBias}]
	Recall the process $Z = \Par{Z_t}_{t \geq 0}$ is defined by: \[Z_t = z_0 + \int_0^t a \Par{Z_{s^-}} dY_s, \quad z_0 \in \R^m\]
	We denote by $\mathcal{Z} = \Par{\mathcal{Z}_t}_{t \geq 0}$ the solution of the integral equation: \[\mathcal{Z}_t = z_0 + \int_0^t a \Par{\mathcal{Z}_{s^-}} dX_s\]
	with $X$ defined in Section \ref{subsec:truncatedProcess}: $X_t = \mu t + \sigma W_t + X^{\lambda}_t$. We further decompose $X^{\lambda}$ into $X^{\lambda} := L'_t + L''_t$, where $L'$ (resp. $L''$) is the process comprising the jumps of $X$ greater (resp. smaller) than $h$.\\
	Recall that the random times $T^h_j$ are defined by $T^h_0 = 0$ and $T^h_{j+1} = \inf \Bracket{t > T^h_j: \Delta X^h_t \neq 0 \text{ or } t = T^h_j + \epsilon}, \; j \in N$ and we denote by $\mathbb{I}$ the set of these random times: $\mathbb{I} := \Bracket{T_j, j \in \N}$.\\
	Let $\epsilon' \in [2\epsilon, 1]$ and define $\mathbb{J}$ the set of random times defined inductively by $T_0 = 0$ and $T_{j+1} = \min \Par{\mathbb{I} \cap \Par{T_j + \epsilon' - \epsilon, +\infty}}$.\\
	Let $l(\cdot)$ and $\eta(\cdot)$ be the functions: $l(t) = \max \Par{\mathbb{I} \cap [0, t]}$ and $\eta(t) = \max \Par{\mathbb{J} \cap [0, t]}$.\\	 
	As in \cite{Dereich2011}, we set $X' = \Par{X_t - L''_t}_{t \geq 0}$  and we introduce the auxiliary processes $\bar{\mathcal{Z}}' = \Par{\bar{\mathcal{Z}}'_t}_{t \geq 0}$, solution to the integral equation
	\[\bar{\mathcal{Z}}'_t = z_0 + \int_0^t a \Par{\bar{\mathcal{Z}}'_{l(s^-)}} dX'_s + \int_0^t a \Par{\bar{\mathcal{Z}}'_{\eta(s^-)}} dL''_{\eta(s)}\]
	and $\bar{\mathcal{Z}} = \Par{\bar{\mathcal{Z}}_t}_{t \geq 0}$ given by
	\[\bar{\mathcal{Z}}_t = \bar{\mathcal{Z}}'_t + a\Par{\bar{\mathcal{Z}}'_{\eta(s^-)}} \Par{L''_t - L''_{\eta(t)}}\]\\
	$\bar{\mathcal{Z}}$ coincides with $\bar{\mathcal{Z}}'$ for all times in $\mathbb{J}$ and satisfies:
	\[\bar{\mathcal{Z}}_t = z_0 + \int_0^t a \Par{\bar{\mathcal{Z}}'_{l(s^-)}} dX'_s + \int_0^t a \Par{\bar{\mathcal{Z}}'_{\eta(s^-)}} dL''_s\]
	Let $\zeta = \Par{\zeta_t}_{t \geq 0}$ be the solution to the integral equation
	\[ \zeta_t = z_0 + \int_0^t a \Par{\zeta_{s^-}} d X^h_s \]
	$\zeta$ is constant on each interval $\left[T^h_j, T^h_{j+1} \right)$ and $\zeta_{l(\cdot)}$ coincide with the Euler approximation process $Z^h$ in \eqref{eq:sdeEulerScheme}.\\
	Next we replace $X$ by the approximation \eqref{Scheme} in the above equations and obtain analogous processes of $\bar{\mathcal{Z}}$ and $\bar{\mathcal{Z}}'$ denoted by $\bar{\zeta}$ and $\bar{\zeta}'$.
	More precisely, $\bar{\zeta}' = \Par{\bar{\zeta}_t'}_{t \geq 0}$ is the solution to the stochastic integral equation
	\[\bar{\zeta}'_t = z_0 + \int_0^t a \Par{\bar{\zeta}'_{l(s^-)}} d\mathcal{X}'_s + \int_0^t a \Par{\bar{\zeta}'_{\eta(s^-)}} c^h dB_{\eta(s)}\]
	where $\mathcal{X}' = \Par{\mathcal{X}'_t}_{t \geq 0}$ is defined by $\mathcal{X}'_t := X^h_t - c^h B_t$; and $\bar{\zeta} = \Par{\bar{\zeta}_t}_{t \geq 0}$ is given by
	\[\bar{\zeta}_t = \bar{\zeta}'_t + a\Par{\bar{\zeta}'_{\eta(s^-)}} \sigma \Par{B_t - B_{\eta(t)}}\]
	By the triangle inequality: \( \NormSup{Z - \zeta_{l(\cdot)}} \leq \NormSup{Z - \mathcal{Z}} + \NormSup{\mathcal{Z} - \zeta_{l(\cdot)}} \).
	The first term on the right hand side is controlled as in Lemma \ref{lemma:MeanBound} and we can use the triangle inequality further for the second term as in \cite{Dereich2011}:
	\[ \NormSup{\mathcal{Z} - \zeta_{l(\cdot)}} \leq \NormSup{\mathcal{Z} - \bar{\mathcal{Z}}} + \NormSup{\bar{\mathcal{Z}} - \bar{\zeta}} + \NormSup{\bar{\zeta} - \zeta} + \NormSup{\zeta - \zeta_{l(\cdot)}} \] 
	Although we have substituted the process $L'$ with the CTMC $X^{h(\lambda)}$ when defining $\bar{\zeta}$ and $\bar{\zeta}'$, the results in \cite{Dereich2011}, which control analogous terms of those of the right hand side of the inequality, still hold; their proof follows the one in \cite{Dereich2011} 
	and is not replicated here.\\
	Following the estimates in Lemma \ref{lemma:MeanBound} and in \cite{Dereich2011}, the following inequality holds:
	\[ Bias(h)^2 \leq K \Par{ \frac{1}{R^2} +  F(h)\epsilon' + \frac{h^2}{\epsilon'} \log \Par{\frac{\epsilon' F(h)}{h^2} \vee e}^2 + \epsilon \log \frac{e}{\epsilon} } \]
	where $F(h) = \int_{A^{2h}_0} \Abs{x}^2 \lambda(dx)$, $\epsilon \in (0, \frac{1}{2}]$, $\epsilon' \in [2\epsilon, 1]$ and $h$ such that $\lambda\Par{\Rd \setminus A^h_0} \leq \frac{1}{\epsilon}$.\\	
	Following Remark \ref{rem:sdeEpsilon}, we let $\epsilon = h^{\beta}$. Note that the assumption $\lambda\Par{\Rd \setminus A^{2h}_0} \leq \frac{1}{\epsilon}$ is verified by \eqref{def:AS2} if we rescale $\epsilon$ by the constant $\aleph$, but for ease of presentation, we keep $\epsilon = h^{\beta}$.\\
	Our aim is now to find a bound for the expression $F(h)\epsilon' + \frac{h^2}{\epsilon'} \log \Par{\frac{\epsilon' F(h)}{h^2} \vee e}^2 + \epsilon \log \frac{e}{\epsilon}$:
	\begin{itemize*}
		\item for the first term: $F(h)\epsilon' \leq F(h)$ and $F(h) = \bigO{h^{2 - \beta}}$ by \eqref{def:AS2}.\\
		\item for the second term, we have: $\frac{h^2}{\epsilon'} \leq \frac{h^2}{2 \epsilon} \leq h^{2 - \beta}$. Furthermore, $\frac{\epsilon' F(h)}{h^2} \leq \frac{2 F(h)}{h^{2 - \beta}}$ and therefore $\log \Par{\frac{\epsilon' F(h)}{h^2} \vee e}^2 \leq \log \Par{\frac{2 F(h)}{h^{2-\beta}} \vee e}^2$, and, given $\frac{\epsilon' F(h)}{h^2} = \bigO{1}$, it yields: $\frac{h^2}{\epsilon'} \log \Par{\frac{\epsilon' F(h)}{h^2} \vee e}^2 = \bigO{h^{2 - \beta}}$.\\
		\item for the third term, with $\beta \in (1, 2)$, $\epsilon \log \frac{e}{\epsilon} = h^{\beta} \Par{1 - \log h^{\beta}} = \bigO{h^{2-\beta}}$.\\
	\end{itemize*}
	From the above estimates, there exists a constant $k_1 > 0$ such that the bias is bounded by:
	\[Bias(h)^2 \leq k_1 \Par{\frac{1}{R^2} + h^{2 -\beta}}\]
	To show that the variance $\Var \Par{f\Par{\zeta}}$ is finite, we write:
	\[ \Var \Par{f\Par{\zeta}} = \Var \Par{f\Par{\zeta} - f(z_0)} \leq \E{\Par{f\Par{\zeta} - f(z_0)}^2} \leq \E{\sup_{t \in [0,T]} \NormSup{\zeta_t - z_0}^2} \]
	and we apply the same idea as in the proof of \cite[Prop. 4.1]{Dereich2011}:\\
	Let $(M_t)_{t \geq 0}$ be the local martingale $M_t = \int_0^t a\Par{\zeta_{s^-}} d\Par{\sigma W_t + c^h B_t + X^{\lambda(h)} - \mu^{h(\lambda)} t}$,
	\[\Abs{\zeta_t - z_0}^2 \leq 2 \Abs{M_t}^2 + 2 \Abs{\int_0^t a\Par{\zeta_{s^-}} (\mu + \tilde{\mu}) ds}^2 \]
	and the second term can be bounded as follow:
	\begin{align}
		\Abs{\int_0^t a\Par{\zeta_{s^-}} (\mu + \tilde{\mu}) ds}^2 
		&\leq 2 \underbrace{\Par{\int_0^t \Abs{a\Par{\zeta_{s^-}} - a\Par{z_0}} \Abs{\mu + \tilde{\mu}} ds}^2}_{\leq 2K T \int_0^t \Abs{\zeta_{s^-} - z_0}^2 ds \text{   by Cauchy Schwarz}} + 2 \Par{\int_0^t \underbrace{\Abs{a\Par{z_0}} \Abs{\mu + \tilde{\mu}}}_{\leq 2K^2} ds}^2 \nonumber \\
		&\leq 4 K T d \int_0^t \sup_{u \in [0,s]} \NormSup{\zeta_u - z_0}^2 ds + 4 K^4 T^2 \label{eq:sdeIntBound}\\
	\end{align}
	Doob's inequality and Lemma A.1 in \cite{Dereich2011} yield:
	\[ \E{\sup_{t \in [0,T]} \Abs{M_t}^2} \leq 4 \E{\int_0^t \Abs{a\Par{\zeta_{s^-}}}^2 d \langle \sigma W_{\cdot{}} + c^h B_{\cdot{}} + X^{\lambda(h)}_{\cdot{}} - \mu^{\lambda(h)} (\cdot{}) \rangle_t }   \]
	where for a multivariate local martingale $(S_t)_{t \geq 0}$, $\langle S \rangle = \sum_j \langle S^{(j)} \rangle$ denotes the predictable compensator of the classical bracket process of the $j$-th coordinate $S^{(j)}$ of $S$.
	\[d \langle \sigma W_{\cdot{}} + c^h B_{\cdot{}} + X^{\lambda(h)}_{\cdot{}} - \mu^{\lambda(h)} (\cdot{}) \rangle_t = \Par{ \Abs{\sigma}^2 + \Abs{c^h}^2 }dt \]
	By assumption $\Abs{\sigma}^2 \leq K^2$ and we can control the last term $\Abs{c^h}$ as in \eqref{eq:inequality_ch} which yields:
	\begin{equation}\label{proof:sde:bracketBound}
		d \langle \sigma W_{\cdot{}} + c^h B_{\cdot{}} + X^{\lambda(h)}_{\cdot{}} - \mu^{\lambda(h)} (\cdot{}) \rangle_t \leq 2K^2 dt
	\end{equation}
	yielding 
	\begin{align*}
		\E{\sup_{t \in [0,T]} \Abs{M_t}^2} 
		&\leq 8 K^2 \E{\int_0^t \Abs{a\Par{\zeta_{s^-}}}^2 dt} \\
		&\leq 16 K^2 \Par{\E{\int_0^t \Abs{a\Par{\zeta_{s^-}} - a\Par{z_0}}^2  dt } + \underbrace{\int_0^t \Abs{a\Par{z_0}}^2  dt}_{\leq K^2 T}}\\
		&\leq 16 K^4 \E{\int_0^t \sup_{u \in [0,s]} \NormSup{\zeta_u - z_0}^2  dt } + 16 K^4 T
	\end{align*}
	Finally, given $\E{\sup_{t \in [0,T]} \NormSup{M_t}^2} \leq \E{\sup_{t \in [0,T]} \Abs{M_t}^2}$ and the bound in \eqref{eq:sdeIntBound}, we have: 
	\[ \E{\sup_{t \in [0,T]} \NormSup{\zeta_t - z_0}^2} \leq 4K(Td + 8K^3) \int_0^T \E{ \sup_{s \in [0,t]} \NormSup{\zeta_{s} - z_0}^2} dt  + 4K^4 T(4 + T)\]
	and by Gr\"onwall's lemma, there exists a constant $k_2 > 0$ such that:
	\[ \E{\sup_{t \in [0,T]} \NormSup{\zeta_t - z_0}^2} \leq k_2 \]	
\end{proof}

\begin{proof}[Proof of Proposition \ref{prop:sdeMLMCCost}]
	With the same notation as above, let $\zeta^h$ denote the solution to the equation:
	\[\zeta^h_t = z_0 + \int_0^t a(\zeta^h_{s^-}) dX^h_s\]
	and $\widetilde{\zeta}^{2h}$ the solution to the equation:
	\[\widetilde{\zeta}^{2h}_t = z_0 + \int_0^t a(\widetilde{\zeta}^{2h}_{s^-}) d\widetilde{X}^{2h}_s\]
	With $f \in Lip(1)$, we can bound the variance as follow:
	\[\Var \Par{f\Par{\zeta^h} - f\Par{\tilde{\zeta}^{2h}}} \leq \E{\Par{f\Par{\zeta^h} - f\Par{\tilde{\zeta}^{2h}}}^2} \leq \E{\sup_{t \in [0,T]} \NormSup{\zeta^h_t - \tilde{\zeta}_t^{2h}}^2} \leq \E{\sup_{t \in [0,T]} \Abs{\zeta^h_t - \tilde{\zeta}_t^{2h}}^2} \]
	and
	\begin{align*}
		\Abs{\zeta^h_t - \tilde{\zeta}^{2h}_t}^2 
		&= \Abs{\int_0^t a(\zeta^h_{s^-}) dX^h_s - \int_0^t a(\widetilde{\zeta}^{2h}_{s^-}) d\widetilde{X}^{2h}_s}^2\\
		&\leq 2 \Abs{\int_0^t \Par{a(\zeta^h_{s^-}) - a(\widetilde{\zeta}^{2h}_{s^-})} dX^h_s }^2 + 2 \Abs{ \int_0^t a(\widetilde{\zeta}^{2h}_{s^-}) d\Par{X^h_s - \widetilde{X}^{2h}_s}}^2\\
	\end{align*}
	The first term on the right hand side can be controlled by \cite[lemma A.1]{Dereich2011}:
	\begin{align*}
		\E{\Abs{\int_0^t \Par{a(\zeta^h_{s^-}) - a(\widetilde{\zeta}^{2h}_{s^-})} dX^h_s }^2} 
		&\leq K^2 \E{\int_0^t \Abs{\zeta^h_{s^-} - \widetilde{\zeta}^{2h}_{s^-}}^2 d \langle X^h_{\cdot} \rangle_s}\\
		&\leq 2 K^3 \int_0^t \E{\Abs{\zeta^h_s - \widetilde{\zeta}^{2h}_s}^2} ds, \quad \text{by the bound in \eqref{proof:sde:bracketBound}}
	\end{align*}
	Let $M$ be the local martingale $M_t := \int_0^t a(\widetilde{\zeta}^{2h}_{s^-}) d\Par{X^h_s - \widetilde{X}^{2h}_s}$, then with the same arguments as in the proof of Proposition \ref{prop:sdeBias} above:
	\[\E{\sup_{t \in [0,T]} \Abs{M_t}^2} \leq 4 \E{\int_0^t \Abs{a(\widetilde{\zeta}^{2h}_{s^-})}^2 d \langle X^h_{\cdot} - \widetilde{X}^{2h}_{\cdot} \rangle_s} \]
	$X^h_s - \widetilde{X}^{2h}_s = (c^h - c^{2h})B + X^{h(\lambda)}_t - \mu^{h(\lambda)}t - \Par{\widetilde{X}^{2h(\lambda)}_t - \mu^{2h(\lambda)}t}$ and $d \langle X^h_{\cdot} - \widetilde{X}^{2h}_{\cdot} \rangle_t = \underbrace{\Abs{c^h - c^{2h}}^2}_{\leq 4 \int_{A^{2h}_0} \Abs{x}^2 \lambda(dx)} dt$	
	which yields:
	\begin{align*}
		\E{\sup_{t \in [0,T]} \Abs{M_t}^2} 
		&\leq 16  \int_{A^{2h}_0} \Abs{x}^2 \lambda(dx) \, \E{ \int_0^t \Abs{a(\widetilde{\zeta}^{2h}_{s^-})}^2 ds}\\
		&\leq 16 \aleph h^{2-\beta} \, \E{ \int_0^t \Abs{a(\widetilde{\zeta}^{2h}_{s^-})}^2 ds}
	\end{align*}
	where the last inequality derives from \eqref{def:AS2}.\\
	The last term on the right hand side can be bounded as in the proof of Proposition \ref{prop:sdeBias} above:
	\begin{align*}
		\E{ \int_0^t \Abs{a(\widetilde{\zeta}^{2h}_{s^-})}^2 ds} 
		&\leq  2 \E{ \int_0^t \Abs{a(\widetilde{\zeta}^{2h}_{s^-}) - a(z_0)}^2 ds} +  2 \E{ \int_0^t \Abs{a(z_0)}^2 ds}	\\
		&\leq  2K^2 \E{ \int_0^t \Abs{\widetilde{\zeta}^{2h}_{s^-} - z_0}^2 ds} +  2 \E{ \int_0^t \Abs{a(z_0)}^2 ds}	\\
		&\leq  2K^2 (1 + k_2)
	\end{align*}
	yielding
	\begin{align*}
		\E{\sup_{t \in [0, T]} \Abs{\zeta^h_t - \tilde{\zeta}^{2h}_t}^2 } \leq 4K^3 \int_0^T \E{\sup_{t \in [0, s]} \Abs{\zeta^h_t - \tilde{\zeta}^{2h}_t}^2 } ds + 32 K^2(1+k_2)\aleph h^{2-\beta}
	\end{align*}
	and the application of Gr\"onwall's lemma finishes the proof.
\end{proof}

\subsection{Controlling the variance of the coupling: proof of Proposition \ref{prop:MLVariance}}\label{proof:prop:MLVariance}

We use the same notation as in Section \ref{proof:mcErrorAnalysis}.
\begin{proof}
	Let \(M = (M_t)_{t \in [0,T]}\) be the martingale given by \(M_t := X^{h(\lambda)}_t - \mu^{h(\lambda)}t - (\widetilde{X}^{2h(\lambda)}_t - \mu^{2h(\lambda)}t)\).\\
	With the same arguments as in Section \ref{proof:subsec:biasMC}, one has:
	\[
	\E{\left(M^{(i)}_T \right)^2} = \Var(M^{(i)}_T) = \Var\left( \left(X^{h(\lambda)}_T - \widetilde{X}^{2h(\lambda)}_T \right)^{(i)} \right) \leq h^2 T \lambda\left(\Rd \setminus\! A^h_0 \right)
	\]
	and
	\[
	\E{\sup_{t \in [0,T]} \NormSup{M_s}^2} \leq \sum_{i=1}^{d} \E{\sup_{t \in [0,T]} \left(M^{(i)}_s\right)^2} \leq 4 \sum_{i=1}^{d} \E{\left(M^{(i)}_s\right)^2} \leq 4dTh^2 \lambda(\Rd \setminus\! A^h_0)
	\]
	In order to control the diffusion part, note the inequality in \eqref{eq:inequality_ch} implies \(\Abs{c^h - c^{2h}}^2 \leq 2 \Abs{c^h}^2 + 2 \Abs{c^{2h}}^2 \leq 4 \int_{A^{2h}_0} \Abs{x}^2 \lambda(dx)\).\\
	Hence:
	\[
	\E{\sup_{t \in [0,T]} \NormSup{(c^h - c^{2h})B_t}^2} \leq 4 \int_{A^{2h}_0} |x|^2 \lambda(dx) \E{\sup_{t \in [0,T]} |B_t|^2} \leq 16 d T\int_{A^{2h}_0} |x|^2 \lambda(dx)
	\]
	The inequality in part (I) now follows by noting that \(X^h - \widetilde{X}^{2h} = (c^h - c^{2h})B + M\). Part (II) follows from Proposition \ref{prop:MeanBound} and assumption \eqref{def:AS2}.
\end{proof}

\subsection{Coupling of jumps}\label{NumApp:subsec:couplingOfJumps}
Fix \(h \in (0,1)\) and assume that the L\'evy measure \(\lambda\) of \(X\) is not zero. The aim in the present section is to define a coupling of CTMCs \(X^{h(\lambda)}\) and \(X^{2h(\lambda)}\).

\subsubsection{Coupling of Poisson point processes}\label{subsubsec:couplingOfPoissonProcess}
The key ingredient in our coupling construction will be a Poisson point process (PPP) (we refer to \cite{Kingman} for a general theory of PPPs) with state space \(\R_+ \times \Rd\) that corresponds to the process of jumps \(\Delta X^{h(\lambda)} = \Par{\Delta X^{h(\lambda)}_t}_{t \in [0, \infty)}\).\\
It is clear that \(\Delta X^{h(\lambda)}_t \in \Zd_h \setminus\! \{0\}\) \(\Proba\)-a.s. and hence the PPP \(\Pi_h^{\lambda}\) consisting of subsets:
\begin{equation}\label{def:pppPih}
	\Pi_h^{\lambda} := \Bracket{\Par{T^h_i, \Delta X^{h(\lambda)}_{T_i^h}}: i \in \N \text{ and } T^h_i \text{ is the i-\textit{th} jump time of } X^{h(\lambda)}}
\end{equation}
in \(\R_+ \times \Rd\) has a \textit{mean measure}:
\[
\nu^{\lambda}_h(dt \otimes dx) = \One{\Zd_h \setminus \{0\}}(x) Q^{h(\lambda)}_{0x} dt
\]
on the Borel \(\sigma\)-field \(\mathcal{B} \Par{\R_+ \times \Rd}\), supported in \(\R_+ \times \Par{\Zd_h \setminus\! \{0\}}\). In particular, for any \(A \in \mathcal{B} \Par{\R_+ \times \Rd}\) the average number of points \(\Pi^{\lambda}_h\) in \(A\) is given by:
\[
\E{\Big | \Pi^{\lambda}_h  \cap A \Big | } = \nu^{\lambda}_h(A) 
\]
where \(\Big | \{\cdot\} \Big |\) is the number of elements in the set \(\{ \cdot\}\) with the convention \(\Big | \emptyset \Big | = 0\). It is clear from this description that jumps of \(X^{h(\lambda)}\) occur with a strictly positive intensity \(-Q_{00}^{h(\lambda)}\) since the L\'evy measure of \(X\) is (in this section) assumed to satisfy \(\lambda \neq 0\) (recall that \(Q_{ss}^{h(\lambda)} = Q_{00}^{h(\lambda)}\) for any \(s \in \Zd_h\)). By the same token, the PPP given by the jumps of \(X^{2h(\lambda)}\).
\begin{equation}\label{def:pppPi2h}
	\Pi_{2h}^{\lambda} := \Bracket{\Par{T^{2h}_i, \Delta X^{2h(\lambda)}_{T_i^{2h}}}: i \in \N \text{ and } T^{2h}_i \text{ is the i-\textit{th} jump time of } X^{2h(\lambda)}}
\end{equation}
possesses a mean measure
\begin{equation}\label{def:meanMeasure2h}
	\nu^{\lambda}_{2h}(dt \otimes dx) = \One{\Zd_{2h} \setminus \{0\}}(x) Q^{2h(\lambda)}_{0x} dt	
\end{equation}

supported in \(\R_+ \times \Par{\Zd_{2h} \setminus\! \{0\}}\). Our task is to couple the PPPs \(\Pi^{\lambda}_h\) from \eqref{def:pppPih} and \(\Pi^{\lambda}_{2h}\) from \eqref{def:pppPi2h}, which will in turn yield a coupling of the CTMCs \(X^{h(\lambda)}\) and \(X^{2h(\lambda)}\).\\

We start by defining a new PPP constructed by marking the PPP \(\Pi^{\lambda}_h\). More precisely define the set \(M_h := \Bracket{-h,0,h}\) and consider the product space:
\begin{equation}\label{def:Mdh}
	M_h^d = \Bracket{m \in \Zd_h : m_j \in \Bracket{-h, 0, h} \text{ for all } j=1,\dots,d} \subset \Rd
\end{equation}
(throughout we identify \(M_d^1\) with \(M_h\)). Define a family of probability mass functions (pmfs):
\[
p^{\lambda}(s, \cdot) : M_h^d \to \SquaredBracket{0,1}, \quad s \in \Zd_h \setminus\! \{0\}
\]
by the formula:
\begin{equation}\label{def:pmf}
	p^{\lambda}(s, m) =
	\left\{
	\begin{array}{ll}
		\frac{\lambda\left(A^{2h}_{s+m} \bigcap A^h_s \right)}{\lambda(A^h_s)}, &\text{if } s + m \in \Zd_{2h} \text{ and } \lambda \Par{A_s^h} > 0\\
		0, &\text{if } s + m \notin \Zd_{2h} \text{ and } \lambda \Par{A_s^h} > 0 \\
		\One{\{0\}}(m),	&\text{if } \lambda \Par{A_s^h} = 0
	\end{array}
	\right.
\end{equation}
where the sets \(A^{2h}_{s+m}\) and \(A^h_s\) are defined in \eqref{def:Ah}.\\
The following observations are immediate:
\begin{itemize}
	\item by the definition of \(\nu^{\lambda}_h\), if \(Q_{0s}^{h(\lambda)} = \lambda \Par{A_s^h} = 0\) for some \(s \in \Zd_h \setminus\! \{0\}\), the jump size \(s\) does not arise \(\Proba\)-a.s. making the final line of the definition of \(p^{\lambda}(s, m)\) arbitrary as far as the marking of the PPP \(\Pi^{\lambda}_h\) is concerned;
	\item if \(s \in \Zd_{2h} \setminus\! \{0\}\), then \(p^{\lambda}(s, m) = \One{\{0\}}(m)\) which is a pmf on \(M_h^d\);
	\item since \(\Bracket{A_s^{2h} : s \in \Zd_{2h}}\) is partition of \(\Rd\) and \(A^{2h}_{s+m} \bigcap A^h_s \neq \emptyset\) if and only if \(s + m \in \Zd_{2h}\), we have \(\sum_{m \in M_h^d} p^{\lambda}(s, m) = 1\) making \(p^{\lambda}(s, \cdot)\) a pmf on \(M_h^d\) for every \(s \in \Zd_h \setminus\! \{0\}\);
	\item if \(supp(\lambda) = \Rd \setminus\! \{0\}\), then for any \(s \in \Zd_{h} \setminus\! \{0\}\) the support of the pmf \(p^{\lambda}(s, \cdot)\) is of the size \(2^{J(s)}\), where \(J(s) := \Big| \Bracket{j \in \Bracket{1, \dots, d} : s_j \notin \Zd_{2h}} \Big|\), in particular, in the case \(d=1\), it holds: if \(s \notin \Zd_{2h} \setminus\! \{ 0 \}\) (resp. \(s \in \Zd_{2h} \setminus\! \{0\}\)), \(p^{\lambda}(s, \cdot)\) is a binomial pmf on \(\Bracket{-h, h} \subset M_h\) (resp. a pmf concentrated at \(0 \in M_h\)).
\end{itemize} 
We can now define the marked PPP \(\Pi_h^{\lambda*}\).
\proposition{\label{prop:markedPPP}  Let \(\Pi^{\lambda}_h\) be the PPP in \eqref{def:pppPih} with the mean measure \(\nu^{\lambda}_h\) and define the random set \(\Pi^{\lambda*}_h\) in \( \R_+ \times \Rd \times \Rd \) by:
	\[
	\Pi^{\lambda*}_h := \Bracket{\Par{T, Z, M(T,Z)} : \Par{T,Z} \in \Pi^{\lambda}_h}
	\]
	where, \textbf{conditional on \(\Pi^{\lambda_h}\)}, the random variables \(M(T,Z), (T,Z) \in \Pi^{\lambda}_h\), satisfy the following:%
	\begin{align}	
		M(T,Z)& \text{ takes values in } M_h^d \text{ and has distribution } p^{\lambda}(Z, \cdot)\; \forall (T, Z) \in \Pi^{\lambda}_h \text{ and} \label{def:pppM}\\
		M(T,Z)&, (T,Z) \in \Pi^{\lambda}_h, \text{ are independent}\label{def:pppTZ}.
	\end{align}
	Then \(\Pi^{\lambda*}_h\) is a PPP with the mean measure (on the Borel \(\sigma\)-field \(\mathcal{B}\Par{\R_+ \times \Rd \times \Rd}\)) given by the formula:
	\begin{align}\label{eq:nuhLambdaStart}
		\nu^{\lambda*}_h \Par{dt \otimes dx \otimes dm} 
		&= \nu^{\lambda}_h \Par{dt \otimes dx} \One{\Zd_h \setminus \{0\}}(x) \One{M^d_h}(m) p^{\lambda}(x, m) \\
		&= \One{\Zd_h \setminus \{0\}}(x) \One{M^d_h}(m) \One{\Zd_{2h}}(x+m) \lambda \Par{A^{2h}_{x+m} \cap A^h_x} dt 
	\end{align}
}
and supported in \(\R_+ \times \Par{\Zd_{h} \setminus\! \{0\}} \times M^d_h\).
\proof{Note that the equality \(Q^{h(\lambda)}_{0s} = \lambda\Par{A_s^h} = 0\) implies that, for any \(m\) such that \(s + m \in \Zd_{2h}\), it must hold \(\lambda\Par{A^{2h}_{s+m} \cap A^h_s}=0\). With this in mind, the proposition follows as a direct consequence of the definitions of the PPP \(\Pi^{\lambda}_h\) in \eqref{def:pppPih}, its mean measure \(\nu^{\lambda}_h\), random variables \(M(s)\), \(s \in \Zd_h \setminus\! \{0\}\), and the Marking Theorem \cite[p.55]{Kingman}.}

In order to define a PPP that has the same law as the PPP \(\Pi^{\lambda}_{2h}\) from \eqref{def:pppPi2h}, we transform the PPP \(\Pi^{\lambda*}_h\) in Proposition \ref{prop:markedPPP} by the map:
\begin{equation}\label{def:pppMapping}
	F: \R_+ \times \Rd \times \Rd \to \R_+ \times \Rd, \quad F(t, x, m) := (t, x+m)
\end{equation}

\proposition{\label{prop:mappedPPP} Let \(\Pi^{\lambda}_h\) and \(\Pi^{\lambda*}_h\) be the PPPs in Proposition \ref{prop:markedPPP} with state spaces \(\R_+ \times \Rd\) and \(\R_+ \times \Rd \times \Rd\) respectively, and let the function \(F\) be as in \eqref{def:pppMapping}. The random subset of \(\R_+ \times \Rd\), given by:
	\[
	F\Par{\Pi^{\lambda*}_h} = \Bracket{\Par{T,Z+M(T,Z)}: (T,Z) \in \Pi^{\lambda}_h},
	\]
	is a PPP with a mean measure given by \eqref{def:meanMeasure2h}, supported in the set \(\R_+ \times \Par{\Zd_{2h} \setminus\! \{0\}}\). The random set
	\[
	\widetilde{\Pi}^{\lambda}_{2h} := F \Par{\Pi^{\lambda*}_h} \cap \Par{\R_+ \times \Rd \setminus\! \{0\}}
	\]
	is a PPP with a mean measure given by \eqref{def:meanMeasure2h}, supported in the set \(\R_+\times \Par{\Zd_{2h} \setminus\! \{0\}}\).
}

\remark{Recall that typically \(T = T^h_i\) and \(Z = \Delta X^{h(\lambda)}_{T^h_i}\) in the definition of \(\Pi^{\lambda*}_h\) are the \textit{i}-th jump time and jump size respectively of the CTMC \(X^{h(\lambda)}\). The random variables \(M(T,Z), (T, Z) \in \Pi^{\lambda}_h\), are defined in \eqref{def:pppM} -- \eqref{def:pppTZ}.
}

\proof{By Proposition \ref{prop:markedPPP}, \(\Pi^{\lambda*}_h\) is a PPP. The aim is to apply the Mapping Theorem \cite[p.18]{Kingman} for PPPs, which will allow us to conclude that \(F(\Pi^{\lambda*}_h)\) is a PPP and analyze its mean measure. The key assumption of the Mapping Theorem requires that the induced measure \(\nu^F\) on the Borel \(\sigma\)-field \(\mathcal{B} \Par{\R_+ \times \Rd}\), defined by the formula:
		\[
		\nu^F \Par{A} := \nu^{\lambda*}_h \Par{F^{-1}\Par{A}}, \quad A \in \mathcal{B} \Par{\R_+ \times \Rd},
		\]
		has no atom. Since the preimage under \(F\) in \eqref{def:pppMapping} of an arbitrary singleton \(\Bracket{(t, x)} \subset \R_+ \times \Rd\) is given by:
		\[
		F^{-1} \Par{\Bracket{(t, x)}} = \Bracket{(t, x', x'') \in \R_+ \times \Rd \times \Rd: x' + x'' = x}
		\]
		it follows by the representation of the measure \(\nu^{\lambda*}_h\) in Proposition \ref{prop:markedPPP} that:
		\[
		\nu^{\lambda*}_h \Par{	F^{-1} \Par{\Bracket{(t, x)}} } = 0
		\]
		Hence the Mapping Theorem \cite[p.18]{Kingman} yields that \(F\Par{\Pi^{\lambda*}_h}\) is a PPP with the mean measure \(\nu^F\) given above. Furthermore, the Restriction Theorem \cite[p.17]{Kingman} implies that the random set \(F\Par{\Pi^{\lambda*}_h} \cap \Par{\R_+ \times \Rd \setminus\! \{0\}}\) is a PPP with a mean measure given by:
		\begin{equation}\label{def:pppNuF}
				A \to \nu^F \Par{A \cap  \Par{\R_+ \times \Rd \setminus\! \{0\}}}, \quad A \in  \mathcal{B} \Par{\R_+ \times \Rd }
			\end{equation}
		In order to prove that the measure in \eqref{def:pppNuF} is given by the formula \eqref{def:meanMeasure2h}, it is sufficient to verify that the following equality holds:
		\begin{equation}\label{eq:nuF}
				\nu^F \Par{[0,t) \times B} = \nu^{\lambda}_{2h} \Par{[0,t) \times B}, \quad \text{for all } t \in [0, \infty) \text{ and } B \in \mathcal{B} \Par{\Rd \setminus\! \{0\}}
			\end{equation}
		With this in mind, define \(f: \Rd \times \Rd \to \Rd\), \(f(x,m) := x+m\) and note that the following identity holds by \eqref{def:Ah} and \eqref{def:Mdh} for any \(z \in \Zd_{2h}\).
		\begin{equation}\label{eq:LambdaMdh}
				\sum_{m \in M^d_h} \lambda \Par{ A^{2h}_z \cap A^h_{z-m}} = \lambda \Par{A^{2h}_z}
			\end{equation}
		with both sides possibly taking infinite value in the case \(z=0\) and finite values otherwise. Furthermore, for any \(t \in [0, \infty)\) and \(C \in \mathcal{B} \Par{\Rd}\) we have:
		\begin{align}
				\nu^F \Par{[0,t) \times C}
				&= \nu^{\lambda*}_h \Par{ [0,t) \times f^{-1}\Par{C}}\nonumber\\
				&= \nu^{\lambda*}_h \Par{ [0,t) \times \Par{ f^{-1}\Par{C} \cap \Par{\Zd_h \setminus\! \{0\} \times M^d_h}  } }\nonumber\\
				&= \nu^{\lambda*}_h \Par{ [0,t) \times  \Bracket{ (s, m) \in \Zd_h \setminus\! \{0\} \times M^d_h: s+m \in C }}\nonumber\\
				&= t \sum_{s \in \Zd_h \setminus \{0\}} \sum_{m \in M^d_h} \One{C}(s+m) \One{\Zd_{2h}}(s+m) \lambda \Par{A^{2h}_{s+m} \cap A^h_s}\nonumber\\
				&= t \sum_{z \in \Zd_{2h} \setminus \{0\}} \One{C}(z)  \sum_{m \in M^d_h} \lambda \Par{A^{2h}_z \cap A^h_{z-m}} + t \One{C}(0) \sum_{m \in M^d_h \setminus \{0\}} \lambda \Par{A^{2h}_0 \cap A^h_{-m}}\nonumber\\
				&= t \sum_{z \in \Zd_{2h} \setminus \{0\}} \One{C}(z)  \lambda \Par{A_z^{2h}} + t \One{C}(0) \sum_{m \in M^d_h \setminus \{0\}} \lambda \Par{A^{2h}_0 \cap A^h_{-m}}\label{eq:nuFDev}
			\end{align}
		where the second equality holds since \(\nu^{\lambda*}_h\) is by Proposition \ref{prop:markedPPP} supported in \(\R_+ \times \Par{\Zd_h \setminus\! \{0\}} \times M^d_h\), the fourth is a consequence of the representation in \eqref{eq:nuhLambdaStart}, the fifth is a change of variable \(z = m+s\) and the sixth follows from \eqref{eq:LambdaMdh}. It now follows directly from \eqref{eq:nuFDev} that identity \eqref{eq:nuF} holds.}

\remark{Note that equality \eqref{eq:nuFDev} in fact describes the mean measure of the PPP \(F \Par{\Pi^{\lambda*}_h}\) in Proposition \ref{prop:markedPPP}, in terms of the L\'evy measure of the process \(X\). Furthermore, even though \(X\) may have infinite activity jumps and the point \((t, 0) \in \R_+ \times \Rd\) may be in the support of \(\nu^F\) for every \(t \in \R_+\), the right-hand side of \eqref{eq:nuFDev} is always finite.}

\subsubsection{Coupling of \(X^{h(\lambda)}\) and \(X^{2h(\lambda)}\)}	
Recall that \(T^h_i\), \(i \in \N\), are the jump times of the chain \(X^{h(\lambda)}\) and define:
\begin{equation}\label{def:nbJumps}
	N_t^{h(\lambda)} := \max \Bracket{i \in \N: T^h_i \leq t}, \quad t \geq 0
\end{equation}
using the convention \(\max \emptyset = 0\), \(N_t^{h(\lambda)}\) is a Poisson process with intensity \(-Q_{00}^{h(\lambda)} = \lambda \Par{\Rd \setminus\! A_0^h}\) counting the number of jumps of \(X^{h(\lambda)}\). For any starting point \(x_0 \in \Zd_{2h}\), it holds:
\[
X^{h(\lambda)}_t = x_0 + \sum_{i=1}^{N_t^{h(\lambda)}} \Delta X^{h(\lambda)}_{T^h_i}, \quad t \geq 0
\]
with the convention \(\sum_{1}^{0} := 0\). The jumps sizes \(\Delta X^{h(\lambda)}_{T^h_i}\), \(i \in \N\), are IID random variables with state space \(\Zd_{h} \setminus\! \{0\}\) and distribution \(\Proba \SquaredBracket{\Delta X^{h(\lambda)}_{T^h_1} = s} = \frac{Q^{h(\lambda)}_{0s}}{-Q^{h(\lambda)}_{00}}\), \(s \in \Zd_{h} \setminus\! \{0\}\).\\
It is clear that the paths of \(X^{h(\lambda)}\) are uniquely determined by the PPP \(\Pi^{\lambda}_h\) defined in Section \ref{subsubsec:couplingOfPoissonProcess} as follows:
\begin{equation}\label{def:XhJumpsDef}
	X^{h(\lambda)}_t = x_0 + \sum_{(T,Z) \in \Pi^{\lambda}_h, T \leq t} \!\!  Z, \quad t \geq 0
\end{equation}

with the convention \(\sum_{\emptyset} = 0\). Since by Proposition \ref{prop:markedPPP}, the PPP \(\widetilde{\Pi}^{\lambda}_{2h}\) has the same law as the PPP \(\Pi^{\lambda}_{2h}\) and the random element \(\Par{\Pi^{\lambda}_{h},\widetilde{\Pi}^{\lambda}_{2h}}\) is well-defined, we can construct a CTMC \(\widetilde{X}^{2h(\lambda)} = \Par{ \widetilde{X}^{2h(\lambda)}_t}_{t \in [0, \infty)}\),
\begin{equation}\label{def:X2hJumpsDef}
	\widetilde{X}^{2h(\lambda)}_t := x_0 + \sum_{(\widetilde{T},\widetilde{Z}) \in \widetilde{\Pi}^{\lambda}_h, \widetilde{T} \leq t} \!\! \widetilde{Z}, \quad t \geq 0
\end{equation}
such that the laws of \(\widetilde{X}^{2h(\lambda)}\) and \(X^{2h(\lambda)}\) coincide and the process \(\Par{X^{h(\lambda)}, \widetilde{X}^{2h(\lambda)}}\) is defined on a single probability space.\\

The difference \(\widetilde{X}^{2h(\lambda)}_t - X^{h(\lambda)}_t\), for any \(t > 0\), takes the form:
\[
\widetilde{X}^{2h(\lambda)}_t - X^{h(\lambda)}_t = \sum_{i=1}^{N^{h(\lambda)}_t} M(T_i, Z_i)
\]
where \(N^{h(\lambda)}_t\) defined in \eqref{def:nbJumps}, counts the jumps of \(X^{h(\lambda)}\), \(T_i\) (resp. \(Z_i\)) is the \textit{i}-th jump time (resp. size) of \(X^{h(\lambda)}\) and \(M(T,Z), (T,Z) \in \Pi^{\lambda}_h\), is defined in \eqref{def:pppM} -- \eqref{def:pppTZ}. By \eqref{def:pppM}, the mark \(M(T_i Z_i)\) (for any \(i \in \N\)) does not depend on the jump time \(T_i\) but only on the jump size \(Z_i\). Since \(Z_i\) are IID, \(M(T_i, Z_i)\), \(i \in \N\), conditionally independent by \eqref{def:pppTZ}, and \(N^{h(\lambda)}_t\) depends only on jump times \(T_i\), \(i \in \N\), by \eqref{def:nbJumps}, we have established the following result:

\theorem{\label{thm:coupling}
	The process \(\Par{X^{h(\lambda)}, \widetilde{X}^{2h(\lambda)}}\) given by \eqref{def:XhJumpsDef} -- \eqref{def:X2hJumpsDef}, is a coupling of the CTMCs \(X^{h(\lambda)}\) and \(X^{2h(\lambda)}\). Moreover, for any \(T > 0\), the difference \(X^{h(\lambda)}_T - \widetilde{X}^{2h(\lambda)}_T\) takes the form:
	\[
	\widetilde{X}^{2h(\lambda)}_T - X^{h(\lambda)}_T= \sum_{i=1}^{N} J_i
	\]
	where the random variable \(N\) is Poisson distributed with parameter \(T \lambda \Par{\Rd \setminus\! A_0^h}\) and \(J_i\) (\(i \in \N^*\)) are IID random variables in \(\Rd\) independent of \(N\), satisfying \(\NormSup{J_i} \leq h\) for all \(i \in \N^*\). More precisely, conditional on the size of the \textit{i}-th jump of \(X^{h(\lambda)}\) being \(s \in \Zd_{h}\), the distribution of \(J_i\) is given by the function \(p^{\lambda}(s, \cdot)\) defined in \eqref{def:pmf} above.}

\remark{\label{rem:couplingThm}The coupling in Theorem \ref{thm:coupling} implicitly prunes the jumps of \(X^{h(\lambda)}\), to obtain the jumps of \(\widetilde{X}^{2h(\lambda)}\), as follows (cf. Figure \eqref{fig:gridExample}): for any \((T,Z) \in \Pi^{\lambda}_h\) we have:
	\[
	(\widetilde{T}, \widetilde{Z}) := (T, Z + M(T,Z)) \in \widetilde{\Pi}^{\lambda}_{2h} \iff Z + M(T,Z) \neq 0
	\]
	where random variables \(M(T,Z), (T,Z) \in \Pi^{\lambda}_h\), are specified in \eqref{def:pppM} -- \eqref{def:pppTZ}. This implies that for any jump \(Z\) of \(X^{h(\lambda)}\), such that \(Z + M(T,Z) = 0\), there is no jump of the chain \(\widetilde{X}^{2h(\lambda)}\).
}

\section{Numerical data and other results}\label{appendix:numData}

\subsection{Benchmark against the series representation of Section \ref{NumApp:subsec:SeriesRepresentation}}\label{appendix:ModelParameters:Series}
In the following tables, prices are truncated to the \nth{6} digit.

\begin{table}[H]
	\centering
	\dataBenchmarkSeries{\DataForPaper/series/std_asian.csv}
	\caption{Data for Asian call options with strong tail dependence}
\end{table}
\begin{table}[H]
	\centering
	\dataBenchmarkSeries{\DataForPaper/series/wtd_asian.csv}
	\caption{Data for Asian call options with weak tail dependence}
\end{table}
\begin{table}[H]
	\centering
	\dataBenchmarkSeries{\DataForPaper/series/std_bestof.csv}
	\caption{Data for Best-of call options with strong tail dependence}
\end{table}
\begin{table}[H]
	\centering
	\dataBenchmarkSeries{\DataForPaper/series/wtd_bestof.csv}
	\caption{Data for Best-of call options with weak tail dependence}
\end{table}

\subsection{Models Parameters}\label{appendix:ModelParameters}
The models described in this section have the following values for the spot \(S_0\), interest rate \(r\) and dividend \(q\):
\begin{itemize}
	\item \(S_0=100\)
	\item \(r=0.02\)
	\item \(q=0.0\)\\
\end{itemize}

\makebox[\linewidth]{
{\small
	\centering
	\begin{tabular}{|l|l|l|c|c|c|}
		\hline
		Model   &  Name		        & Parameters 		                  	            		      & \(\beta\) & Finite activity & Finite variation  \\ \hline \hline
		Cai-Kou &\textit{HEM}		& \(p=0.6, \,\sigma=0.05, \,\eta_1=20, \,\eta_2=25, \,\lambda=3\) &  0        & \ding{51} 	    & \ding{51} 		\\
		Variance-Gamma & \textit{VG}& \(\sigma=0.1, \,\nu = 0.06, \,\theta=0.1\)					  &  0	      & \xmark          & \ding{51}		    \\
		CGMY    &\textit{\cgmyO}	& \(c=1.23,\; \,g=15, \,m=20, \,y=0.2\)	   					      &  0.2      & \xmark    	    & \ding{51} 		\\
		CGMY    &\textit{\cgmyI}	& \(c=0.70,\; \,g=15, \,m=20, \,y=0.4\)	   					   	  &  0.4      & \xmark    	    & \ding{51} 		\\
		CGMY    &\textit{\cgmyIV}	& \(c=0.025, \,g=2,\;\; m=4,\;\:\, y=1.1\)	   				   	  &  1.2      & \xmark   	    & \xmark 			\\
		CGMY    &\textit{\cgmyVII}	& \(c=0.007, \,g=2,\;\; m=4,\;\:\, y=1.5\)	   				   	  &  1.5      & \xmark   	    & \xmark 			\\
		\hline
	\end{tabular}
}
}

\bibliography{mlmc_ams_formatting.bbl}

\section*{Acknowledgments}
A. Mijatovi\'c is supported by EPSRC under grants EP/V009478/1 and EP/P003818/2, by the Turing Fellowship funded by the Programme on Data-Centric Engineering of Lloyd’s Register Foundation, and by The Alan Turing Institute under the EPSRC grant EP/N510129/1.

\end{document}